\documentclass[twoside,12pt]{article}
\usepackage{times}
\usepackage{epsfig}
\usepackage{marvosym}
\usepackage{lineno}
\usepackage{amssymb}
\usepackage{url}

\def\Journal#1#2#3#4{{#1} {#2} (#4) #3 }

\def\NPB{{\em Nucl. Phys.} B}

\def\PL{{\em Phys. Lett.}}
\def\PRL{\em Phys. Rev. Lett.}

\def\PRD{{\em Phys. Rev.} D}
\def\PRC{{\em Phys. Rev.} C}

\def\RMP{{\em Rev. Mod. Phys.}}

\def\JINST{\em J. Instrumentation}
\def\epj{\em Eur. Phys. J.}
\def\NIM{\em Nucl. Instr. and Meth.}
\def\ieee {\em IEEE Trans. Nucl. Sci.~}
\def\APF{\em Astropart. Phys.}
\def\nat{\em Nature}
\def\EA{\em Exp. Astr.}
\def\JPCS{\em J. Phys. Conf. Ser.}
\def\RSI{\em Rev. Sci. Instr.}
\def\CP{\em Chin. Phys.}

\def\hb{\hfill\break}

\def\etal{{\it et al.}} 
\def\eg{{\it e.g.,~}}
\def\ie{{\it i.e.,~}}
\def\cf{{\it cf.~}}
\def\etc{{\it etc.}}
\def\vs{{\it vs.~}}

\newcommand{\be}{\begin{equation}}
\newcommand{\ee}{\end{equation}}
\newcommand{\bea}{\begin{eqnarray}}
\newcommand{\eea}{\end{eqnarray}}

\begin{document}

\title{ \vspace{1cm} New Developments in Calorimetric Particle Detection}
\author{Richard Wigmans\\
\\
Department of Physics and Astronomy, Texas Tech University, Lubbock, U.S.A.}
\maketitle
\begin{abstract} 
In nuclear, particle and astroparticle physics experiments, calorimeters are used to measure the properties of particles with kinetic energies that range from a fraction of 1 eV to 10$^{20}$ eV or more. These properties are not necessarily limited to the energy carried by these particles, but may concern the entire four-vector, including the particle mass and type.
In many modern experiments, large calorimeter systems play a central role, and this is expected to be no different for experiments that are currently being planned/designed for future accelerators.

In this paper, the state of the art as well as new developments in calorimetry are reviewed. The latter are of course inspired by the perceived demands of future experiments, and/or the increasing demands of the current generation of experiments, as these are confronted with, for example, increased luminosity. These demands depend on the particles to be detected by the calorimeter. In electromagnetic calorimeters, radiation hardness of the detector components is a major concern. 
The generally poor performance of the current generation of hadron calorimeters is considered inadequate for future experiments, and a lot of the R\&D in the past decade has focused on improving this performance. The root causes of the problems are investigated and different methods that have been exploited to remedy this situation are evaluated.

In the past two decades, experiments in astroparticle physics have started to make major contributions to our fundamental understanding of physics and of a variety of processes that are inaccessible in laboratory experiments here on Earth. These experiments typically make use of calorimetric particle detection.
At the extreme low end of the energy spectrum, ingenious instruments are used to study phenomena involving energy transfers of the order of 1 eV using calorimetric methods. In separate sections, some salient aspects of this work are reviewed. 

\vskip -5mm
\end{abstract}
%

\section{Introduction}

In the past half-century, calorimeters have become very important components of the detector system at almost every experiment in high-energy physics. This is especially true for $4\pi$ experiments at high-energy particle colliders, such as the S$\bar{p}p$S, LEP and the LHC at CERN, the Tevatron at Fermilab and RHIC at Brookhaven. Experiments at proposed future colliders such as the FCC (CERN), CEPC (China) and ILC (Japan) would be designed around a powerful central calorimeter system, should these proposed projects be realized. Calorimetric particle detection also plays a crucial role in many astroparticle physics experiments. 

A calorimeter is a detector in which the particles to be detected are completely absorbed. The detector provides a signal that is a measure for the energy deposited in the absorption process. In {\sl homogeneous} calorimeters, the entire detector volume may contribute to the signals. In {\sl sampling} calorimeters, the functions of particle absorption and signal generation are exercised by different materials, called the {\sl passive} and the {\sl active} medium, respectively. Almost all calorimeters operating in collider experiments are of the latter type, with the PbWO$_4$ crystal calorimeter of CMS as a notable exception. The passive medium is usually a high-density
material, such as iron, copper, lead or uranium. The active medium generates the light or the electric charge
that forms the basis for the signals from such a calorimeter.
In some non-accelerator experiments, the calorimeter is also the {\sl source} that generates the particles
to be detected. Examples include the large water \v{C}erenkov counters built to search for proton decay and the high-purity $^{76}$Ge crystals \cite{Ge76} or $^{136}$Xe detectors \cite{Xe136} that are being used to study $\beta \beta$ decay.

Reasons for the increased emphasis on calorimetric particle detection in modern experiments include:
\begin{itemize}
\item Calorimeters can provide important information on the particle collisions, in particular information on the {\sl energy flow}
in the events (transverse energy, missing energy, jet production, \etc) 
\item Calorimeters can provide this information {\sl very fast}, almost instantaneously. In modern experiments, \eg at the LHC, it has become possible to decide whether an event is worth retaining for offline inspection on a time scale of the order of $10^{-6}$ seconds. Since the LHC experiments have to handle event rates at the level of 10$^9$ each second, this triggering possibility is a crucial property in these experiments.
\item Calorimeter data can be very helpful for {\sl particle identification}.   
\item Important aspects of the calorimeter performance, such as the energy and position resolutions, tend to {\sl improve with energy}.
\end{itemize}

Calorimetric detection of $\gamma$s and electrons has a long tradition, which goes back to the early days of nuclear spectroscopy,
when scintillating crystals such as NaI(Tl) were the detectors of choice. In high-energy physics, detection of electromagnetic (em) showers 
is nowadays routinely performed with an energy resolution at the 1\% level, both in homogeneous \cite{bgo} and sampling \cite{na48} calorimeters. 
The success of experiments at a future $e^+e^-$ collider will also depend critically on the quality of the hadron calorimetry. Unfortunately, the performance of hadron calorimeters leaves much to be desired.

In this paper, I review the state of the art in the field of calorimetric particle detection and describe recent developments intended to improve the performance, with the needs of envisaged future particle physics experiments in mind. These needs also, and especially, include the harsh conditions in terms of radiation levels and event rates faced by the LHC experiments when the luminosity of the collider is substantially increased (2024).
The discussion concentrates mainly on calorimeters that are used in experiments at particle accelerators. 

In Section 2, some basic facts about the absorption of high-energy particles in matter are presented, and their consequences for calorimetry explained.
Electromagnetic shower detectors are covered in Section 3, hadronic ones in Sections 4 and 5. In Section 6, I discuss some 
(unexpected) practical features that have come to light as a result of the operating experience with large calorimeter systems, in the hope that designers of future experiments will take notice. Section 7 is dedicated to the PFA\footnote{Particle Flow Analysis.} approach, which has been adopted as a remedy to improve the quality of jet detection in some experiments. To conclude this part of the review, Section 8 contains some general comments about energy resolution, which is often considered one of the most important performance characteristics of calorimeters.\hb
The very large calorimetric detectors that are used in non-accelerator experiments, or in long-baseline neutrino experiments installed at hundreds of kilometers from the accelerator that produces the neutrino beam, are subject to boundary conditions and requirements that are completely different from those in a colliding-beam environment. Recent developments in that field are discussed in Section 9. The ingenious instruments that use calorimetric methods to study phenomena that either involve energy transfers of the order of 1 eV, or require truly exceptional energy resolution, are briefly reviewed in Section 10. In Section 11, I conclude with my assessment of the current status of this important experimental technique and an outlook on its future.

\section{The Basics of Calorimetry\hb
\it \large ``Everything about calorimetry is obvious, once you understand it"}

The absorption of high-energy particles in a block of matter proceeds typically through a complicated process that is usually referred to as {\sl shower development}. Eventually, the entire energy carried by the absorbed particle, including its rest energy (except when that particle is a proton, neutron or electron)
is used to heat up the absorber\footnote{The Latin word {\sl calor} means heat.}. However, even for the highest-energy particles studied in (astro)particle physics experiments, the resulting 
increase in temperature is typically negligibly small\footnote{1 calorie  is equivalent to $\sim 10^7$ TeV. However, some cryogenic calorimeters are based on
phenomena that are associated with an increase in temperature (Section 10).}. Calorimeters used in these experiments thus exploit other features of the absorption process to measure the properties of the showering particles, such as the production of light or electric charge.

In the absorption process, the energy carried by the incoming particle is eventually distributed in a typically large number of intermediate steps to electrons and (constituents of) nuclei that are part of the absorber structure. In this process, the fundamental laws of physics, \eg conservation of energy, electric charge, baryon and lepton number are rigorously respected. This may seem like a trivial statement, but it leads in practice to fundamental and substantial differences between the absorption of high-energy protons and pions (Figure \ref{pdif}), which at first sight may not seem all that trivial.

Eventually, all the available energy of the incoming particle is thus shared among a very large number of shower particles, each of which carries so little kinetic energy that all that is left for them to do is to ionize the medium they traverse or (in the case of neutrons) scatter elastically off atomic nuclei before being captured by one.
And even though calorimeters are often intended to measure energy deposits at the level of 10$^9$ eV and up, their performance
is in practice determined by what happens at the MeV, keV and sometimes eV levels, simply because the particles that carry these low energies are so abundantly produced in the absorption process.

\begin{figure}[hbt]
\epsfysize=8.5cm
\centerline{\epsffile{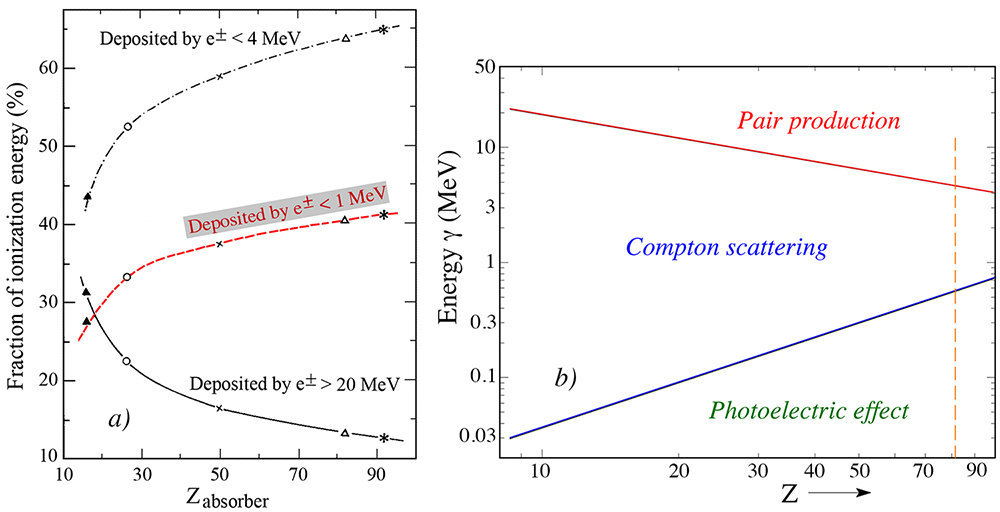}}
\caption{\footnotesize
The composition of em showers. Shown are the
percentages of the
energy of 10 GeV electromagnetic showers deposited through shower particles
with energies below 1 MeV (the dashed curve), below 4 MeV (the dash-dotted curved) or above 20 MeV (the solid
curve), as a function of the $Z$ of the absorber material ($a$). 
The energy domains in which photoelectric effect, Compton scattering and pair production are the 
most likely processes through which $\gamma$s are absorbed, as a function of the $Z$ value of the absorber material ($b$). Results of EGS4 simulations \cite{Wig17}.}
\label{showen}
\end{figure}

\subsection{\it Electromagnetic showers}

To illustrate the importance of the last stages of the shower development, Figure \ref{showen}a shows what happens to the energy carried by a 10 GeV electron that is absorbed in a block of material,
as a function of the $Z$-value of that material. If we take as an example lead, which is often used as absorber material in calorimeters intended for detecting such electrons, it turns out that almost 70\% of the entire energy is deposited by shower electrons that carry less than 4 MeV kinetic energy. Electrons of less than 1 MeV even carry 40\% of the total energy, which means that there must thus be {\sl at least} 4,000 such electrons, and probably many more.
Figure \ref{showen}b shows that these soft electrons are almost exclusively the result of Compton scattering and the photoelectric effect, since these processes are the most likely ones to occur when $\gamma$s with energies below 4 MeV interact with the absorber material.

To understand the peculiarities of electromagnetic calorimeters, one thus needs to understand the relevant characteristics of Compton scattering and the photoelectric effect. I mention three examples.
\begin{enumerate}
\item The {\sl radiation length} ($X_0$) was introduced as a parameter to describe electromagnetic shower development in a material independent way.
Longitudinal shower profiles should thus be material independent when expressed in terms of $X_0$.
Figure \ref{emlong} shows that this is by no means the case.

\begin{figure}[hbt]
\epsfysize=6.0cm
\centerline{\epsffile{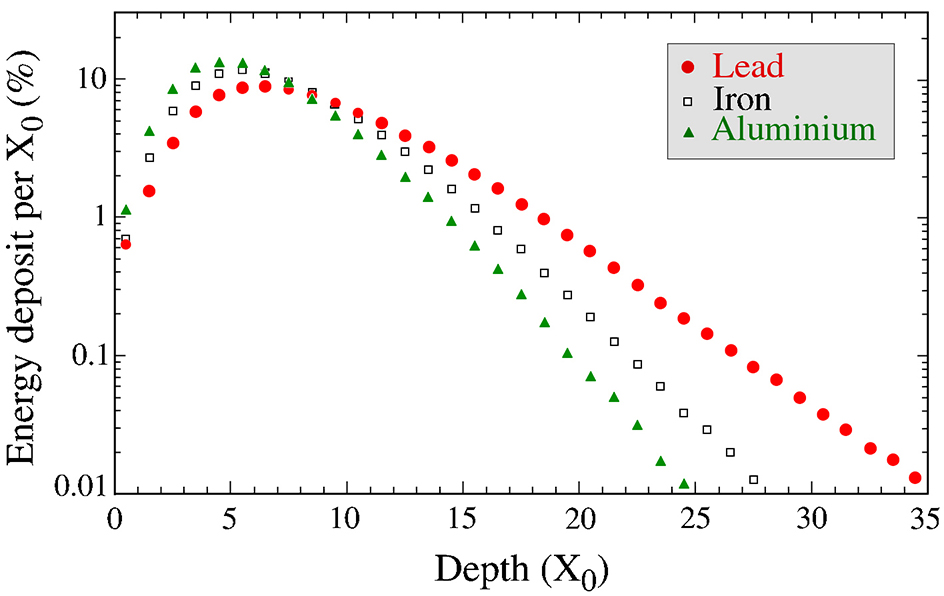}}
\caption{\footnotesize
Energy deposit as a function of depth, expressed in radiation lengths, for 10 GeV electron showers developing in aluminium, iron and lead.
Results of EGS4 calculations \cite{Wig17}.}
\label{emlong}
\end{figure}
Figure \ref{emcont}a also illustrates the inadequacy of this parameter for determining the calorimeter depth needed to contain em showers at the level of 99\%. For 10 GeV
electrons, this depth ranges from $16 X_0$ to $22 X_0$, depending on the chosen absorber material.
These discrepancies are a result of the fact that $X_0$ is defined based on the properties of the high-energy component of the showers, where electrons predominantly lose energy by radiation (Bremsstrahlung) and $\gamma$s interact by producing $e^+e^-$ pairs. These processes do not play an important role in the last stages of the shower development during which, as shown above, a major fraction of the total energy is deposited.
\item 
There is no preferential direction for the electrons produced in Compton scattering and by the photoelectric effect.
Because of the dominant role of these processes in em shower development, 
a large fraction of {\sl the shower particles thus travel in random directions}
with respect to the particle that initiated the shower and whose properties are being measured
with the calorimeter (Figure \ref{egsdist}). This means that the orientation of the active layers of 
a sampling calorimeter can be chosen as desired,
without serious implications for the calorimetric performance of the 
detector.\hb
\begin{figure}[hbt]
\epsfysize=6.5cm
\centerline{\epsffile{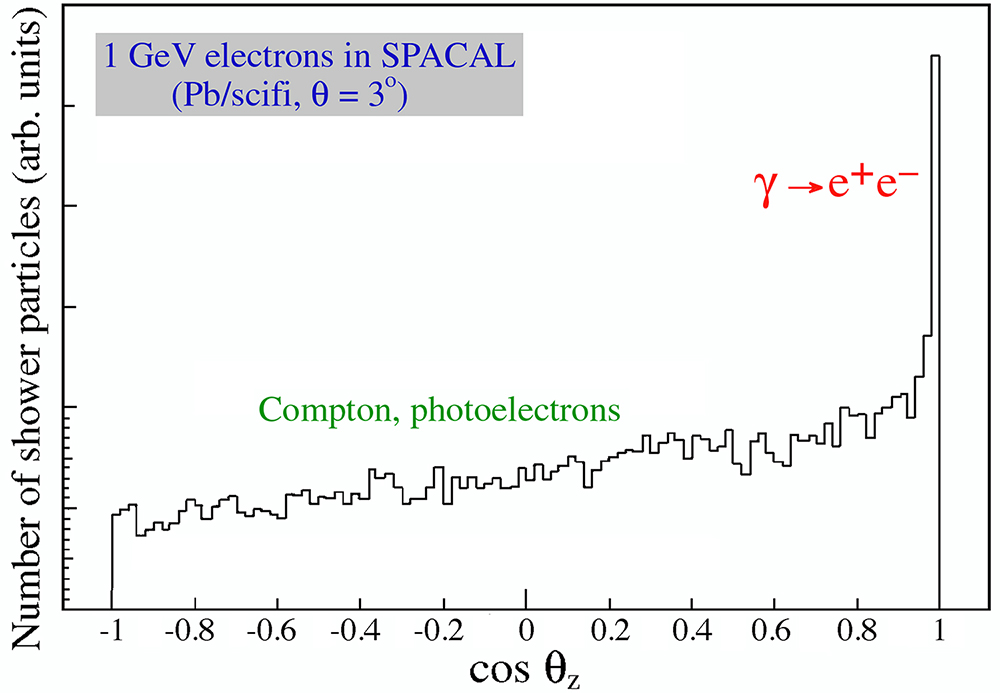}}
\caption{\footnotesize
Angular distribution of the shower particles (electrons and positrons) 
through which the energy of a 1 GeV electron is absorbed in a lead-based fiber calorimeter.
The angle between the direction of the shower particles and the fiber axis ($\theta_z$) was chosen to be 3$^\circ$ in these
EGS4 Monte Carlo simulations \cite{Aco90}.}
\label{egsdist}
\end{figure}
The first generation of sampling calorimeters used in particle physics experiments
consisted almost exclusively of instruments of the ``sandwich type,'' \ie detectors
composed of alternating layers of absorber and active material, oriented
perpendicular to the direction of the particles to be detected. Although
this, from an intuitive point of view, may seem to be the only correct choice, the R\&D
with fiber calorimeters has proven that there is no need for such a
geometry \cite{Liv95}.\hb 
The notion that the active calorimeter layers do not necessarily have to be oriented
perpendicular to the direction of the incoming particles has had a considerable
impact on the design of detectors for new experiments. Other orientations may offer 
considerable advantages in terms of detector hermeticity, readout, granularity, \etc~Apart from the ``spaghetti'' type of calorimeters built for a number
of experiments, this development is also illustrated by the liquid-argon
calorimeters with an ``accordion'' geometry and the 
tile/fiber hadron calorimeter of the ATLAS experiment at the LHC \cite{Aad08}.
\item 
{\sl The cross section for the photoelectric effect is extremely $Z$-dependent ($\propto Z^5$)}. This has very important consequences 
for sampling calorimeters that consist of high-$Z$ absorber material and low-$Z$ active layers. Low-energy $\gamma$s produced in the shower development
will, for all practical purposes, {\bf only} interact in the absorber material, and the photoelectrons produced in this process will only contribute to the signals
if they are produced very close to a boundary layer. In practice, they are much less efficiently sampled than the high-energy electron/positron pairs produced in the early stages of the shower development. As a result, the sampling fraction in such calorimeters decreases as the shower develops in depth (see Figure \ref{emipz}). Also, the fact that $e/mip \ne 1$ in such calorimeters\footnote{See Section 4.2 for the definition of $e/mip$.} is the result of this phenomenon. Some consequences for calorimetry are discussed in Section 6.1.
\end{enumerate}

\begin{figure}[hbt]
\epsfysize=5.8cm
\centerline{\epsffile{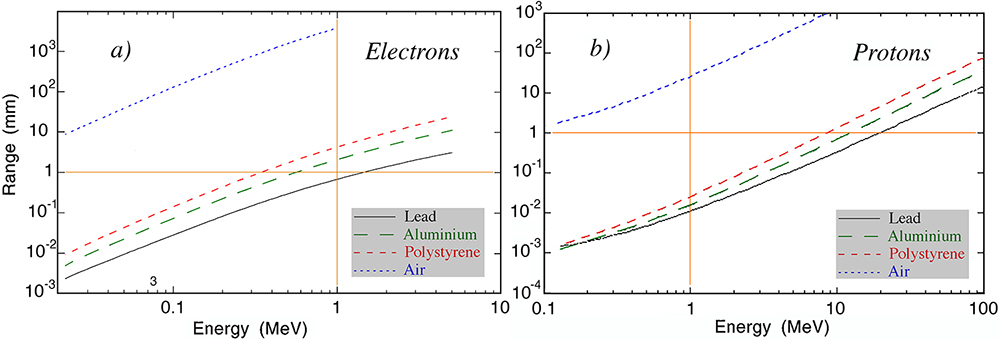}}
\caption{\footnotesize
Average range of electrons ($a$) and protons ($b$) in various absorber materials, as a function of energy \cite{Ber93}.}
\label{range}
\end{figure}

The fact that a large fraction of the em shower energy is deposited by very soft electrons also has implications for the energy resolution of em sampling calorimeters. This may be concluded from Figure \ref{range}a, which shows the range of electrons with energies from 20 keV to 10 MeV in several materials. For example, 1 mm of aluminium is enough to stop electrons with a kinetic energy up to 590 keV. For silicon, a material that is increasingly being considered as active material for sampling calorimeters, electrons with energies up to 115 keV stop in 100 $\mu$m thick layers, while 330 keV electrons have a range of 500 $\mu$m \cite{Ber93}.
If one replaced in the same absorber structure 500 $\mu$m thick Si sensors by 100 $\mu$m thick ones, shower electrons that escaped from the absorber layers carrying a kinetic energy between 
115 keV and 330 keV would thus always produce the same signals in the thicker sensors, while the signals in the thinner ones would depend on the angle of incidence. Because of this additional source of fluctuations, the energy resolution would deteriorate as the sensors were made thinner and thinner. The fact that this does indeed happen (Section 3.1) is an indication of the important contributions of shower electrons from this very-low-energy bracket.

\subsection{\it Hadron showers}

All the comments made in the previous subsection are primarily (and most of the time only) important for sampling calorimeters. However, for a variety of practical reasons, these are in practice the only calorimeters considered for future applications in $4\pi$ collider experiments. That is especially true for calorimeters intended for hadron detection. Also in hadron showers, the energy of the incoming particle is eventually deposited by a very large number of low-energy shower particles. There are, however, several major differences with em showers. The most important difference, at least for what concerns the implications for calorimetry, is the fact that some fraction of the initial energy is used to break up atomic nuclei in the numerous nuclear interactions that take place in the absorption process. This fraction, usually referred to as the {\sl invisible energy} component since it does not contribute to the calorimeter signals, varies wildly from one event to the next. This has two important consequences: 
\begin{enumerate}
\item The calorimeter signal is typically substantially smaller compared to that from showers initiated by electrons or $\gamma$s of the same energy.
\item Event-to-event variations in the calorimeter signal are much larger than for em showers of comparable energy, since this phenomenon has no equivalent in the latter.
\end{enumerate}
These consequences are exacerbated by the fact that in the absorption of high-energy hadrons typically also some fraction of the energy is used for the production of $\pi^0$s. These decay almost instantaneously into two $\gamma$s, which develop em showers. The problems resulting from these features for calorimetry are discussed in Section 4.

In order to understand the properties of hadron calorimeters, it is also important to realize that most of the energy in the {\sl non-$\pi^0$} shower component is
deposited by nucleons released in the nuclear reactions. The contribution of short-lived hadrons such as pions and kaons is relatively minor. This was already concluded from pre-GEANT Monte Carlo simulations more than 30 years ago. Gabriel \cite{Gab85} found, for example, that in the absorption of 5 GeV pions in lead, on average, 13\% of the energy contained in the {\sl non-$\pi^0$} shower component is deposited in the form of ionization by pions \vs 33\% by protons. The invisible energy component represents 39\% of the energy and neutrons carry the remaining 15\%. In lead nuclei, the average binding energy per nucleon amounts to $\sim 8$ MeV. Based on these numbers, one would thus expect that each nucleon released in nuclear breakup carries, on average, 
$8\times (33+15)/39 \approx 10$ MeV kinetic energy. However, that conclusion is overly simplistic. 

Neutrons outnumber protons by a factor of ten in the nuclear reactions in lead absorber. This is mainly because a proton faces a Coulomb barrier of $\sim 13$ MeV on its way out. The (spallation) protons that do escape carry typically a much larger kinetic energy than the mentioned average value. On the other hand, the kinetic energy of the overwhelming majority of the neutrons is much less than 10 MeV. They are produced in the {\sl evaporation stage} of the nuclear reactions, in which the highly excited compound nucleus gets rid of most of its excitation energy by releasing neutrons. The spectrum of such neutrons resembles a Boltzmann-Maxwell distribution with a temperature of 2 -- 3 MeV. 
In lower-$Z$ absorber materials, the discrepancy between the numbers and kinetic energies of the protons and neutrons produced in the absorption process are  smaller. For example, in iron the average binding energy per nucleon and the Coulomb barrier are $\sim 9$ MeV and $\sim 6$ MeV, respectively. 

Just as the last stages of em shower development are important for the characteristics of electromagnetic calorimeters, the signals of hadron calorimeters are determined by the processes in which the numerous nucleons produced in the shower development are absorbed in the calorimeter structure.
The fastest spallation protons may induce nuclear reactions themselves. For example, the nuclear interaction length in iron is 16.8 cm, which corresponds to the range of a 430 MeV proton. Protons with a kinetic energy less than that are more likely to lose that energy by ionizing the medium they traverse. The neutrons may cause nuclear reactions down to energies of a few MeV, and they create in such processes additional nucleons, plus $\gamma$ rays from the de-excitation of the compound nucleus. 

The low-energy evaporation neutrons, which are the most abundantly produced nucleons, lose their kinetic energy either by elastic or inelastic scattering off nuclei in the absorber medium. In inelastic scattering, some fraction of the neutron's kinetic energy is used to bring the nucleus into an excited state, from which it returns to the ground state through $\gamma$-ray emission. Elastic scattering off high-mass absorber nuclei is a very inefficient way to lose energy. However, if the calorimeter contains even a very small fraction of hydrogen atoms, the neutrons will find those and transfer most of their kinetic energy in elastic neutron-proton collisions. Compensating calorimeters make use of this phenomenon to selectively boost the calorimeter response to the non-$\pi^0$ shower component (Section 5.1). 

Eventually, the low-energy neutrons either escape the detector or are captured by a nucleus. In the latter process, the binding energy originally needed to release them is gained back. However, the thermalization that is needed before the cross section becomes adequately large takes such a long time ($\mu$s) that this energy is in practice not useful for calorimetry \cite{Kru92}.
\vskip 2mm
As a final remark about the physics of shower development, it should be stressed that the particles to be detected by the calorimeters do not know or care that the detector system may consist of separate sections that are optimized for em or hadronic showers. In practice, hadron showers start most of the time in the em section of the calorimeter system and deposit often a sizable fraction of their energy in that section. Differences between the characteristic features of the em and hadronic section may lead to a situation in which the quality of the calorimetric detection of single hadrons or jets is worse for the calorimeter system as a whole than for the hadronic section in stand-alone mode. The reasons for this, as well as examples taken from practice, are presented in Section 4.1.

\subsection{\it The calorimeter signals}
Calorimeter signals consist typically of electric pulses produced either directly by collecting electric charge (\eg in liquid-argon calorimeters or resistive plate chambers), or by photons that produce such pulses as a result of the photoelectric effect (\eg in photomultiplier tubes, silicon photomultipliers or photodiodes). 
Since calorimeters are intended for measuring energy, the total signal should be a good measure of that energy. In practice, there are a number of {\sl instrumental effects} that may spoil the relationship. These effects should be clearly distinguished from the {\sl physics} effects described in the previous subsections (invisible energy, inefficient sampling of certain types of shower particles, or in late stages of the shower development, \etc)

One of the most worrisome instrumental effects is {\sl non-linearity}, \ie the fact that the size of the signal is not proportional to the amount of energy that causes it. We should distinguish two types of non-linearity, which have quite different consequences for the calorimeter performance.
\begin{figure}[hbt]
\epsfysize=7cm
\centerline{\epsffile{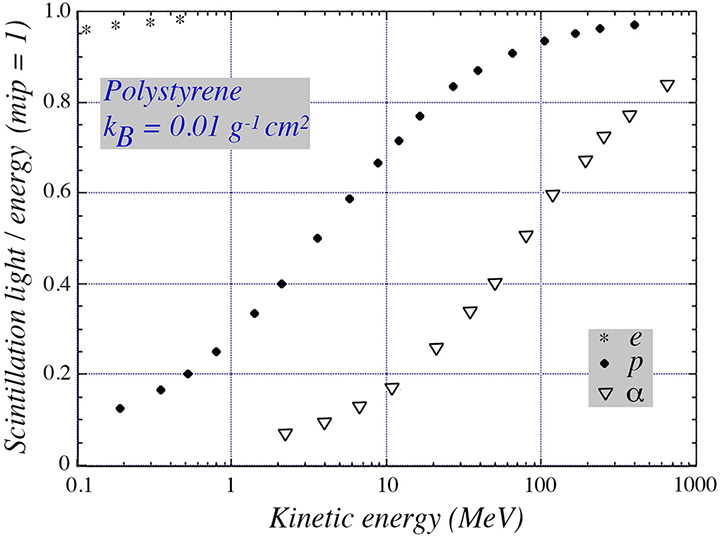}}
\caption{\footnotesize
Suppression of the scintillation light signals for densely ionizing particles because of quenching effects. Results are given for electrons, protons and $\alpha$ particles
in polystyrene as a function of the kinetic energy of the particles.}
\label{birks1}
\end{figure}

The first type of non-linearity is caused by {\sl quenching} effects in the signal producing medium. It affects the signals from densely ionizing shower particles.
Detector media that are susceptible to this effect include scintillators and noble liquids. The latter are based on ionization charge drifting over a rather long distance to 
an electrode that converts the collected charge into a measurable signal. If the ionization density in the liquid is large, then the probability for recombination of electrons and ions into atoms along the track increases and the charge collected at the electrode decreases as a result. A similar phenomenon occurs in scintillators, where the effects are well described by Birks' law:
\begin{equation}
{dL\over dx}~=~S{{dE/dx}\over {1 + k_{\rm B}\cdot dE/dx}}
\label{birk}
\end{equation}
where $L$ is the amount of light produced by a particle of energy $E$, $S$ a proportionality
constant and $k_{\rm B}$ a
material property known as Birks' constant \cite{Bir64}.
This constant is typically of the order of
0.01 g cm$^{-2}$ MeV$^{-1}$, whereas the specific ionization ($dE/dx$, also known as the stopping power) of a minimum ionizing particle (mip) is of the order of
1 MeV g$^{-1}$ cm$^2$.
Figure \ref{birks1} illustrates the effects of this quenching in polystyrene-based scintillators. It shows that signals from 10 MeV protons are suppressed by $\sim 30\%$ compared to the signals from mips. For 1 MeV protons, the signal reduction is a factor of three. The figure also shows that the effects are much larger for $\alpha$ particles, and negligible for electrons. 
\begin{figure}[htbp]
\epsfysize=7cm
\centerline{\epsffile{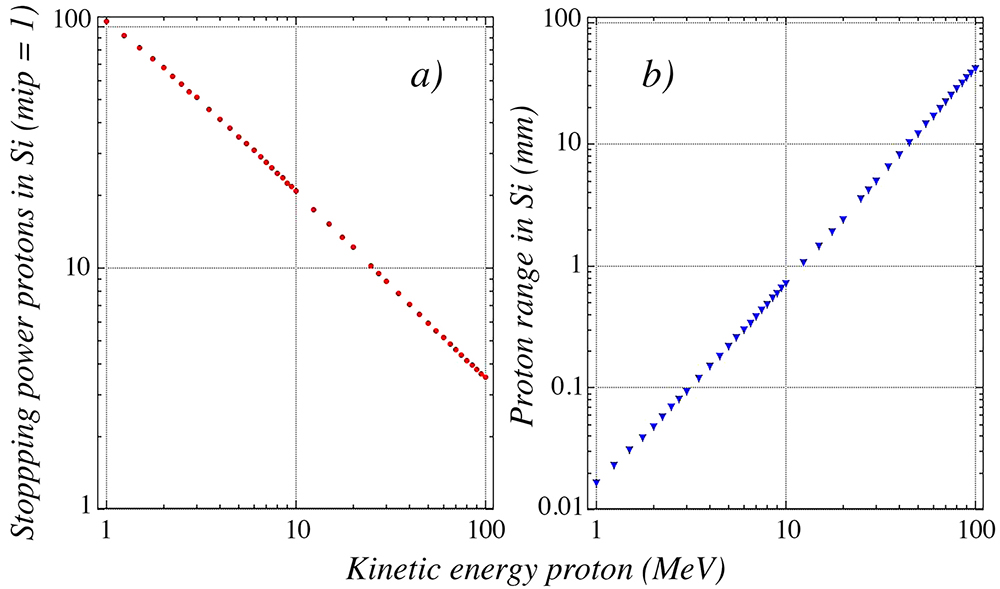}}
\caption{\footnotesize
The stopping power, normalized to the mip value ($a$) and the range ($b$) of densely ionizing protons in silicon \cite{Ber93}.}
\label{pinSi}
\end{figure}

Such quenching effects do not play a role in certain other detector media, in particular silicon semiconductors and gases. This may lead to some very troublesome effects in sampling calorimeters that use such detectors as active material. Figure \ref{range}b shows that 1 MeV protons have a range of about 2 cm in air. The range in gases used in proportional wire chambers is very similar. Such a (recoil) proton may be produced in elastic $n - p$ scattering in that gas and deposit its entire 1 MeV energy inside the wire chamber. On the other hand, a mip traversing the same wire chamber typically deposits at least three orders of magnitude less energy in 
it \cite{Ber93}. 
This leads to a phenomenon known as the {\sl Texas tower effect}, in which one individual shower particle can mimick a local deposit of an enormous amount  
of energy. It occurs in calorimeters with a very small sampling fraction and a non-quenching active medium. In Section 6.2, several practical examples of this effect are discussed. These examples include calorimeters in which thin silicon sensors are used to produce the signals.

Figure \ref{pinSi}a shows that the stopping power of silicon is larger than that for mips by a factor ranging from 20 to 100 in the relevant energy range (10 - 1 MeV) of protons produced in hadron shower development.  
The amplification factor of a proton signal in such calorimeters is also determined by the range of these particles in the sensors (Figure \ref{pinSi}b). The thinner the silicon, the more prone the detector will be to local spikes in the measured energy deposit pattern.
\vskip 2mm 
The Texas tower effect does not play a role in calorimeters that use plastic scintillator or liquid-argon as active material, because of the combined effects of signal quenching and the
larger thickness of the active layers. And since densely ionizing shower particles represent a more or less energy independent fraction of the (non-$\pi^0$) shower component, the quenching effects do not adversely affect the performance characteristics of such calorimeters. That is different for the second type of non-linearity, which is the result of signal {\sl saturation}. Saturation is a consequence of the intrinsic limitations of the device that generates the signals. 

\begin{figure}[htbp]
\epsfysize=6cm
\centerline{\epsffile{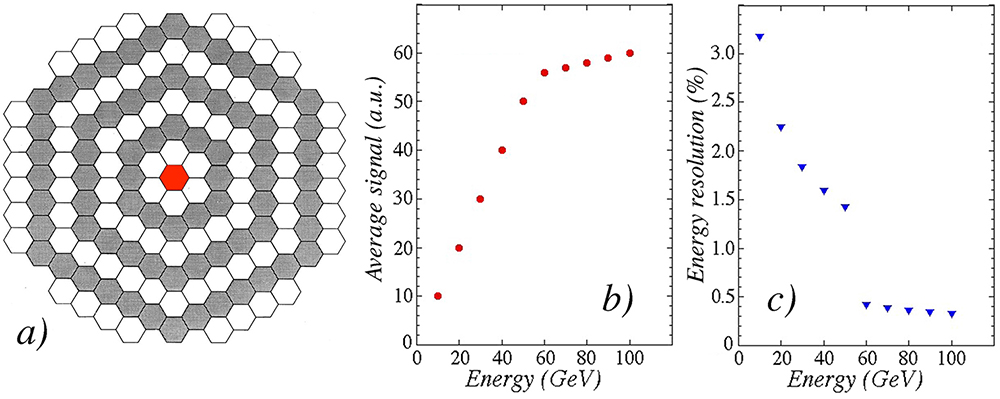}}
\caption{\footnotesize Saturation effects in one of the towers of the SPACAL calorimeter ($a$). Shown are the average signal ($b$) and the energy resolution ($c$) as a function of energy, measured when a beam of electrons was sent into this tower.}
\label{spasat}
\end{figure}

Let me use data from one of my own experiments to illustrate the effects of signal saturation.
The SPACAL calorimeter (Figure \ref{spasat}a) consisted of 155 hexagonal towers. The scintillation light produced in each of these towers was read out by a photomultiplier tube (PMT). Each tower was calibrated by sending a beam of 
40 GeV electrons into its geometric center. Typically, 95\% of the shower energy was deposited in that tower, the remaining 5\% was shared among the six neighbors. The high-voltage settings of the PMT were chosen such that the maximum energy deposited in each tower during the envisaged beam tests would be well within the dynamic range of that tower. For most of the towers (except the central septet), the dynamic range of the PMT signals was chosen to be 60 GeV. 

When we did an energy scan with electrons in one of these non-central towers, the results shown in Figures \ref{spasat}b
and \ref{spasat}c were obtained. Up to 60 GeV, the average calorimeter signal increased proportionally with the beam energy, but above 60 GeV a non-linearity became immediately apparent (Figure \ref{spasat}b). The PMT signal in the targeted tower had reached its maximum value, and would from that point onward produce the same value for every event. Any increase measured in the total signal was due to the tails of the shower, which developed in the neighboring towers. A similar trend occurred for the energy resolution (Figure \ref{spasat}c). Beyond 60 GeV, the energy resolution suddenly improved dramatically. Again, this was a result of the fact that the signal in the targeted tower was the same for all events at these higher energies. The energy resolution was thus completely determined by event-to-event fluctuations in the energy deposited in the neighboring towers by the shower tails. These were relatively small because of the small fraction of the total energy deposited in these towers.

This example illustrates two of the most important consequences of signal saturation: 
\begin{itemize}
\item Non-linearity of the calorimeter response, \ie the total calorimeter signal is not proportional to the energy of the detected particle (Figure \ref{spasat}b).
\item Overestimated energy resolution. The energy resolution is determined by the combined effects of all fluctuations that may occur in the shower development. Signal saturation leads to the suppression of a certain source of fluctuations, and the actual resolution is thus worse than measured (Figure \ref{spasat}c). 
\end{itemize}

In the above example, these consequences could be easily avoided, by decreasing the high voltage and thus the gain of the PMT that converted the scintillation light into electric signals. This is very different for the devices that produce the signals in so-called {\sl digital} calorimeters. An example of such a calorimeter is discussed in Section 7.3. In that calorimeter, the active elements are (500,000!) small resistive plate chambers (RPCs) with a cross section of $1 \times 1$ cm$^2$ each.
These devices, which operate in the saturated avalanche mode, produce a signal when they are traversed by a charged particle. The calorimeter in question
produces pretty energy deposit patterns (Figure \ref{CAL_evt}), but is otherwise not a very good calorimeter. This is because the RPC does not make a difference between 1, 3, 17 or 53 charged particles. It is a digital device, with two options: {\sl yes} or {\sl no}. In Section 7.3, attempts are described to get some more information out of the RPC signals: the {\sl semi-digital option}.

It is clear that this type of calorimeter exhibits the effects outlined above: response non-linearity and overestimated energy resolution. These effects are also worse for em showers compared to hadronic ones because of the larger spatial density of shower particles.
The described effects could of course be mitigated by reducing the size of the RPCs, but the number of readout channels that would have to be handled in that case might exceed the limit of what is reasonably possible in that respect.   

Similar, albeit probably easier to solve, problems are faced when silicon photomultipliers (SiPM) are used to detect the light signals produced by calorimeters. 
A SiPM consists of a large number of very tiny pixels, each of which is a photon detector operating in the Geiger mode \cite{Ren09,Gar11}. At the moment of this writing, SiPMs with up to 40,000 pixels per mm$^2$ are the state of the art \cite{Ace18}. Since these pixels are also digital detectors (yes/no), saturation effects will occur when two or more photons hit the same pixel within the time used to collect the light signals, with the consequences described above.
In this case, these consequences might be mitigated by a further reduction of the pixel size and/or a decrease of the intensity of the light signals to which the SiPM is exposed. 

\section{Electromagnetic calorimetry}

\subsection{\it Shower containment and energy resolution}

Electromagnetic calorimeters are specifically intended for the detection of energetic electrons and 
$\gamma$s, but produce usually also signals when traversed by other types of particles. They are used over a very wide energy range, from the semiconductor crystals that measure $X$-rays down to a few keV to shower counters such as AGILE, PAMELA and FERMI, which orbit the Earth on satellites in search for electrons, positrons and $\gamma$s with energies $>10$ TeV \cite{GLAST}.

Because of the peculiarities of em shower development, these calorimeters don't need to be very deep, especially when high-$Z$ absorber material is used. For example, when 100 GeV electrons enter a block of lead, $\sim 90\%$ of their energy is deposited in only 4 kg of material.

\begin{figure}[htb]
\epsfysize=6.5cm
\centerline{\epsffile{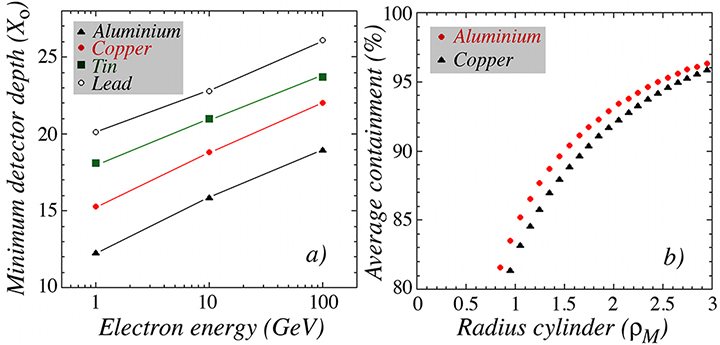}}
\caption{\footnotesize Size requirements for electromagnetic shower containment. The depth of a calorimeter needed to
contain electron showers, on average, at the 99\% level, as a function of the electron energy. Results are given for four different absorber media ($a$). Average lateral containment of electron-induced showers in a copper and an aluminium based calorimeter, as a function of the radius of an infinitely deep cylinder around the shower axis ($b$). From Reference \cite{Wig11}.}
\label{emcont}
\end{figure}

\begin{figure}[htbp]
\epsfysize=6cm
\centerline{\epsffile{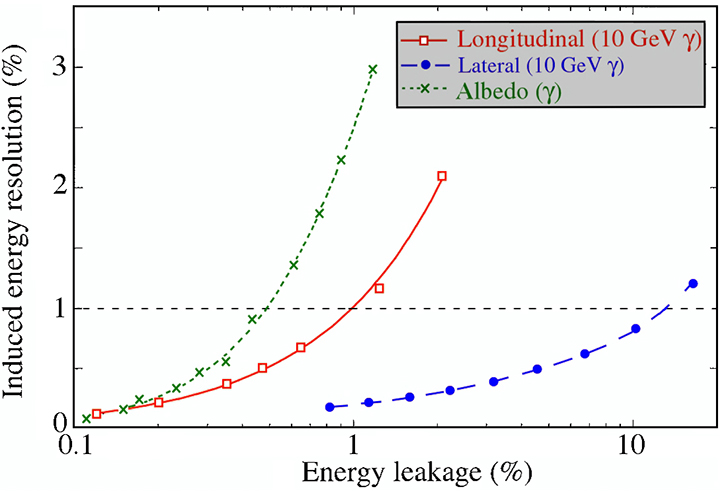}}
\caption{\footnotesize A comparison of the effects caused by different types of shower leakage. Shown are the induced energy resolutions resulting from albedo, longitudinal and lateral shower leakage as a function of the average energy fraction carried by particles escaping from the detector. The longitudinal and lateral leakage data concern 10 GeV $\gamma$s, the albedo data are for $\gamma$-induced showers of different (low) energies. Results from EGS4 Monte Carlo calculations in which the detector was represented by a block of tin\cite{Zeyrek}.}
\label{eleakage5}
\end{figure}

Good energy resolution can only be achieved when the shower is, on average, sufficiently contained. Figure \ref{emcont}a shows the required depth of the calorimeter, as a function of the electron energy, needed for 99\% longitudinal containment. The figure shows this depth requirement, needed for energy resolutions of 1\% (\cf Figure \ref{eleakage5}), for calorimeters using Pb, Sn, Cu and Al as absorber material. The fact that the four curves are not identical indicates that the radiation length ($X_0$), which is commonly used to describe the longitudinal development of em showers, is not a perfect scaling variable (\cf Figure \ref{emlong}).
It should also be noted that $\gamma$-induced showers require about one radiation length more material
in order to be contained at a certain level than do  electron showers of the same energy \cite{Zeyrek}. 

The effects of lateral shower leakage on the energy resolution are much smaller than for longitudinal leakage. 
According to Figure \ref{eleakage5}, as much as 10\% leakage can be tolerated before the induced energy resolution resulting from leakage {\sl fluctuations} exceeds 1\%.
Figure \ref{emcont}b shows the average lateral leakage fraction as a function of the radius of an infinitely deep calorimeter centered on the shower axis. It turns out that a radius of two Moli\`ere radii ($\rho_M$) is more than adequate for 90\% lateral containment. This number is almost independent of the shower energy and the absorber material, in stark contrast with the depth requirements. The figure also indicates that em showers tend to have considerable radial tails. In order to capture 99\% instead of 90\% of the shower energy, the detector mass has to be increased by an order of magnitude.

\begin{figure}[htbp]
\epsfysize=8cm
\centerline{\epsffile{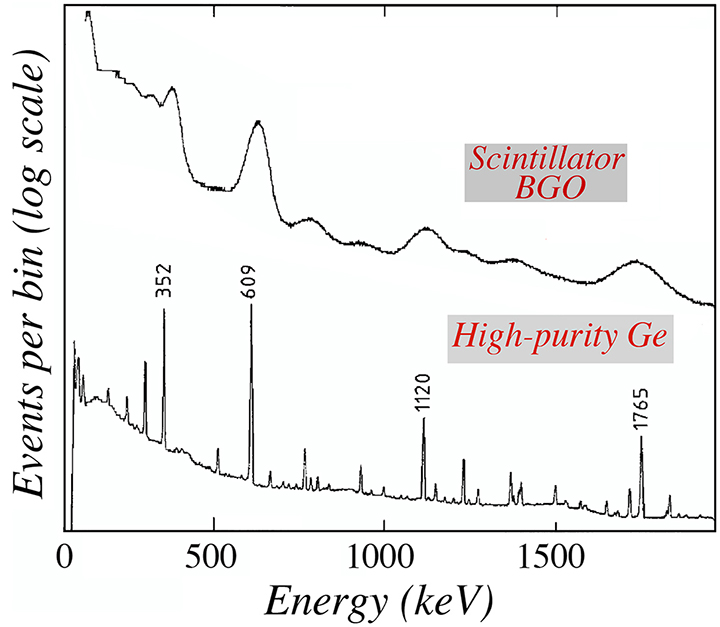}}
\caption{\small
Nuclear $\gamma$-ray spectrum of decaying uranium nuclei, measured with a bismuth germaniumoxide
scintillation counter ({\sl upper curve}) and with a high-purity germanium crystal ({\sl lower
curve}). From Reference \cite{Wig17}.}
\label{eres}
\end{figure}

Spectacular energy resolutions have been obtained with large semiconductor crystals, and in particular high-purity germanium. These are the detectors of choice in nuclear $\gamma$ ray spectroscopy, and routinely obtain resolutions ($\sigma/E$) of 0.1\% in the 1 MeV energy range. An example is shown in 
Figure \ref{eres}, which also makes a comparison with the next best class of detectors, scintillating crystals. 
Such excellent energy resolutions are critically important for experiments searching for evidence of neutrinoless $\beta\beta$ decay, since the signature for this process
is a mono-energetic signal (equal to the sum of the energies of the two electrons) at precisely the $Q$-value of the decay process. One of the most interesting candidates 
is $^{136}$Xe, since this nuclide offers the possibility to be both the source and the detector. Current searches are based on a high-pressure xenon time projection chamber (TPC) \cite{next}, a cryogenic liquid-xenon TPC using enriched $^{136}$Xe \cite{exo} or enriched $^{136}$Xe dissolved in liquid scintillator \cite{zen}.
The TPC experiments detect both the ionization and the scintillation light produced in the decay in an attempt to obtain sub-1\% energy resolution at the $Q_{\beta\beta}$
value of 2.459 MeV.  

Scintillating crystals are often the detectors of choice in experiments involving $\gamma$ rays in the energy range from 1 -- 20 GeV, which can be measured with energy resolutions of the order of 1\%. Examples of experiments for which such performance is crucially important include BES-III at the BEPC $\tau$/charm factory \cite{Don08} and BELLE-II at the KEK SuperB collider \cite{Abe10}. The experiments at these $e^+e^-$ colliders are both using CsI(Tl) crystals.

\begin{table}[htbp]
\centering
\caption{\footnotesize Relevant properties of a variety of light-based detectors that are used as homogeneous electromagnetic calorimeters in particle physics experiments. }
\vskip 3mm
\renewcommand{\arraystretch}{1.2}
\setlength\tabcolsep{5pt}
\begin{tabular} {lcccccccc}\hline
{\sl Detector}&$\rho$&$X_0$&$\lambda_{\rm int}$&$\lambda_{\rm emis}$&$\tau$&$\sigma/ E$&{\sl Reference} \\
~~~~~~$\downarrow$&(g cm$^{-3}$)&(cm)&(cm)&(nm)&(ns)&(1 GeV)\\ \hline
{\sl Scintillating crystals}\\
NaI(Tl)&3.67&2.59&41.4&410&230&2.7\%&\cite{Kor17}\\
CsI(Tl)&4.51&1.85&37.0&560&1300&2.7\%&\cite{Abe10}\\
BGO&7.13&1.12&21.8&480&300&2.1\%&\cite{bgo}\\
PbWO$_4$&8.30&0.89&18.0&420&10&3.1\%&\cite{Aam08}\\
{\sl Scintillating liquids}\\
LKr&2.41&4.7&61&147&$\sim 85$&1.9\%&\cite{na48,Dok90}\\
LXe&2.95&2.77&57&174&$\sim 30$&1.6\%&\cite{Saw07}\\
{\sl \v{C} calorimeters}\\
Lead glass&4.06&2.5&33&$\lambda^{-2}$&0&5\%&\cite{Aut96}\\
BaF$_2$&4.89&2.06&29.9&$\lambda^{-2}$&0&6\%&\cite{Fie15}\\
Water&1.0&36.1&84.9&$\lambda^{-2}$&0&2.6\%&\cite{SuperK}\\
\hline
\end{tabular}
\label{tabcal:1}
\end{table}
Excellent performance in this energy range has also been reported for liquid krypton and xenon detectors, which are bright (UV)
scintillators. Other homogeneous detectors of em showers are based on \v{C}erenkov light, in particular 
lead glass. Very large water \v{C}erenkov calorimeters (\eg SuperKamiokande \cite{SuperK}) should also be mentioned in this category.

Table \ref{tabcal:1} lists materials commonly used as homogeneous calorimeters in particle physics experiments and some of their relevant properties.
Not mentioned in this table is the light yield of these materials. The for practical purposes relevant strength of the signals depends  sensitively on the effects of self-absorption and other factors that cause light losses, as well as on the quantum efficiency of the detector that converts the light into electrical pulses. These effects may vary strongly for different detectors based on materials listed in the table. Of these materials, NaI(Tl) is the brightest source of photons: $\approx 50,000$ per MeV deposited energy. CsI(Tl) and the scintillating liquids generate light at the same order of magnitude. The light yield in BGO is one order of magnitude less, while PbWO$_4$ generates three orders of magnitude less light than the brightest scintillators. The light yield deriving from the \v{C}erenkov mechanism is four orders of magnitude less than that of sodium iodide, and photoelectron statistics tends to dominate the electromagnetic energy resolution of detectors based on this mechanism. As illustrated by the values listed in this table, light yield is not the dominant limiting factor for the energy resolution ($\sigma/E$) of the mentioned scintillators, at least not in the GeV domain.

Sampling calorimeters, which are typically much cheaper, become competitive at higher energies.
In properly designed instruments of this type, the energy resolution is determined by {\sl sampling fluctuations}. 
\begin{figure}[htb]
\epsfysize=8cm
\centerline{\epsffile{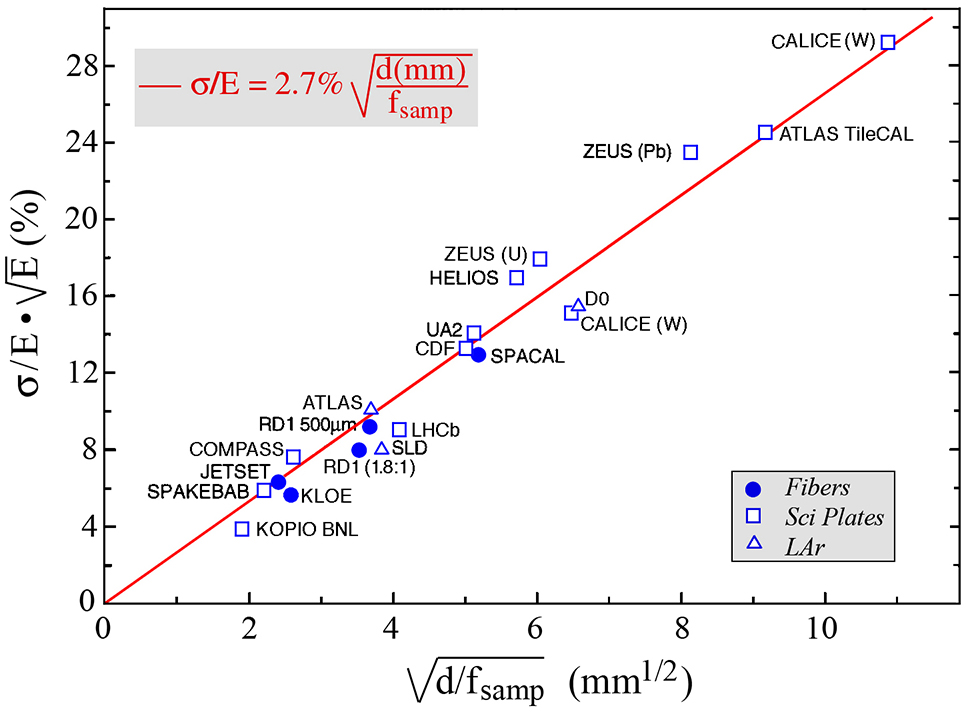}}
\caption{\footnotesize The em energy resolution of a variety of sampling calorimeters as a function of the parameter $(d/f_{\rm samp})^{1/2}$, in which $d$ is the thickness (in mm) of an active sampling layer (\eg the diameter of a fiber or the thickness of a liquid-argon gap), and $f_{\rm samp}$ the sampling fraction for mips. The energy $E$ is expressed in units of GeV \cite{Liv95}.}
\label{sfrac}
\end{figure}
These represent fluctuations in the number of different shower particles that contribute to the  calorimeter signals, convolved with fluctuations in the amount of energy deposited by individual shower particles in the active calorimeter layers. They depend both on the {\sl sampling fraction}, which is determined by  the ratio of active and passive material, and on the {\sl sampling frequency}, determined by the number of different sampling elements in the region where the showers develop. 
Sampling fluctuations are stochastic and their contribution to the energy resolution is thus described by 
\begin{equation}
(\sigma/E)_{\rm samp}~=~{a_{\rm samp}\over \sqrt{E}}, ~~~~~{\rm with}~~~~~a_{\rm samp}~=~0.027 \sqrt{d/f_{\rm samp}}
\label{eq:samp}
\end{equation}
in which $d$ represents the thickness of individual active sampling layers (in mm), and $f_{\rm samp}$ the sampling fraction for minimum ionizing particles ({\sl mip}s).
This expression describes the em energy resolution obtained with a variety of different sampling calorimeters based on plastic scintillator or liquid argon as active material reasonably well (Figure \ref{sfrac}) \cite{Liv95}.
\begin{table}[b!]
\centering
\caption{\footnotesize A representative selection of electromagnetic sampling calorimeters used in past and present particle physics experiments.  The energy $E$ is expressed in GeV, the sampling fraction $f_{\rm samp}$ refers to minimum ionizing particles. For the energy resolution, only the $E^{-1/2}$ scaling term, which dominates in the practically important energy range for these experiments, is listed. }
\vskip 3mm
\renewcommand{\arraystretch}{1.2}
\setlength\tabcolsep{5pt}
\begin{tabular}{lccccc} \hline\noalign{\smallskip}
Experiment &Calorimeter structure&$X_0$ (cm)&$f_{\rm samp}$&$\sigma/E$&Reference\\ \hline\noalign{\smallskip} 
KLOE (Frascati)&Pb/fibers&1.6 &17\%&$4.7\%/\sqrt{E}$&\cite{Ant95}\\
ZEUS (DESY)&$^{238}$U/scintillator&0.7&9\%&$18\%/\sqrt{E}$&\cite{Beh90}\\
NA48 (CERN)&Pb/LKr&1.5&23\%&$3.5\%/\sqrt{E}$&\cite{na48}\\
ATLAS (LHC)&Pb/LAr&$\approx 3$&$\approx 25\%$&$10\%/\sqrt{E}$&\cite{Aha06}\\
PHENIX (RHIC)&Pb/scintillator&3.1&29\%&$7.8\%/\sqrt{E}$&\cite{Kis94}\\
AMS-02 (ISS)&Pb/fibers&1.3&19\%&$10.4\%/\sqrt{E}$&\cite{Gal15} \\ \hline
\end{tabular}
\label{tabcal:2}
\end{table}

Table \ref{tabcal:2} lists some characteristics of a representative selection of sampling calorimeters used in partlcle physics experiments. 
Above 100 GeV, the resolution of all calorimeters mentioned above is $\sim 1\%$, and systematic factors,
such as stability of the electronic components, the effects of light attenuation, or temperature variations of the light yield, tend to dominate the performance.

The validity of Equation \ref{eq:samp} is limited to calorimeters with plastic scintillator or liquid argon/krypton as active material. When the active layers are very thin (in terms of stopping power), as in calorimeters with gaseous or silicon readout, an additional factor contributes to the energy resolution: {\sl pathlength fluctuations}. In such calorimeters, the energy deposited by a typical shower electron depends on its trajectory inside the active material. For example, the energy loss of electrons in 100 $\mu$m silicon amounts to $\sim 115$ keV \cite{Ber93}. The signal from shower electrons with energies larger than 115 keV produced in Compton scattering or photoelectric effect therefore depends on the angle at which they traverse an active layer. The larger the angle with the shower axis, the larger the contribution of these particles to the signal. For 500 $\mu$m silicon, the same is true for shower electrons with energies larger than 330 keV. And since the sampling fluctuations are determined by fluctuations in the {\sl total energy deposited} by the shower particles, and since these soft electrons are an important component of the developing em showers, these pathlength fluctuations are important for calorimeters with very thin active layers. 

\begin{figure}[htb]
\epsfysize=8cm
\centerline{\epsffile{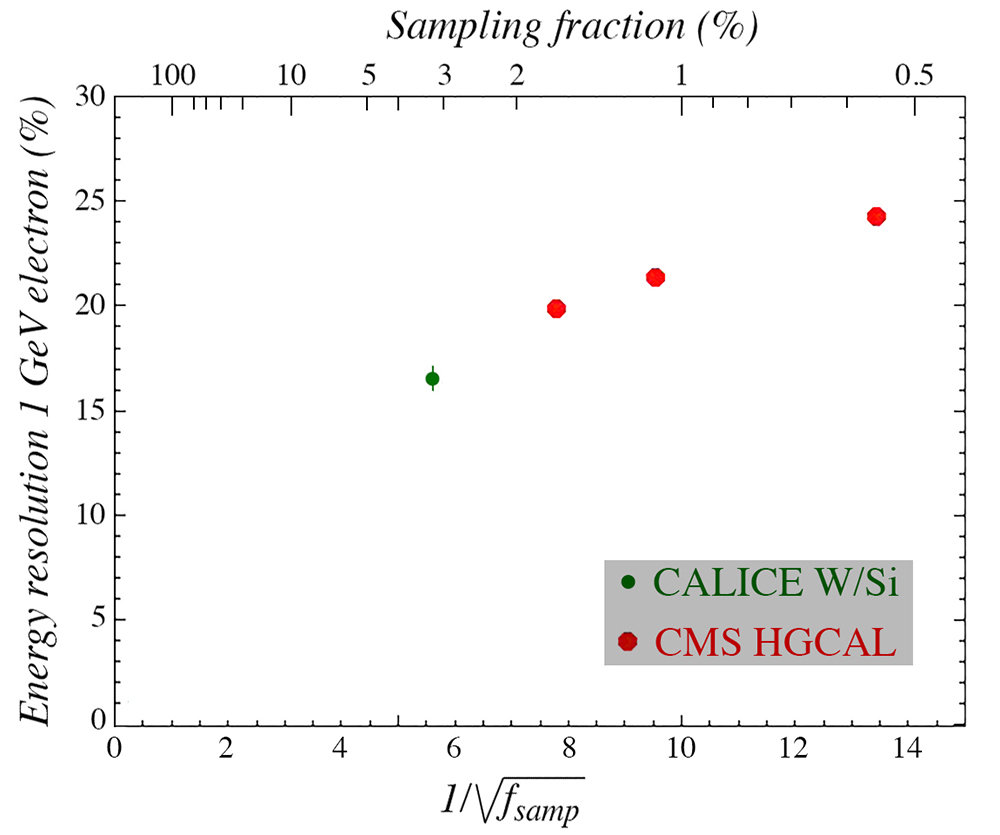}}
\caption{\footnotesize The em energy resolution of sampling calorimeters with silicon layers as active material. Data from \cite{Sef15,CMS15b}.}
\label{hgcalres2}
\end{figure}

This is illustrated by Figure \ref{hgcalres2}, which shows the energy resolution for calorimeters with thin silicon layers as active material. All calorimeters have approximately the same structure. Absorber layers of tungsten with a thickness that increases with depth from 1.5 mm to 4.5 mm are interleaved with thin layers of silicon. In the four configurations of which the em energy resolution is displayed, the thickness of the silicon is 100 $\mu$m, 200 $\mu$m, 300 $\mu$m and 525 $\mu$m, respectively. These calorimeters thus have different sampling fractions, but the ratio $d/f_{\rm samp}$ used in Eq. \ref{eq:samp} is approximately the same. The experimental data point comes from CALICE \cite{Sef15}, which used 525 $\mu$m silicon, and obtained a resolution of $16.5\%/\sqrt{E}$
(Equation \ref{eq:samp} gives $\sim 12\%/\sqrt{E}$ for this configuration).
The other data points concern GEANT4 simulation results for the HGCAL upgrade calorimeter for the CMS endcap region \cite{CMS15b}, which has about the 
same $d/f_{\rm samp}$ value as the CALICE one, but uses thinner silicon layers. The expected energy resolution for this device ranges from $19.9\%/\sqrt{E}$
for 300 $\mu$m silicon to $24.3\%/\sqrt{E}$ for 100 $\mu$m silicon. These results illustrate the importance of the contribution of soft shower particles to the signals from sampling calorimeters.  

This importance is also illustrated by another crucial
characteristic of these sampling calorimeters. The sampling fraction, \ie the fraction of the energy that is deposited in the active calorimeter layers, {\sl decreases} as the shower develops. This is because the spectrum of the $\gamma$s that produce the shower particles that constitute the signals (electrons, positrons) becomes much softer.
In the early stages, these $\gamma$s mainly convert into energetic $e^+e^-$ pairs, but beyond the shower maximum Compton scattering and photoelectron production become more and more dominant.
Because of the $Z$ dependence of the latter processes ($Z^5$ for the photoelectric effect), the soft $\gamma$s 
almost exclusively react in the high-$Z$ absorber material (lead or uranium), and the resulting electrons are thus much less efficiently sampled by the signal producing calorimeter layers than the $e^+e^-$ pairs that dominate the early part of the shower. This phenomenon may cause large, non-trivial problems for the calibration of longitudinally segmented calorimeters, as illustrated in Section 6.1.

\subsection{\it Shower Profiles}

It is commonly assumed that the radial energy deposit profile of em showers scales with the Moli\`ere radius. For this reason, experiments typically choose the granularity of their em calorimeter in terms of that parameter.  For example, if a cell size with an effective radius of $1 \rho_M$ is used, em showers deposit typically $\sim 80\%$ of their energy in one cell, if the particle enters that cell in its central region (Figure \ref{emcont}b). In order to increase that percentage to 90\%, the effective radius of the 
cell has to be doubled, \ie the number of cells is reduced by a factor of four. Recent measurements with a calorimeter that used a much finer granularity showed that the energy deposit profile is very strongly concentrated near the shower axis \cite{rd52sipm}. 

\begin{figure}[htbp]
\epsfysize=9cm
\centerline{\epsffile{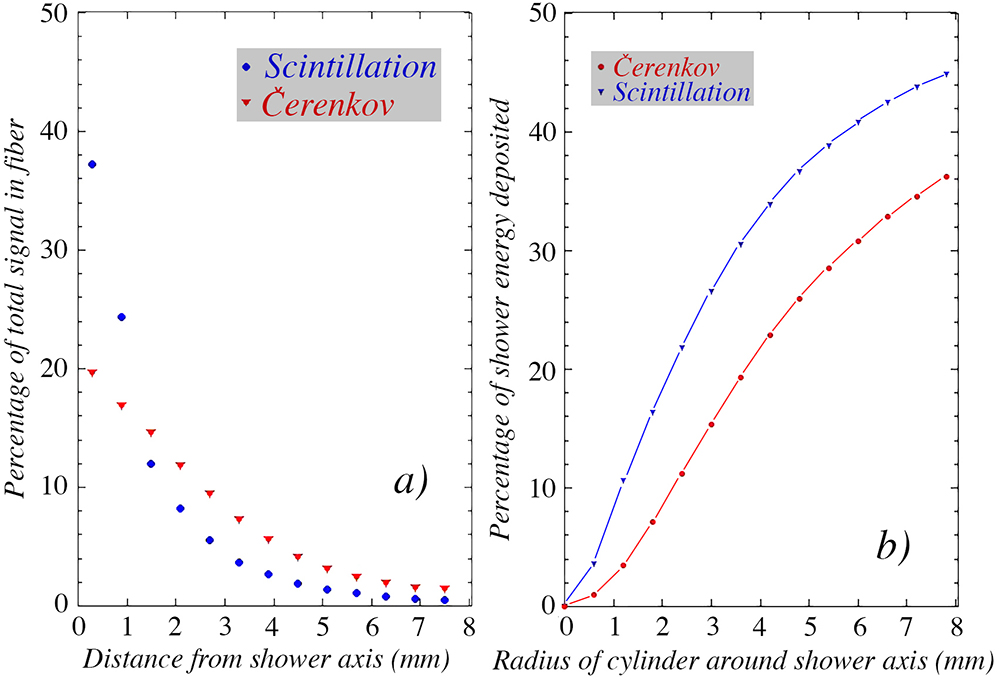}}
\caption{\footnotesize{Lateral profiles of electromagnetic showers in the  brass-fiber dual-readout SiPM calorimeter, measured separately with the \v{C}erenkov
and the scintillation signals ($a$). The fraction of the shower energy deposited in a cylinder around the shower axis as a function of the radius of that cylinder, measured separately with the \v{C}erenkov and the scintillation signals ($b$). One mm corresponds to $0.031 \rho_M$ in this calorimeter \cite{rd52sipm}.}}
\label{emprofiles}
\end{figure}
For example, when the granularity was increased by a factor of 20 (reducing the effective radius of a detector cell from $1 \rho_M$ to $0.22 \rho_M$), one cell contained $\sim 45\%$ of the shower energy, and if the granularity was increased by another factor of 30 (to cells with a radius of $0.04 \rho_M$), $\sim$10\% of the total shower energy was still deposited in one cell. Figure \ref{emprofiles} shows the profiles measured with this detector, a dual-readout calorimeter with fibers read out by a silicon photomultiplier array. The signals from each individual fiber were sensed by a 1 mm$^2$ SiPM \cite{rd52sipm}.

Figure \ref{emprofiles}a shows a remarkable difference between the profiles measured by the two types of fibers that constitute the active material of this calorimeter. The \v{C}erenkov light is much less concentrated near the shower axis than the scintillation light. This is a consequence of the fact that the early, extremely collimated component of the developing shower does not contribute to the \v{C}erenkov signals, since the \v{C}erenkov light falls outside the numerical aperture of the fibers. This phenomenon has interesting consequences for the detection of muons with a device of this type, since a
comparison between the two signals makes it possible to distinguish the ionization and radiative components of the energy loss by muons traversing this calorimeter (see Section 5.2 \cite{DREAMmu}).

\subsection{\it Radiation Damage}

As the luminosities at colliding-beam machines have increased (\eg to compensate for the fact that the cross sections for the interesting processes rapidly decrease as $\sqrt{s}$ increases), radiation damage of crucial detector components has become more and more a point of concern. At hadron colliders, scintillation based calorimeters have become less favored because of the sensitivity of active media based on light production to ionizing radiation. Even CMS, which chose lead tungstate crystals based on the (advocated and) supposed radiation hardness, had to face the fact that the endcap region of their calorimeter system is fast becoming unusable, after having received only a small fraction of the envisaged total integrated luminosity of the Large Hadron Collider.

\begin{figure}[htb]
\epsfysize=7.5cm
\centerline{\epsffile{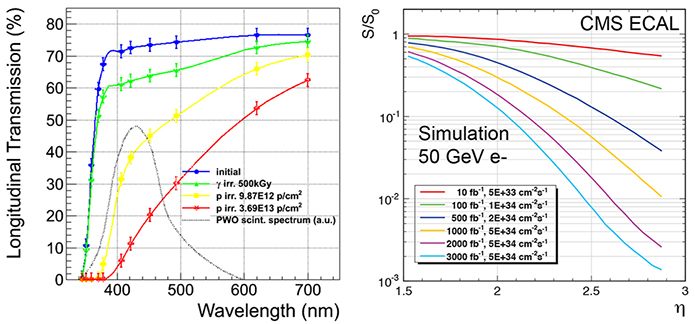}}
\caption{\footnotesize Effects of radiation damage on the performance of the CMS em calorimeter. The left diagram shows the light transmission in the PbWO$_4$ crystals after irradiation with $\gamma$s and protons. The overlaid black dotted line represents the lead tungstate light emission spectrum. The right diagram shows the (simulated) effect on the scintillation signals from 50 GeV showers in the crystals as a function of the pseudorapidity, for various 
values of the integrated luminosity \cite{CMS15b}.}
\label{CMSraddam}
\end{figure}

This is illustrated in Figure \ref{CMSraddam} \cite{CMS15b}. 
The left diagram shows the effect of ionizing radiation on the light transmission, as a function of wavelength. Especially the short wavelength range, which is crucial for the transmission of the scintillation light generated by the crystals, is affected.
The dominating source of damage are the hadrons. Unlike the effects caused by photons, the damage induced by hadrons cannot be undone by annealing at room temperature. The right diagram in Figure \ref{CMSraddam} shows the reduction of the PbWO$_4$ signals as a function of the pseudorapidity at which the crystals are installed in the endcap region. At the present time, the signals are already down by a factor of two in the regions close to the beam pipe, and the
effects are projected to become much worse over time (up to a reduction by a factor 1000 after the envisaged total integrated luminosity).
\begin{figure}[htbp]
\epsfysize=8cm
\centerline{\epsffile{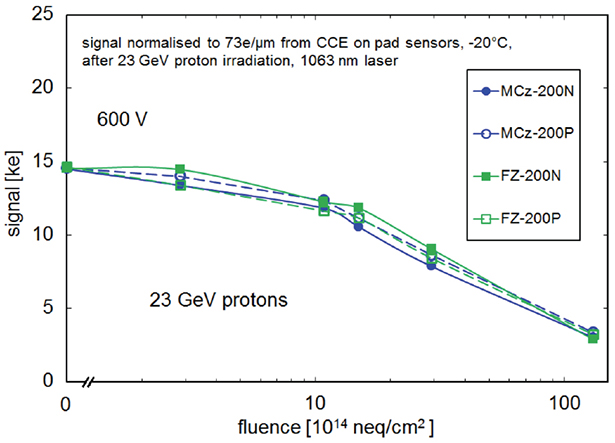}}
\caption{\footnotesize Effects of exposure to neutrons on the signals from different types of 200 $\mu$m thick silicon detectors, as a function of the total neutron fluence. The signals are expressed in kilo-electrons, the neutron fluence in 1 MeV neutron equivalent per cm$^2$ \cite{CMS15b}.}
\label{nraddam}
\end{figure}

In order to better deal with high radiation levels, modern calorimeters are based on liquid active material, such as liquid argon, which can be relatively easily replaced, or on intrinsically (more) radiation hard materials, such as silicon. The CMS Collaboration has chosen a silicon based sampling calorimeter as a replacement for the crystals in the endcap section of their detector. Contrary to light based systems, radiation damage by ionizing particles is not the main problem in silicon. Rather, neutrons that scatter off the silicon nuclei and alter the lattice structure of the semiconductor material are the main source of concern in this case.

In the environment that the CMS endcap calorimeter has to face in the high-luminosity LHC, it is expected that the detectors will have to deal with a total neutron fluence of up to $10^{16}$/cm$^2$ close to the beam pipe, for the integrated total luminosity of 3000 fb$^{-1}$.  Figure \ref{nraddam} shows the effects of exposure to neutrons on the signals from a number of different silicon detectors, as a function of the fluence. At the maximum expected rates, the signals are reduced by a factor of about four. These results are for silicon sensors with a thickness of 200 $\mu$m. Measurements have shown that the relative signal reduction decreases with the thickness of the depletion layer. For this reason, CMS contemplates using thinner sensors near the beam pipe \cite{CMS15b}.

Not only the active calorimeter layers are subject to the effects of radiation. Damage of silicon based electronics embedded in the absorber structure is a point of concern as well. As an example, I mention the case of the ATLAS Liquid Argon Calorimeter \cite{ATL12}. In preparation for the increased LHC luminosities foreseen for the future, ATLAS will replace the ASICs\footnote{Application-Specific Integrated Circuit.} that handle the calorimeter signals by more radiation hard ones. ASICs based on IBM's 130 nm CMOS (8RF) technology meet the requirements in that respect. The new ASICs will also be better adapted to handle the trigger rates expected at the higher luminosities.
The analog on-detector Level-1 pipeline will be replaced by a system in which all calorimeter signals are digitized at 40 MHz and sent to the off-detector front-end electronics. This approach is expected to remove all constraints imposed by the calorimeter readout on the ATLAS trigger system. 

\subsection{\it Pileup}

Radiation damage is not the only problem LHC experiments face as a result of the increasing luminosity. At the present time, ATLAS and CMS have to deal with, on average, 55 events per 25 ns bunch crossing (at a luminosity of $2\cdot 10^{34}$ cm$^{-2}$s$^{-1}$). This rate is increasing proportionally with the luminosity and in the High Luminosity LHC era (after 2024), one will have to cope with a luminosity of $7\cdot 10^{34}$ cm$^{-2}$s$^{-1}$, at which point
interesting events accompanied by more than 200 ``pileup'' events will be no exception. 
Of course, the overwhelming majority of these ``underlying events'' are uninteresting, and involve relatively few high-p$_\perp$ particles that contribute to the detector 
signals. Yet, pileup induced background is expected to be a factor that seriously 
deteriorates the detector performance. High-precision timing is considered one of very few options to mitigate these effects, and this has given rise to a number of dedicated experimental studies \cite{Ada16,Bar18,Whi17}.

The elapsed time for an LHC bunch crossing has an rms spread of 170 picoseconds, which means that the 50 -- 100 ps time resolution commonly achieved in the time-of-flight systems used for particle identification purposes is not adequate for solving this problem. One expects to need time resolutions of at least 20 -- 30
ps to make a significant difference in this respect. A major complicating factor is that this performance has to be achieved in a very-high-rate environment.
The approach followed by the mentioned R\&D projects focuses on instantaneous light signals, such as those produced by the \v{C}erenkov mechanism, combined with ultrafast photo detectors, such as Avalanche Photo Diodes (APDs) \cite{Ada16}, Microchannel Plates (MCPs) \cite{Bar18}, or micromegas \cite{Whi17}.
\begin{figure}[htbp]
\epsfysize=7cm
\centerline{\epsffile{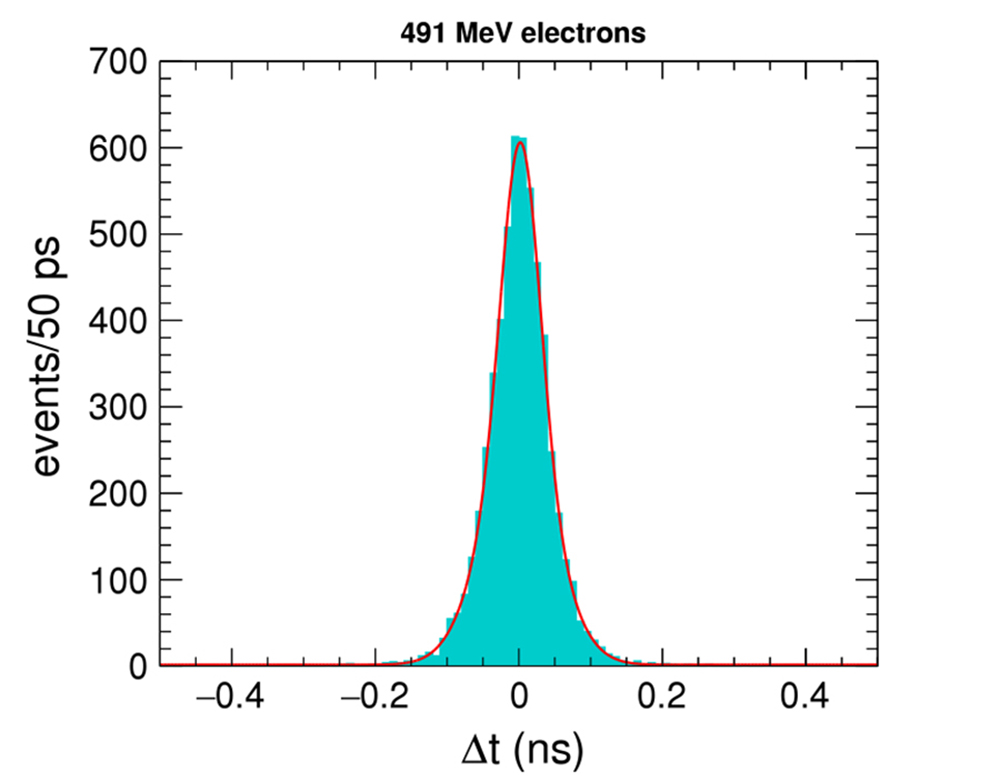}}
\caption{\footnotesize Distribution of the time difference between the signals from 491 MeV electrons measured in two MCPs, one operated in $i$-MCP mode and the other one in PMT-MCP mode \cite{Bar18}.}
\label{mcptime}
\end{figure}

Figure \ref{mcptime} shows results recently obtained with MCPs.
These detectors are either operated in the standard PMT-MCP mode, or in the $i$-mode, in which the photocathode is removed and the signal is produced by secondary emission of electrons from the MCP layers crossed by the ionizing particles \cite{Win12}. The figure shows the distribution of the time difference between signals from 491 MeV electrons measured in two MCPs, one operated in $i$-mode and the other in PMT-MCP mode. This Gaussian distribution has a width ($\sigma$) of 17$\pm$2 ps. A time resolution of 75 ps was reported {\sl for single photoelectrons} by \cite{Whi17}.
These are encouraging developments, but there is of course still a long way to go before systems capable of assigning signals to different events occurring in the same bunch crossing will be  available for implementation in the extremely high-rate environment of LHC-type experiments. 

In a more traditional approach to the pileup problem, it is treated as an additional source of electronic noise. Using a series of signal samples collected at intervals of 25 ns, the properties of the true signal are estimated on the basis of a variance minimization of the noise covariance matrix, a method known as optimal filtering \cite{Cle94}.
The bipolar pulse shaping, introduced by Radeka \cite{Rad88b}, is crucial for the success of this method with the LAr signals from the ATLAS calorimeters. This method works best for Gaussian noise. Pile-up has the tendency to add positive or negative tails to the noise distribution, which thus becomes non-Gaussian. The extent of these effects depends on the number of underlying events, and thus leads to a luminosity dependence.

\begin{figure}[htbp]
\epsfysize=7cm
\centerline{\epsffile{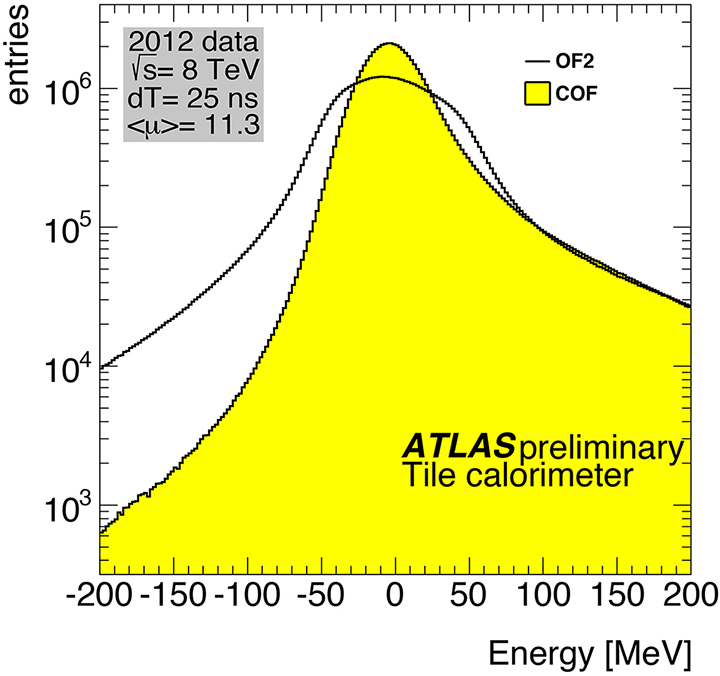}}
\caption{\small
Cell energy distribution reconstructed by the Constrained Optimal Filter (COF) and by the Optimal Filtering (OF2) algorithms, using 2012 ATLAS $pp$ collision data at $\sqrt{s}$ = 8 TeV and 25 ns bunch spacing. The average number of interactions per bunch crossing was 11.3 for this event sample (around 25 millions entries). The COF method is resilient to out-of-time signals and, therefore, leads to a better energy resolution than OF2. Its design is luminosity independent and requires only the information of the pulse shape and pedestal value to compute the 7 amplitudes associated to the 7 samples of the read-out. In this figure only the central sample reconstruction is shown \cite{Fil15}.}   
\label{seixas}
\end{figure}

Recently, a method has been proposed that is in principle independent of the luminosity. It is based on a deconvolution process of the same type used in digital processing for communication channel equalization, and aims to fully recover the target signal, rather than estimate its amplitude from pulse sampling \cite{Fil15}. This method has been tested with experimental data obtained with the ATLAS TileCal hadronic calorimeter. Figure \ref{seixas} shows some results from this work.

\section{The fundamental problems of hadron calorimetry}

\subsection{\it The $e/h$ ratio}

The development of hadronic cascades in dense matter differs in essential ways from that of electromagnetic ones, with important consequences for calorimetry.
Hadronic showers consist of two distinctly different components:
\begin{enumerate}
\item An {\sl electromagnetic} component; $\pi^0$s and $\eta$s generated in the absorption process
decay into $\gamma$s which develop em showers.
\item A {\sl non-electromagnetic} component, which combines essentially everything else that
takes place in the absorption process. 
\end{enumerate}
For the purpose of calorimetry, the main difference between these components is that
some fraction of the energy contained in the non-em component does {\sl not} contribute to the signals. This {\em invisible energy}, which mainly consists of the binding energy of nucleons released in the numerous nuclear reactions, may represent up to 40\% of the total non-em energy, with large event-to-event fluctuations. For this reason, homogeneous hadron calorimeters offer no particular advantage. As a matter of fact, sampling calorimeters hold all the records for best hadronic performance. 
\begin{figure}[htbp]
\epsfxsize=10cm
\centerline{\epsffile{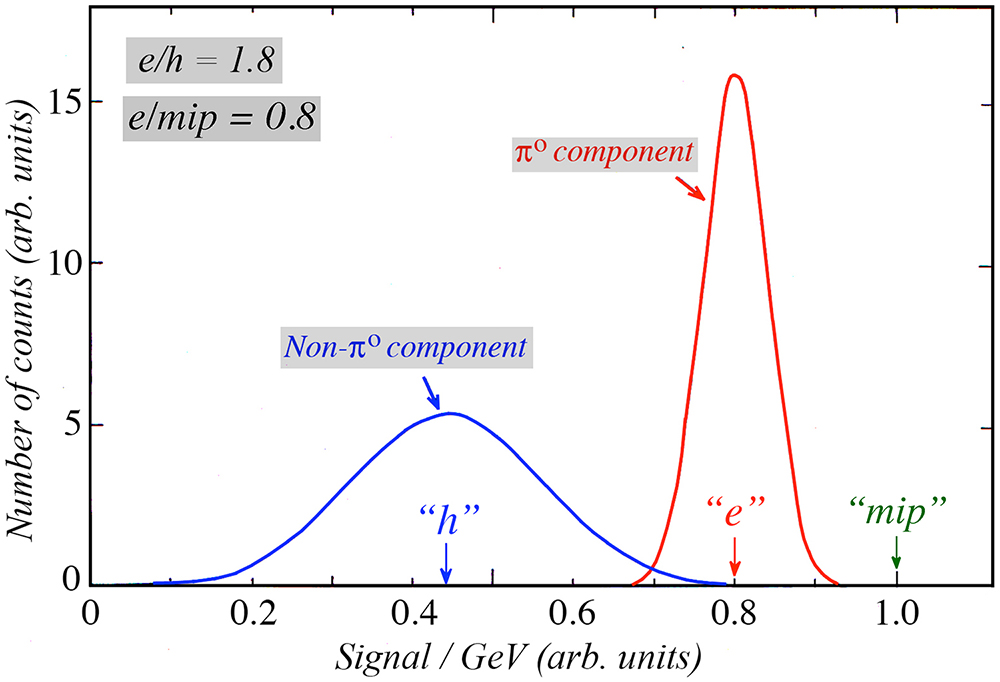}}
\caption{\small
Illustration of the meaning of the $e/h$ and $e/mip$ values of a calorimeter. Shown are distributions of the signal per unit deposited energy for the electromagnetic and non-em components of hadron showers. These distributions are normalized to the response for minimum ionizing particles ($``mip"$). The average values of the em and non-em distributions are the em response ($``e"$) and non-em response ($``h"$) , respectively.}  
\label{ehprinciple}
\end{figure}

Let us define the calorimeter {\em response} as the conversion efficiency from deposited energy to generated signal, and normalize it to minimum ionizing particles. If the average signal is proportional to the deposited energy, \ie if the calorimeter is linear, this definition implies that the response of this calorimeter is energy independent. The responses of a given calorimeter to the em and non-em hadronic shower components, $e$ and $h$, are usually not the same, as a result of invisible energy and a variety of other effects. I will call the distribution of the signal per unit deposited energy around the mean value ($e$ or $h$) the {\sl response function}.

Figure \ref{ehprinciple} illustrates the different aspects of the calorimeter response schematically. The em response is larger than the non-em one, and the non-em response function is broader than the em one, because of event-to-event fluctuations in the invisible energy fraction. Both $e$ and $h$ are smaller than the calorimeter response for minimum ionizing particles, because of inefficiencies in the shower sampling process \cite{Liv17}. The calorimeter is characterized by the $e/h$ and $e/mip$ ratios, which in this example have values of 1.8 and 0.8, respectively.
Calorimeters for which $e/h \ne 1$ are called {\sl non-compensating}.

\begin{figure}[htbp]
\epsfxsize=15cm
\centerline{\epsffile{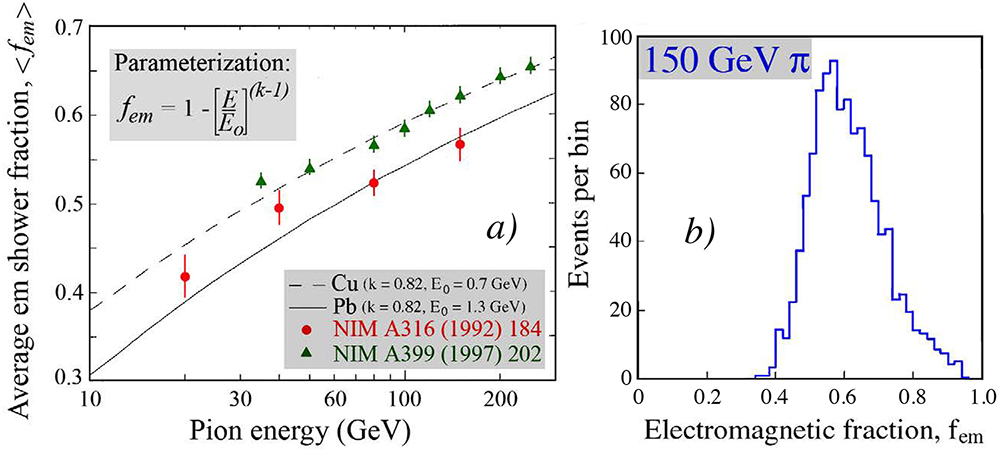}}
\caption{\small
Properties of the electromagnetic fraction of hadron showers. Shown are the results of measurements of the average value of that fraction as a function of energy, for showers developing in lead or copper ($a$) and the distribution of $f_{\rm em}$ values measured for 150 GeV $\pi^-$ showers developing in lead ($b$). The curves in diagram $a$ represent Equation \ref{femE}. Experimental data from \cite{Aco92b,Akc97}.}.  
\label{femprops}
\end{figure}

\begin{figure}[htbp]
\epsfxsize=15cm
\centerline{\epsffile{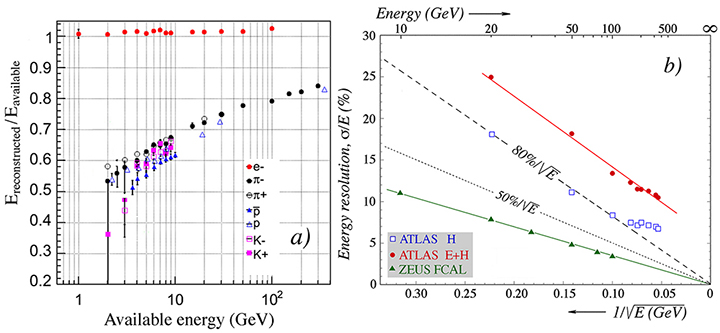}}
\caption{\small
Experimental consequences of non-compensation for the hadronic calorimeter performance. The non-linearity reported by CMS \cite{Akc12a} ($a$) and the energy resolution reported by ATLAS \cite{Aba10}, both for the Tilecal in stand-alone mode and for the combination of the em and hadronic calorimeter sections ($b$). For comparison, the hadronic energy resolution reported for the compensating ZEUS calorimeter \cite{Beh90} is shown as well. See text for details.}  
\label{detperformance}
\end{figure}

The properties of the em shower component have important consequences for the hadronic {\em energy resolution}, signal {\em linearity} and {\em response function}.
The average fraction of the total shower energy contained in the em component, $\langle f_{\rm em} \rangle$, was measured to increase
with energy following a power law \cite{Aco92b,Akc97}, confirming an induction argument made to that effect \cite{Gab94}:

\begin{equation}
\langle f_{\rm em} \rangle ~=~1 - \biggl[ \biggl({E\over E_0}\biggr)^{k-1}\biggr]
\label{femE}
\end{equation}
where $E_0$ is a material-dependent constant related to the average multiplicity in hadronic interactions (varying from 0.7 GeV to 1.3 GeV for $\pi$-induced reactions on Cu and Pb, respectively), and $k \sim 0.82$ (Figure \ref{femprops}a). 
A direct consequence of the energy dependence of $\langle f_{\rm em} \rangle$ is that calorimeters for which $e/h \ne 1$ are by definition {\sl non-linear} for hadron detection, since the response to hadrons is determined by 
$ \langle f_{\rm em} \rangle + \bigl[1 -  \langle f_{\rm em} \rangle\bigr] h/e$, and thus energy dependent.
This is confirmed by many sets of experimental data, for example the ones reported for CMS \cite{CMS07} shown in Figure \ref{detperformance}a.

Event-to-event fluctuations in $f_{\rm em}$ are large and non-Poissonian \cite{Aco92b}, as illustrated in Figure \ref{femprops}b. If $e/h \ne 1$, these fluctuations 
tend to dominate the hadronic energy resolution and their asymmetric characteristics are reflected in the 
response function \cite{Liv17}.
It is often assumed that the effect of non-compensation on the energy resolution is energy independent (``constant term''). This is incorrect, since it implies that the effect is insignificant at low energies, \eg 10 GeV, which is by no means the case. The measured effects of {\em fluctuations} in $f_{\rm em}$ can be described by a term that is very similar to the one used for its energy dependence
(\ref{femE}). 
This term should be added in quadrature to the $E^{-1/2}$ scaling term which accounts for all Poissonian fluctuations:
\begin{equation}
{\sigma\over E}~=~{a_1\over \sqrt{E}} \oplus a_2 \biggl[ \biggl({E\over E_0}\biggr)^{l-1}\biggr]
\label{lognoncomp}
\end{equation}
where the parameter $a_2 = |1 - h/e|$ is determined by the degree of non-compensation \cite{deg07}, and $l \sim 0.72$.
It turns out that in the energy range covered by the current generation of test beams, \ie up to 400 GeV, 
Equation \ref{lognoncomp} leads to results that are very similar to those obtained with an expression of the type
\begin{equation}
{\sigma\over E}~=~{c_1 \over \sqrt{E}} + c_2
\label{linres}
\end{equation}
\ie a {\em linear sum} of a stochastic term and a constant term. Many sets of experimental hadronic energy resolution data exhibit exactly this characteristic, for example the results reported for ATLAS \cite{Aba10} shown in Figure \ref{detperformance}b. In this figure, the energy resolution is plotted 
on a scale linear in $-E^{-1/2}$. Scaling with $E^{-1/2}$ is thus represented by a straight line through the bottom right corner in this plot, and the experimental ATLAS data are located on a line that runs parallel to such a line, indicating that the stochastic term ($c_1$) is $\approx 80\%$ and the constant term ($c_2$) is $\approx 5\%$ in this case.

Figure \ref{detperformance}b also shows another interesting phenomenon, namely that the ATLAS hadronic energy resolution was actually measured to be {\sl better} when
the hadronic calorimeter section (Tilecal) was used in stand-alone mode, rather than in combination with the LAr ECAL. The reason for this is the fact that these calorimeter sections have different $e/h$ values. For the Tilecal, an $e/h$ value of $1.336 \pm 0013$ has been reported \cite{Adr09} , while the value for the Pb/LAr 
ECAL, which unlike the Tilecal is very insensitive to the neutrons produced in the shower development, is estimated at $\sim 1.5$ \cite{Wig17}. Typically, the energy deposited by showering hadrons and jets is shared between these compartments, and the large event-to-event fluctuations in this energy sharing translate into an additional contribution to the hadronic energy resolution. This contribution is absent when the showers develop entirely in the Tilecal, hence the better energy resolution.
\begin{figure}[htbp]
\epsfysize=7cm
\centerline{\epsffile{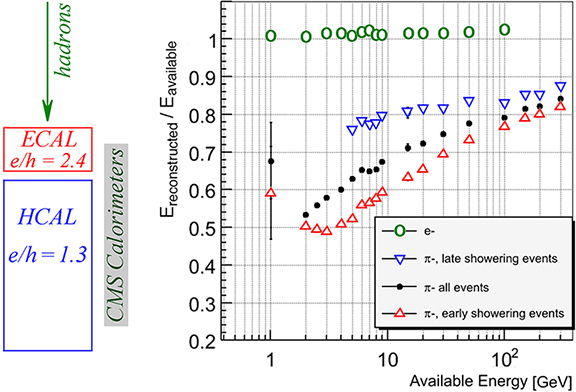}}
\caption{\footnotesize The response to electrons and pions as a function of energy, for the CMS barrel calorimeter. The pion events are subdivided into two samples according to the starting point of the shower, and the pion response is also shown separately for these two samples  \cite{Akc12a}.} 
\label{kazim}
\end{figure}

The discrepancy described above is even much larger for the CMS calorimeter, where the crystal em section has a value of 2.4, while $e/h = 1.3$ for the hadronic section.
The hadronic performance of this calorimeter system was systematically studied with various 
types of particles ($e,\pi,K,p,\bar{p}$), covering a momentum range from 1 -- 300 GeV/$c$ \cite{Akc12a}. Figure \ref{kazim} shows that the response strongly depends on the starting point of the showers.
The figure shows results for two event samples, selected on that basis: showers starting in the em section ($\bigtriangleup$) or in the hadronic section ($\bigtriangledown$). At low energies, the response is more than 50\% larger for the latter (penetrating) events.
In practice in an experiment, it is often 
hard/impossible to determine where the shower starts, especially if these pions are traveling in close proximity to other jet fragments (\eg photons from $\pi^0$ decay) which develop showers in the em section. 

\vskip 2mm
A different calorimeter response to the em and non-em components of hadron showers also leads to differences in the response functions for different types of hadrons. The absorption of different types of hadrons in a calorimeter may differ in very fundamental ways, as a result of 
applicable conservation rules. For example, in interactions induced by a proton or neutron, conservation of baryon number has important consequences.
The same is true for strangeness conservation in the absorption of kaons.
This has implications for the way in which the shower
develops. For example, in the first interaction of a proton, the leading particle has to be a baryon. This precludes the production
of an energetic $\pi^0$ that carries away most of the proton's energy. Similar considerations apply in the absorption of 
strange particles. On the other hand, in pion-induced showers it is not at all uncommon that most of the energy carried by the
incoming particle is transferred to a $\pi^0$. The resulting shower is in that case almost completely electromagnetic. This phenomenon
is the reason for the asymmetric distribution seen in Figure \ref{femprops}b. It also leads to a smaller value of $\langle f_{\rm em} \rangle$ in baryon induced showers, compared to pion induced ones (Equation \ref{femE}).
\begin{figure}[htbp]
\epsfysize=6.5cm
\centerline{\epsffile{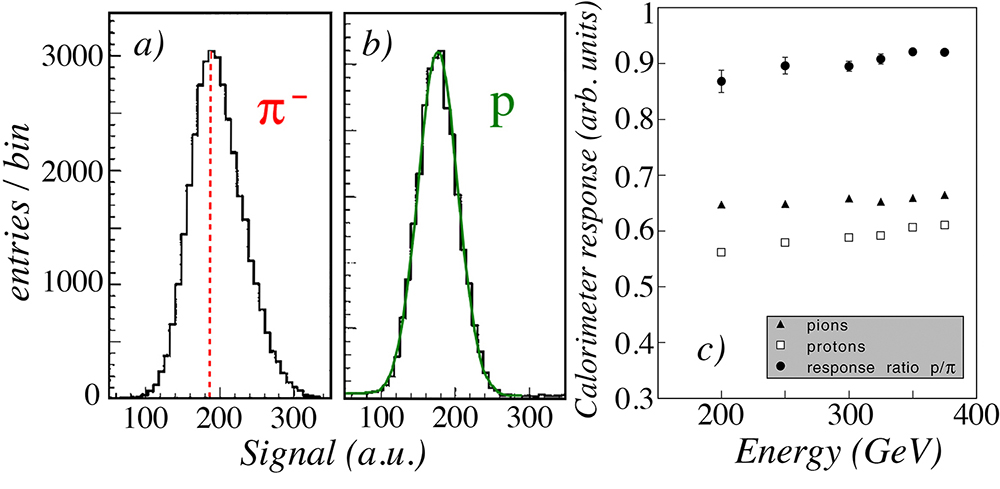}}
\caption{\small
Signal distributions for 300 GeV pions ($a$) and protons ($b$) in the CMS forward calorimeter. Average signals per GeV for protons and 
pions as well as the ratio of these response values in this detector, as a function of energy ($c$) \cite{Akc98}.}
\label{pdif}
\end{figure}

Experimental studies have confirmed these effects \cite{Akc98,Adr10}. Figure \ref{pdif} shows the signal distributions measured for 300 GeV pions ($a$) and protons ($b$), respectively.
The signal distribution for protons is much more symmetric, as indicated by the Gaussian fit. This is because the em component of proton-induced showers is typically populated by $\pi^0$s that share the energy contained in this component more evenly than in pion-induced showers.The figure also shows that the rms width of the proton 
signal distribution is significantly smaller (by $\sim 20\%$) than for the pions. Figure \ref{pdif}c shows that the average signal per GeV deposited energy is smaller for the protons than for the pions, by about 10\%. This is also a consequence of the limitations on $\pi^0$ production that affect the proton signals in this non-compensating calorimeter ($e/h > 1$). So while the response to protons is smaller in this calorimeter, the energy resolution is better. Similar effects are expected to play a role for the detection of kaons, where $\pi^0$ production is limited as a result of strangeness conservation in the shower development.

\subsection{\it The $e/mip$ ratio}

Figure \ref{ehprinciple} shows that the response of typical calorimeters to the em shower component not only differs from that to the non-em component, but it also differs from the response to minimum ionizing particles. The reasons for this are discussed in Section 6.1.
This effect, which I will refer to as $e/mip \ne 1$, may have important consequences for some aspects of the calorimeter performance, even if the calorimeter can be made compensating ($e/h = 1.0$). For example, at low energies, the probability that a hadron is stopped in the calorimeter before it has an opportunity to initiate a nuclear reaction, and thus start a shower, rapidly increases. The entire energy of such hadrons is used to ionize the calorimeter material. There are thus no losses due to invisible energy and, as a result, the calorimeter response to such hadrons is larger than that to hadrons that do develop showers while being absorbed. The response is actually similar to that for muons, which deposit their energy in the same way.
\begin{figure}[b!]
\epsfysize=7cm
\centerline{\epsffile{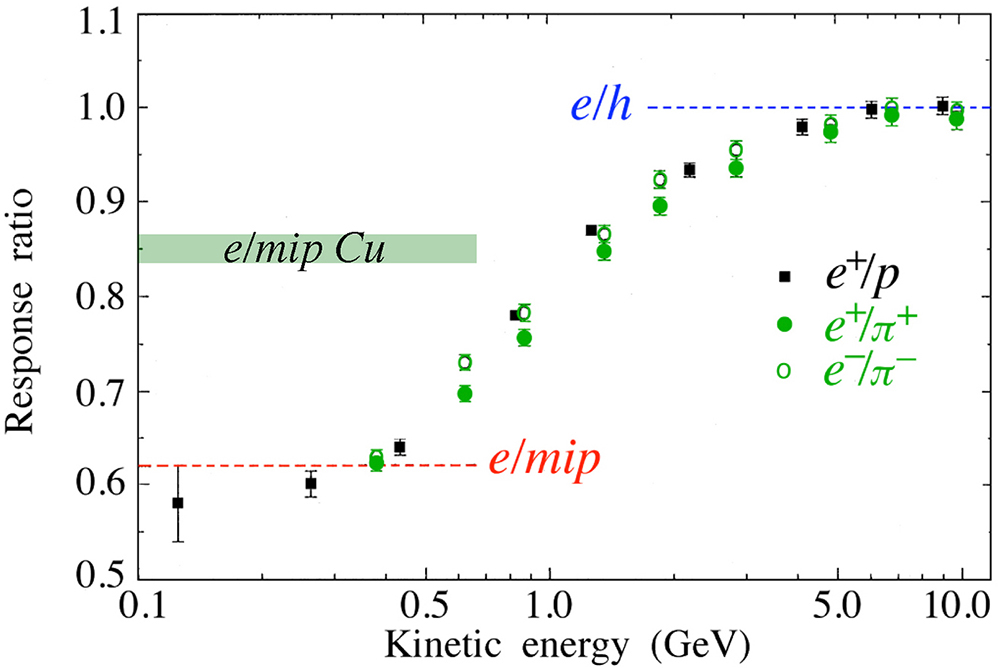}}
\caption{\small
The ratio of the responses of the (compensating) ZEUS calorimeter to electrons and (low-energy) hadrons. This ratio equals 1.0 for energies above  $\sim 10$ GeV. At low energies, the hadron response increases because of the absence of nuclear interactions, and the associated losses in nuclear binding energy. 
Data from \protect\cite{And90}.}
\label{zeuslin}
\end{figure}

\begin{figure}[htbp]
\epsfysize=6.5cm
\centerline{\epsffile{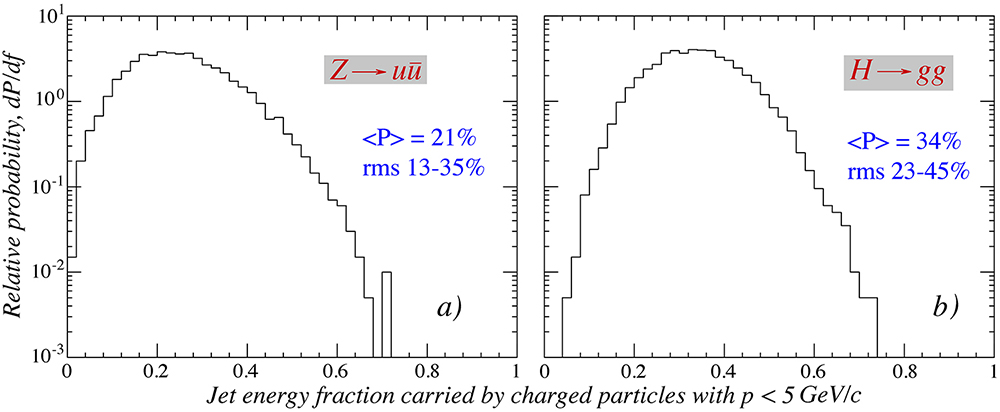}}
\caption{\footnotesize Distribution of the fraction of the energy released by hadronically decaying $Z^0$ ($a$) and $H^0$ ($b$) bosons
at rest that is carried by charged final-state particles with a momentum less than 5 GeV/$c$ \cite{Web15}.} 
\label{webber}
\end{figure}

This is illustrated in Figure \ref{zeuslin}, which shows the response of the 
$^{238}$U/plastic-scintillator ZEUS calorimeter
to low-energy charged hadrons \cite{And90}.
This calorimeter had an $e/h$ value very close to 1.0, but since the $e/mip$ value
was about 0.6, the hadronic response increased for energies below a few GeV, reflecting the increasingly
{\sl mip}-like absorption process. This calorimeter was thus quite non-linear for hadrons with energies less than 5 GeV, which had important consequences for the performance for jet detection. A jet is a collection of particles (mainly pions and $\gamma$s) produced in the fragmentation of a quark or gluon. 
Relatively low-energy fragments account for a significant fraction of the energy of high-energy jets, such as the ones produced in the hadronic decay of the $W$ and $Z$ intermediate vector bosons. Figure \ref{webber} shows the distribution of the energy released by $Z^0$s (decaying through the process $Z^0 \rightarrow u\bar{u}$) and Higgs bosons (decaying into a pair of gluons) at rest that is carried by charged final-state particles with a momentum less than 5 GeV/$c$ \footnote{I would like to thank Dr. Bryan Webber for carrying out these calculations at my request.}. The figure shows that, most probably, 21\% of the energy equivalence of the $Z^0$ mass is carried by such particles, and the event-to-event fluctuations are such that this fraction varies between 13\% and 35\% (for a $1 \sigma_{\rm rms}$ interval). For Higgs bosons decaying into a pair of gluons, the average fraction is even larger, 34\%, with rms variations between 23\% and 45\%.

As a result of the important contribution from soft jet fragments, and the large event-to-event fluctuations in this contribution, the energy resolution for intermediate vector bosons measured with the compensating ZEUS uranium calorimeter turned out to be worse than expected on the basis of the single-pion resolution. 

\section{Methods to improve hadronic calorimeter performance }

The root cause of the poor performance of hadron calorimeters is thus the invisible energy. Because some fraction of the energy carried by the hadrons 
and released in the absorption process does not contribute to the signal, the response to the non-em shower component is typically smaller than that to the em shower component. And the characteristic features of the energy sharing between these two components lead to hadronic signal non-linearity, a poor energy resolution and a non-Gaussian response function.

To mitigate these effects, one has thus to use a measurable quantity that is correlated to the invisible energy. The stronger that correlation, the 
better the hadronic calorimeter performance may become. In this section, two such measurable quantities are discussed: the kinetic energy released by neutrons in the absorption process (Section 5.1) and the total non-em energy (Section 5.2). In Section 5.3, the beneficial effects of both methods are compared.

\subsection{\it Compensation}

The first successful attempt to mitigate the effects described in the previous section involved a calorimeter that used depleted uranium as absorber material. The underlying idea was that the fission energy released in the absorption process would compensate for the invisible energy losses. By boosting the non-em calorimeter response this way, the $e/h$ ratio would increase and, as a matter of good fortune, reach the (ideal) value of 1.0. This is the reason why calorimeters with $e/h = 1.0$ have become known as {\sl compensating} calorimeters. 

Indeed, it turned out that the mentioned effects  of non-compensation on energy resolution, linearity and line shape, as well as the associated calibration problems \cite{CMS07}, are absent in compensating calorimeters. 
However, it also turned out that fission had nothing to do with this, and that the use of uranium was neither necessary nor sufficient
for reaching the compensation condition. The crucial element was rather the active material of the sampling calorimeter, which 
had to be very efficient in detecting the numerous neutrons produced in the shower development process. Hydrogenous active material may
meet that condition, since in a sampling calorimeter with high-$Z$ passive material, MeV type neutrons lose most of their kinetic energy in elastic neutron-proton scattering, whereas the charged particles are sampled according to $dE/dx$.
 
Compensation can thus be achieved in sampling calorimeters with high-$Z$ absorber material and hydrogenous active material. It requires a very specific sampling fraction, so that the response to shower neutrons is boosted by the precise factor needed to equalize $e$ and $h$. For example, in Pb/scintillating-plastic structures, this sampling fraction is $\sim 2\%$ for showers \cite{Bern87,Aco91c,Suz99}. This small
sampling fraction sets a lower limit on the contribution of sampling fluctuations, while the need for efficient detection of MeV-type neutrons requires signal integration over a relatively large volume and at least 30 ns. Yet, the experiment that holds the current world record in hadronic energy resolution (ZEUS, $\sigma/E \sim 35\%/\sqrt{E}$) used a calorimeter of this type \cite{Beh90}). The experimental energy resolution data reported for this calorimeter are shown in Figure \ref{detperformance}b. Especially at high energies, this resolution is much better than that of the calorimeters currently operating in the LHC experiments.

In compensating calorimeters, the total kinetic energy of the neutrons produced in the hadronic shower development thus represents the measurable quantity correlated to the invisible energy. The relative magnitude of the signal provided by these neutrons can be tuned to achieve equality of the electromagnetic and non-electromagnetic calorimeter responses ($e/h = 1.0$), by means of the sampling fraction. This mechanism works because the calorimeter response to charged shower particles is much more sensitive to a change in the sampling fraction than the response to neutrons. 

\subsection{\it Dual-readout calorimetry}

The dual-readout approach \cite{LLW18} aims to achieve the advantages of compensation without the
disadvantages mentioned in the previous section:
\begin{itemize}
\item The need for high-$Z$ absorber material, and the associated small $e/mip$ value, which causes non-linearities at low energy
and deteriorates the jet performance,
\item A small sampling fraction, which limits the em energy resolution,
\item The need to detect MeV-type neutrons efficiently, which implies integrating the signals over relatively large detector volumes and long times.
\end{itemize}

The purpose of the dual-readout technique is to measure the em shower fraction ($f_{\rm em}$) {\sl event by event}. If successful, this would make it possible to diminish/eliminate the effects of fluctuations in $f_{\rm em}$ on the hadronic calorimeter performance.
This was in itself not a new idea. Starting around 1980, attempts have been made to disentangle the energy deposit profiles of hadronic showers with the goal to identify the em components, which are typically characterized by a high localized energy deposit \cite{Abr81}.
Such methods are indeed rather successful for isolated high-energy hadron showers, but fail at low energy, and in particular when a number of particles develop showers in the same vicinity, as is typically the case for jets.  

The dual-readout method exploits the fact that
the energy carried by the non-em shower components is mostly deposited by non-relativistic shower particles (protons), and therefore does not contribute to the signals of a 
\v{C}erenkov calorimeter. By measuring simultaneously $dE/dx$ and the \v{C}erenkov light generated in the shower absorption process, it is possible to determine $f_{\rm em}$ event by event and thus eliminate
(the effects of) its fluctuations. The correct hadron energy can be determined from a combination of both signals.

\begin{figure}[htbp]
\epsfxsize=16cm
\centerline{\epsffile{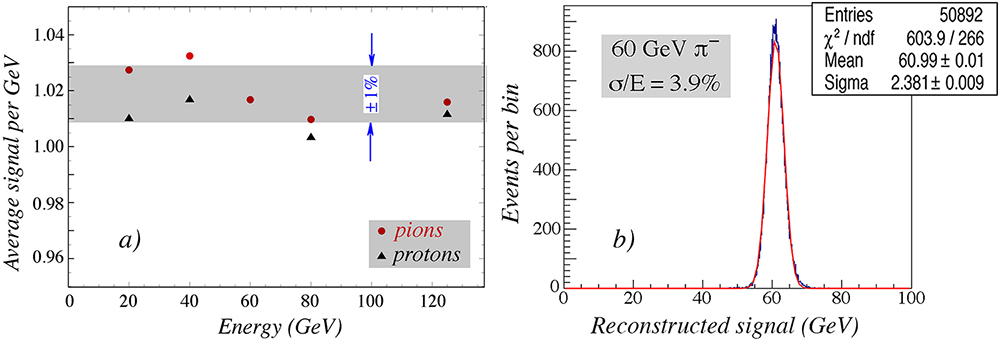}}
\caption{\small
Results obtained with the RD52 lead-fiber dual-readout calorimeter \cite{slee17}. Shown are the average signal per unit deposited energy as a function of energy, separately for pions and protons ($a$), and the measured signal distribution for 60 GeV $\pi^-$ ($b$).} 
\label{rd52}
\end{figure}

This principle was first demonstrated by the DREAM Collaboration \cite{Akc05a}, with a Cu/fiber calorimeter. Scintillating fibers measured $dE/dx$, quartz fibers the 
\v{C}erenkov light.
The response ratio of these two signals was related to $f_{\rm em}$ as
\begin{equation}
{C\over S} ~=~ {f_{\rm em} + 0.21~(1 - f_{\rm em})\over {f_{\rm em} + 0.77~ (1 - f_{\rm em})}}
\label{eq2}
\end{equation}
where 0.21 and 0.77 represent the $h/e$ ratios of the \v{C}erenkov and scintillator calorimeter structures, respectively.  The hadron energy ($E$) could be derived directly from the two signals $S$ and $C$ \cite{deg07}:
\begin{equation}
E~=~ {{S - \chi C}\over {1 - \chi}}
\label{eq5}
\end{equation}
in which $\chi$ is constant, independent of energy and of the particle type, determined solely by the $e/h$ values of the scintillation and \v{C}erenkov calorimeter structures.

Some of the merits of this method are illustrated in Figure \ref{rd52}, which shows that the dual-readout calorimeter is very linear and produces the same response for pions and protons (Figure \ref{rd52}a), that the response function is well described by a Gaussian and, most importantly,
that the hadronic energy was correctly reproduced in this way (Figure \ref{rd52}b). This was true both for single hadrons as well as for multiparticle events \cite{slee17}. 
A major advantage of this method is that the reproduced energy does not depend on the type of hadron. 
Measurements with conventional calorimeters have clearly shown significant differences between the response functions of protons and pions 
(Figure \ref{pdif}). Response differences of $\sim 5\%$ have been reported by ATLAS \cite{Adr09}, while differences in the CMS Forward Calorimeter exceeded 10\% for energies below 100 GeV \cite{Akc98}. This feature translates into a systematic uncertainty in the hadronic energy measurement, unless one knows what type of hadron caused the shower (which at high energies is, in practice, rarely the case). 

In dual-readout calorimeters, the total non-em energy, which can be derived from the measured total energy (Equation \ref{eq5}) and the em shower fraction (Equation \ref{eq2}), thus represents the measurable quantity correlated to the invisible energy. The limitations that apply for compensation do not apply in this case. Any absorber material may be used, as a matter of fact the dual-readout method may even be applied for {\sl homogeneous} calorimeters, such as BGO crystals \cite{Akc09c}. The sampling fraction is not restricted and neutron detection is not a crucial ingredient for this method.
Therefore, one is considerably less constrained when designing a calorimeter system of this type than in the case of a system based on compensation.

\begin{figure}[htbp]
\epsfysize=7cm
\centerline{\epsffile{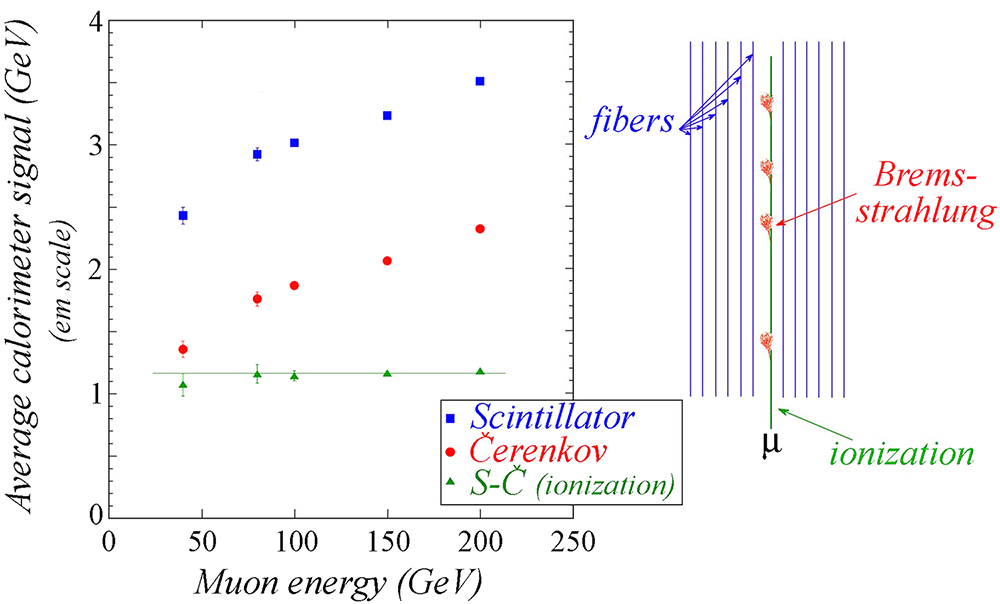}}
\caption{\small Average values of the scintillation and \v{C}erenkov signals from muons traversing the
DREAM calorimeter, as a function of the muon energy. Also shown is the difference between
these signals. All values are expressed in units of GeV, as determined by the electron
calibration of the calorimeter \cite{DREAMmu}.} 
\label{DREAMmuon}
\end{figure}

Simultaneous detection of the scintillation and \v{C}erenkov light produced in the shower development turned out to have other, unforeseen beneficial aspects as well. One such effect concerns the detection of muons. Figure \ref{DREAMmuon} shows the average signals from muons traversing the DREAM calorimeter along the fiber direction \cite{DREAMmu}. The gradual increase of the response with the muon energy is a result of the increased contribution of radiative energy loss (bremsstrahlung) to the signals. The \v{C}erenkov fibers are {\em only} sensitive to this energy loss component, since the primary \v{C}erenkov radiation emitted by the muons falls outside the numerical aperture of the fibers. The constant (energy-independent) difference between the total signals observed in the scintillating and \v{C}erenkov fibers thus represents the non-radiative component of the muon's energy loss. Since the signals from both types of fibers were calibrated with em showers, their responses to the radiative component were equal.  This is a unique example of a detector that separates the energy loss by muons into radiative and non-radiative components.

\subsection{\it Dual-readout \vs compensation}

Compensating calorimeters and dual-readout calorimeters both try to eliminate/mitigate the effects of fluctuations in the invisible energy on the signal distributions by means of a measurable variable that is correlated to the invisible energy. As mentioned in the previous subsections, the variables used for this purpose are different in compensating and dual-readout calorimeters. However, with both methods a very significant improvement of the hadronic calorimeter performance is obtained, compared to the standard non-compensating calorimeters used in the current generation of particle physics experiments:
the hadronic response is energy independent (\ie the calorimeter is linear), the hadronic response function Gaussian, the hadronic energy resolution much better and, most importantly, a calibration with electrons also provides the correct energy 
for hadronic showers.

\begin{figure}[htbp]
\epsfysize=8.5cm
\centerline{\epsffile{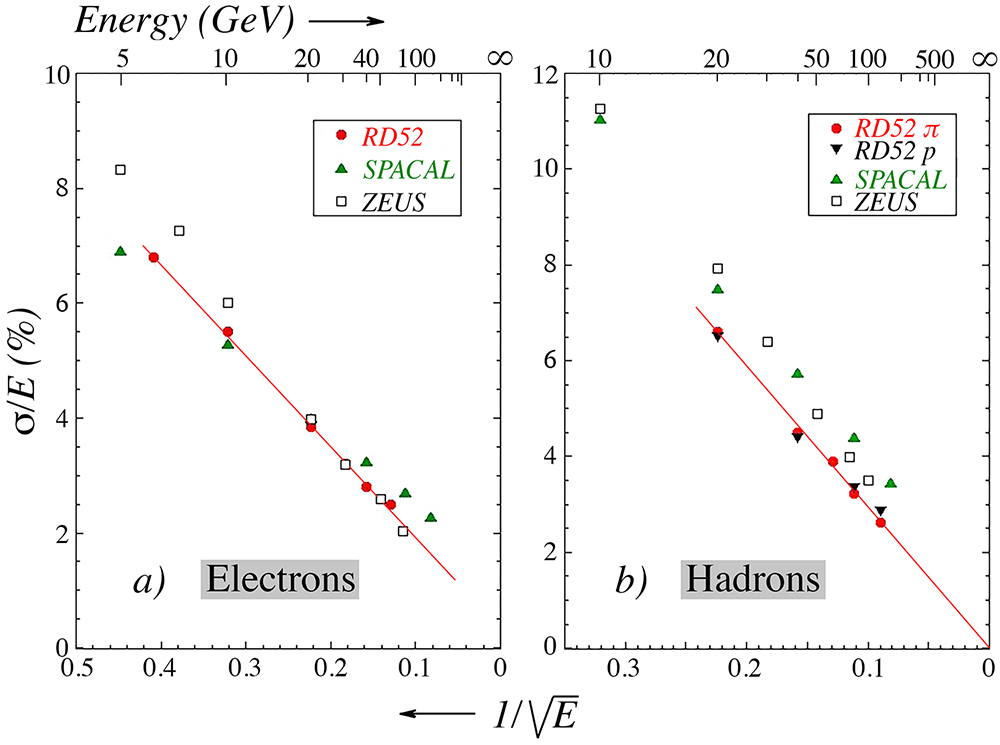}}
\caption{\small
Energy resolutions reported for the detection of electrons($a$) and hadrons ($b$) by RD52 \cite{slee17,Akc14}, SPACAL \cite{Aco91c} and ZEUS \cite{Beh90}.}
\label{Erescomp} 
\end{figure}

In Figure \ref{Erescomp}, the energy resolutions obtained with the best compensating calorimeters, ZEUS \cite{Beh90} and SPACAL \cite{Aco91c}, are compared with the results obtained with the RD52 dual-readout fiber calorimeter. Figure \ref{Erescomp}b shows that the hadronic RD52 values are actually better than the ones reported by ZEUS and SPACAL, while Figure \ref{Erescomp}a shows that the RD52 em energy resolution is certainly not worse.

In making this comparison, it should be kept in mind that
\begin{enumerate}
\item The em energy resolutions shown for RD52 were obtained with the calorimeter oriented at a much smaller angle with the beam line 
($\theta,\phi = 1^\circ ,1.5^\circ$) than the ones for SPACAL ($\theta,\phi = 2^\circ ,3^\circ$) \cite{Akc14}. It has been shown that
the em energy resolution is extremely sensitive to the angle between the beam particles and the fiber axis when this angle is very small \cite{Car16}.
\item The instrumented volume of the RD52 calorimeter (including the leakage counters) was less than 2 tons, while both SPACAL and ZEUS
obtained the reported results with detectors that were sufficiently large ($> 20$ tons) to contain the showers at the 99+\% level. The hadronic resolutions shown for RD52 are dominated by fluctuations in lateral shower leakage, and a larger instrument of this type is 
thus very likely to further improve the results.
\end{enumerate}
The comparison of the hadron results (Figure \ref{Erescomp}b) seems to indicate that the dual-readout approach offers better opportunities to achieve superior hadronic performance than compensation. 
Apparently, in hadronic shower development the correlation with the total nuclear binding energy loss is thus stronger for the 
total non-em energy (derived from the em shower fraction) than for the total kinetic neutron energy. Intuitively, this is not a surprise, since the total non-em energy consists of other components than just neutrons, and the total kinetic energy of the neutrons is 
not an exact measure for the {\sl number} of neutrons (which is the parameter expected to be correlated to the binding energy loss).
\begin{figure}[htbp]
\epsfysize=7.8cm
\centerline{\epsffile{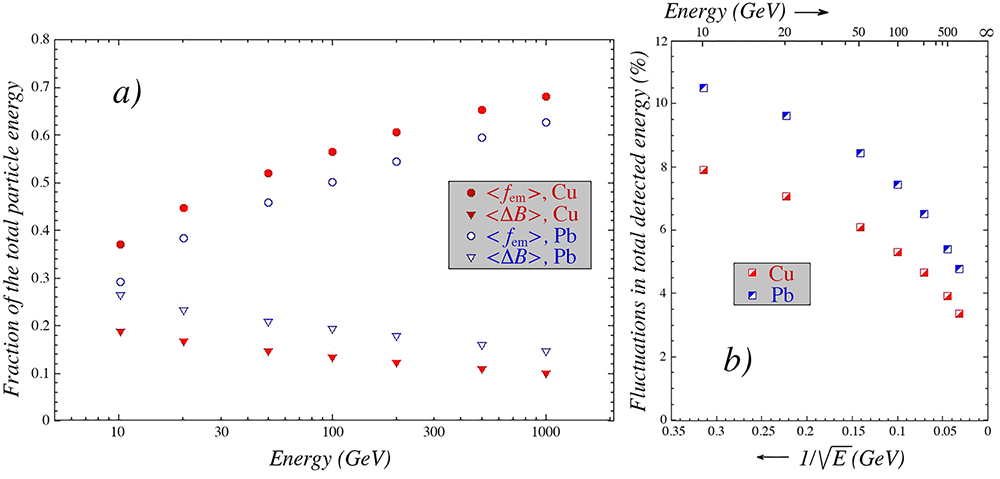}}
\caption{\small
The average value of the em shower fraction and the average fraction of the energy represented by nuclear binding energy losses, for pions absorbed in large blocks of copper and lead, as a function of the pion energy ($a$). Event-to-event fluctuations in the nuclear binding energy losses, expressed as a fraction of the total detected energy ($b$) \cite{Lee17}.}
\label{sehwook5}
\end{figure}

In order to investigate the validity of this interpretation of the experimental results, Lee \etal ~performed Monte Carlo simulations of shower development in a block of matter that was sufficiently large to make the effects of shower leakage insignificantly small. Large blocks of copper or lead were used for this purpose \cite{Lee17}.
The simulations were carried out with the GEANT4 Monte Carlo package \cite{geant} for pions of 10, 20, 50, 100, 200, 500 and 1000 GeV, using the default physics list used in simulations for the CMS and ATLAS experiments at CERN's Large Hadron Collider~\cite{g4_pl}.

\begin{figure}[b!]
\epsfysize=6cm
\centerline{\epsffile{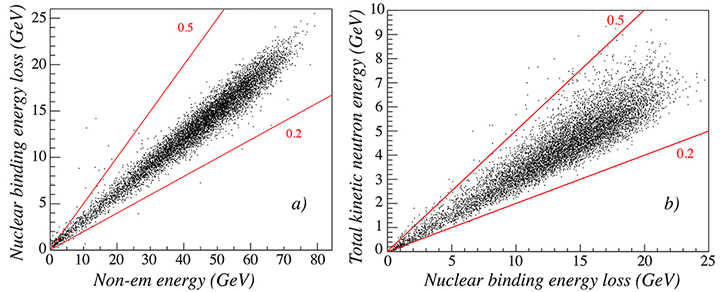}}
\caption{\small
Scatter plots in which the nuclear binding energy losses for 100 GeV pions absorbed in copper are compared event by event to the non-em energy measured with the dual-readout method ($a$) and the total kinetic energy of the neutrons produced in the absorption process, which is the essential ingredient of compensating calorimeters ($b$). The straight lines represent constant values (0.2 or 0.5) of the ratio between the parameters plotted on the vertical and horizontal axes. Results of GEANT4 Monte Carlo simulations \cite{Lee17}.}
\label{Cu100}
\end{figure}

\begin{figure}[htbp]
\epsfysize=6cm
\centerline{\epsffile{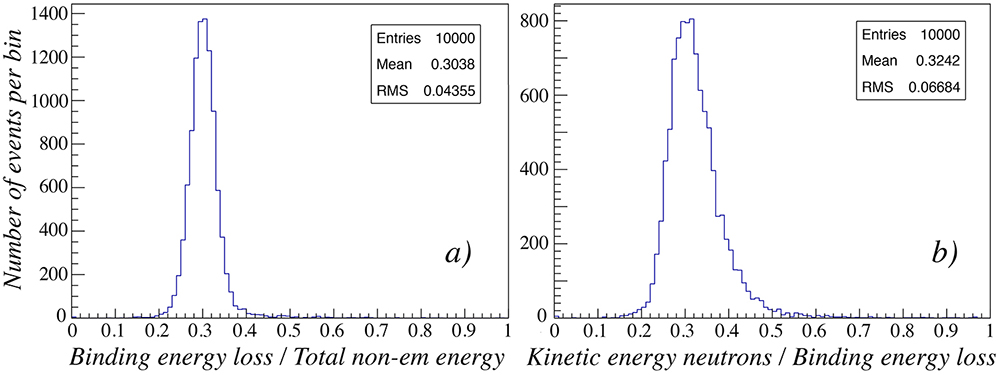}}
\caption{\small
Distributions of the ratio of the non-em energy and the nuclear binding energy loss ($a$) and the ratio of the total kinetic energy carried by neutrons and the nuclear binding energy loss ($b$) for hadron showers generated by 50 GeV $\pi^-$ in a massive block of copper. Results from GEANT Monte Carlo 
simulations \cite{Lee17}.}
\label{sehwook2}
\end{figure}

\begin{figure}[b!]
\epsfysize=8.2cm
\centerline{\epsffile{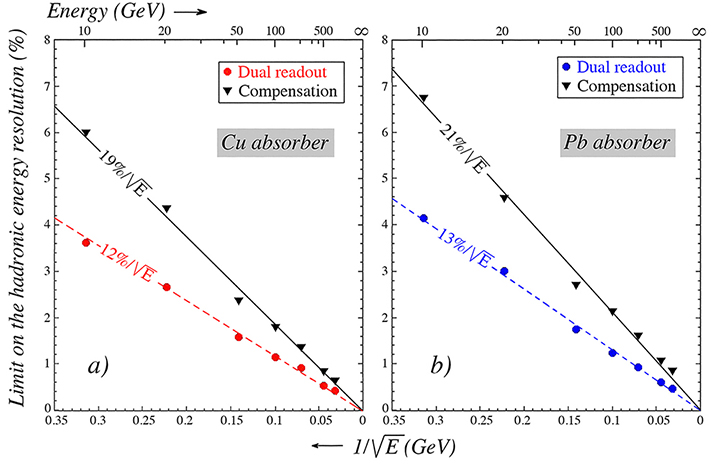}}
\caption{\small
The limit on the hadronic energy resolution derived from the correlation between nuclear binding energy losses and the parameters measured in dual-readout or compensating calorimeters, as a function of the particle energy. The straight lines represent resolutions of $20\%/\sqrt{E}$ and $10\%/\sqrt{E}$, respectively, and are intended for reference purposes.
Results from GEANT Monte Carlo simulations of pion showers developing in a massive block of copper ($a$) or lead ($b$).}
\label{sehwook4}
\end{figure}

Some results of these simulations are shown in Figures 
\ref{sehwook5} - \ref{sehwook4}. Figure \ref{sehwook5}a shows the average value of the em shower fraction and the average fraction of the pion energy represented by nuclear binding energy losses as a function of the pion energy. There are clear differences between copper and lead. The em shower fraction 
increases with energy and is larger for copper (in agreement with the measurement results shown in Figure \ref{femprops}a), while the average binding energy losses decrease with energy and are larger for lead. The event-to-event fluctuations in the binding energy loss are also larger for lead, as illustrated in Figure \ref{sehwook5}b. This figure represents the ultimate precision with which the energy of the pions can be measured in a copper- or lead-based calorimeter in which no effort is made to mitigate the effects of these fluctuations on the 
hadronic energy resolution.

However, both the non-em energy (which follows directly from $f_{\rm em}$) and the total kinetic neutron energy turned out to be clearly correlated with the nuclear binding energy loss (Figure \ref{Cu100}).
To examine the degree of correlation, event-by-event ratios were determined. 
Histograms of these ratios are shown in Figure \ref{sehwook2} for 50 GeV $\pi^-$ showers in copper.

These figures confirm that the correlation between the total non-em energy and the nuclear binding energy loss is better than the correlation between
the total kinetic neutron energy and the nuclear binding energy loss. This was found to be true both for copper and for lead. 
%

These simulations were also used to estimate the effects of the correlations discussed above on the energy resolution for hadron 
calorimeters that are based on dual-readout or compensation. The results are summarized in Figure \ref{sehwook4}, for pions in copper ($a$) and lead ($b$). These resolutions should be considered Monte Carlo predictions for the {\sl ultimate hadronic energy resolution} that can be achieved with calorimeters
using either dual-readout or compensation as the method to mitigate the effects of (fluctuations in) invisible energy. 
A comparison of these results with those from Figure \ref{sehwook5}b shows to what extent these methods are successful in that respect, especially at increasing energy.

The resolution limits scale remarkably well with $E^{-1/2}$, in the energy range considered here (10 - 1000 GeV).
Both for lead and for copper absorber, the limits are considerably better for dual-readout calorimeters  than for compensating ones, in this entire energy range.
The ultimate limit for the energy resolution that can be achieved with calorimetric detection of hadron showers seems to be $\sim 12\%/\sqrt{E}$.

\begin{figure}[htbp]
\epsfysize=8cm
\centerline{\epsffile{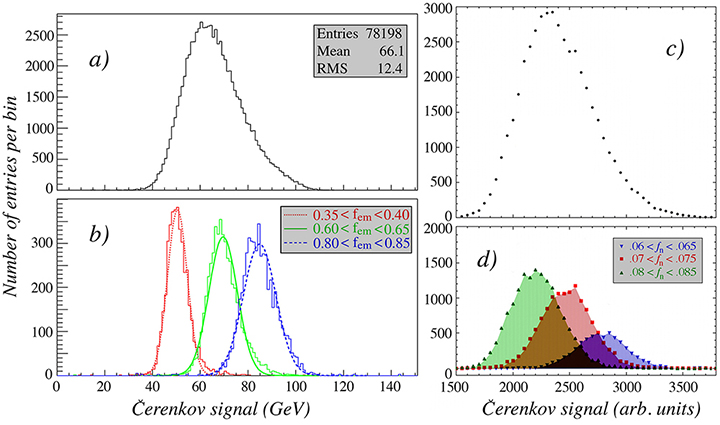}}
\caption{\small
Distribution of the total \v{C}erenkov signal for 100 GeV $\pi^-$ ($a$) and the distributions for three subsets of events selected on the basis of the electromagnetic shower fraction ($b$). Data from \cite{Akc05a}.
Distribution of the total \v{C}erenkov signal for 200 GeV multiparticle events ($c$) and the distributions for three subsets of events selected on the basis of the fractional contribution of neutrons to the scintillator signal ($d$). Data from \cite{Akc09a}.}
\label{femnextract}
\end{figure}

Experimental data obtained by the RD52 Collaboration also support the conclusion that the correlation exploited in dual-readout calorimeters provides a more accurate measurement of the invisible energy. Figures \ref{femnextract}a,b show that the (\v{C}erenkov) signal from the DREAM
fiber calorimeter is actually a superposition of many rather narrow, Gaussian signal distributions. Each sample in Figure \ref{femnextract}b contains events with (approximately) the same $f_{\rm em}$ value, \ie with the same total non-em energy. 
The dual-readout method combines all these different subsamples and centers them around the correct energy value. The result
is a relatively narrow, Gaussian signal distribution with the same central value as for electrons of the same energy.

Figure \ref{femnextract}d shows that the DREAM (\v{C}erenkov) signal is also a superposition of Gaussian signal distributions of a different type. In this  case, each sample consists of events with (approximately) the same total kinetic neutron energy. The dual-readout method may combine all these different subsamples in the same way as described above.
In doing so, the role of the total non-em energy is taken over by the total kinetic neutron energy, and the method becomes thus very similar to the one used in compensating calorimeters.

A comparison between Figures \ref{femnextract}b and \ref{femnextract}d shows that the signal distributions from the event samples are clearly wider when the total kinetic neutron energy is chosen to dissect the overall signal. This is consistent with our assessment that dual-readout is a more effective way to reduce the effects of fluctuations in invisible energy on the hadronic energy resolution.

Apart from that, dual-readout offers also several other crucial advantages:
\begin{itemize}
\item Its use is not limited to high-$Z$ absorber materials.
\item The sampling fraction can be chosen as desired.
\item The performance does not depend on detecting the neutrons produced in the absorption process. Therefore, there 
is no need to integrate the calorimeter signals over a large detector volume.
\item The signal integration time can be limited for the same reason. 
\end{itemize}
\begin{figure}[htb]
\epsfxsize=15cm
\centerline{\epsffile{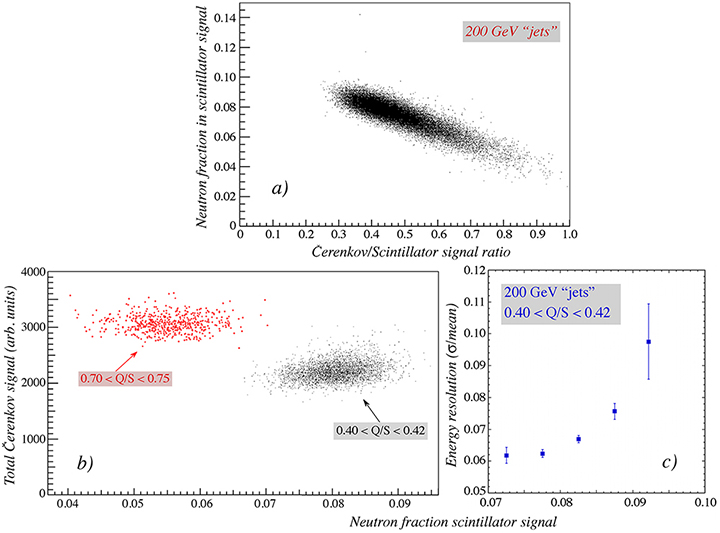}}
\caption{\small
Results obtained with the DREAM copper-fiber dual-readout calorimeter, in which the time structure of the signals was measured \cite{Akc09a}.
A scatter plot in which the measured contribution of neutrons to the signals ($f_n$) is plotted versus the measured ratio ($Q/S$) of the \v{C}erenkov and scintillation signals ($a$). A scatter plot in which the measured \v{C}erenkov signal is plotted versus $f_n$, for two different bins of the $Q/S$ distribution ($b$). The energy resolution as a function of  $f_n$, for events with (approximately) the same $f_{\rm em}$ value ($c$). }  
\label{femfncor}
\end{figure}

This is not to say that there is no advantage in detecting the neutrons produced in the shower development. In fact, this may further improve the hadronic calorimeter resolution, since $f_{\rm em}$ and $f_n$ are correlated with the nuclear binding energy losses in different ways, and thus may offer complementary benefits. Figure \ref{femfncor}a shows that a decrease in the \v{C}erenkov/scintillation signal ratio (from which $f_{\rm em}$ can be derived) corresponds to an increase of the neutron component ($f_n$) of the scintillation signal. However, as shown in Figure \ref{femfncor}b, this correlation is not perfect. In this scatter plot, the $f_n$ values are plotted for two narrow bins in the distribution of the \v{C}erenkov/scintillation signal ratio. In both cases, the $f_n$ values cover a much larger range than the $\pm 2\%$ range of the $f_{\rm em}$ values. Figure \ref{femfncor}c shows that the energy resolution depends
rather strongly on the chosen $f_n$ value, for a given value of $f_{\rm em}$. RD52 has shown that the complementary information provided by measurements of $f_{\rm em}$ and $f_n$ leads to a further improved hadronic energy resolution \cite{Akc09a}. However, even without explicitly determining $f_n$, which involves measuring the time structure of each and every signal, the hadronic energy resolution that can be obtained with the dual-readout method is already superior to what has been achieved by the best compensating calorimeters.

\subsection{\it Exploiting the time structure of the signals}

The availability of ultrafast electronics at a reasonable price has opened new possibilities for applications in calorimetry. 
In Section 3.4, the mitigation of pileup effects in the LHC experiments by means of measurements with a time resolution of 10 ps is discussed. In this subsection, I give some examples of how much more modest time resolutions could help with particle identification and with the elimination of systematic sources of error in calorimetric measurements.

There is a deeply rooted belief that calorimeter systems for high-energy collider experiments should be longitudinally subdivided into several sections.
As a minimum, one will usually want to have an electromagnetic and a hadronic section. A major reason for this belief is that such a subdivision is needed
for recognizing em showers, and thus identify electrons and $\gamma$s entering the calorimeter. 
\begin{figure}[htbp]
\epsfysize=10cm
\centerline{\epsffile{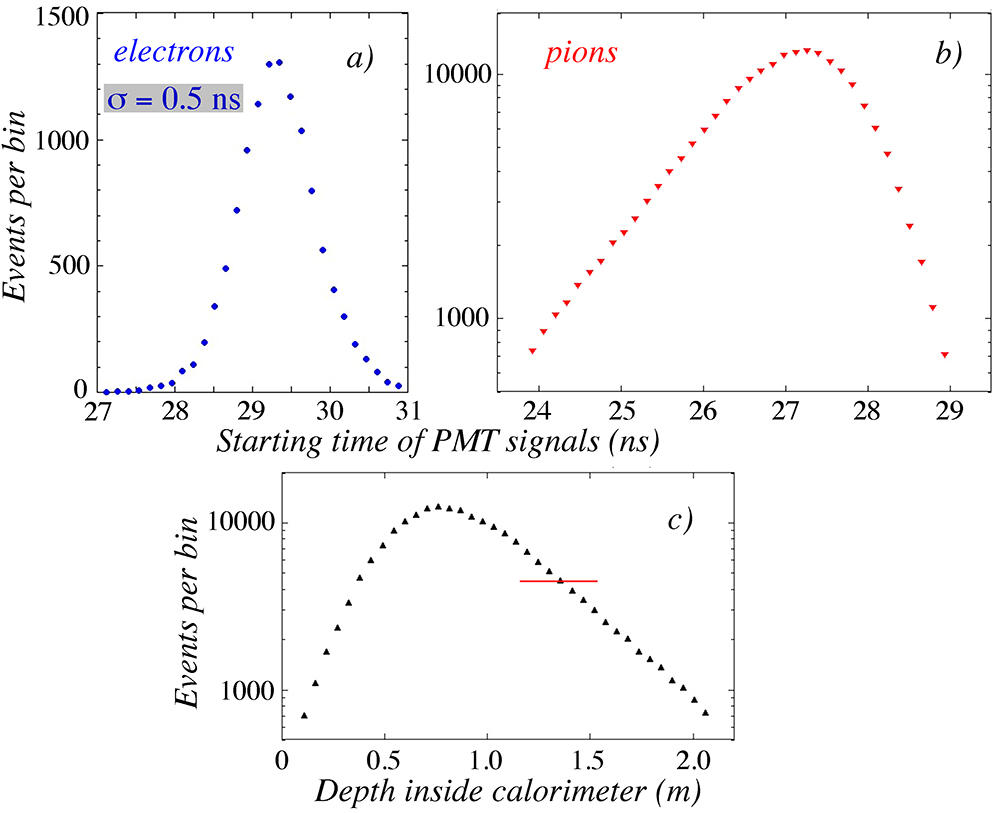}}
\caption{\footnotesize The measured distribution of the starting time of the calorimeter's scintillation signals produced by 60 GeV electrons ($a$) and 60 GeV pions ($b$). This time is measured with respect to the moment the beam particle traversed a trigger counter  installed upstream of the calorimeter. These data were also used to determine the distribution of the average depth at which the light was produced in the hadron showers ($c$). The horizontal line represents the position resolution of this measurement \cite{Akc14b}.}
\label{timing}
\end{figure}

This is a myth. It has been demonstrated repeatedly that there are several ways to identify em showers in longitudinally {\sl unsegmented} calorimeters, and a good time resolution can be a wonderful tool in that respect.
Figure \ref{timing} illustrates one of these methods, developed by the RD52 Collaboration, which uses the starting time of the calorimeter signals, measured with respect to the signal produced in an upstream detector \cite{Akc14b}.
This method is based on the fact that light in the optical fibers travels at a lower speed ($c/n$) than the particles that generate this light ($\sim c$).
The deeper inside the calorimeter the light is produced, the earlier the calorimeter signal starts. For the polystyrene fibers used in this detector, the effect amounted to 2.55 ns/m. 

Figure \ref{timing} shows the measured distribution of the starting time of the signals from 60 GeV $e^-$ (Figure \ref{timing}a) and $\pi^-$ (Figure \ref{timing}b). 
The starting time of the electron signals has a standard deviation of 0.5 ns. Since all em showers generated light at approximate the same depth inside the calorimeter, this time resolution translates into a longitudinal position resolution of $\sim 20$ cm. 
This pion distribution peaked $\sim 1.5$ ns earlier than that of the electrons, which means that the light was, on average, produced 60 cm deeper inside the calorimeter. The distribution is also asymmetric, it
has an exponential tail towards early starting times, \ie light production deep inside the calorimeter. This signal distribution was also used to reconstruct the average depth at which the light was produced for individual pion showers. The result, depicted in Figure \ref{timing}c, essentially shows the longitudinal profile of the 60 GeV pion showers in this calorimeter. 

Apart from particle identification, the measurement of the depth of the light production in a longitudinally unsegmented calorimeter may also turn out to be useful for other purposes. For example, it may be used to correct for the effects of light attenuation in the fibers on the calorimeter signals.
Even though the attenuation lengths are typically (considerably) longer than 5 m, uncertainties in the depth at which the light is produced are not completely negligible in high-resolution hadron calorimeters. The position resolution of 20 cm (indicated by the red line in Figure \ref{timing}c) limits the contribution of light attenuation effects to the energy resolution. It can of course be further improved when better time resolution is available.
\begin{figure}[htbp]
\epsfysize=7.5cm
\centerline{\epsffile{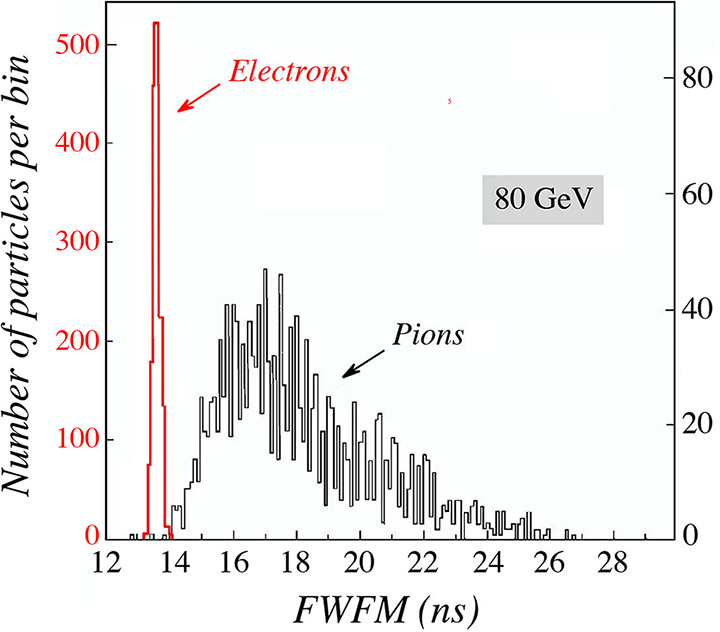}}
\caption{\footnotesize Distribution of the Full Width at one Fifth of the Maximum (FWFM) of the signals produced by 80 GeV electrons and pions in a longitudinally unsegmented fiber calorimeter \cite{Aco91b}.}
\label{pid}
\end{figure}

It is also possible to use the time structure of the calorimeter signals themselves, for example for particle identification. This was demonstrated by the SPACAL Collaboration, who used the pulse width at 20\% of the amplitude (FWFM) to this end \cite{Aco91b} and measured very significant differences between the distributions of this variable for electrons and pions. In their case, the differences were considerably increased by the fact that the
upstream ends of their fibers were made reflective. Therefore, the deeper inside the calorimeter the (scintillation) light was produced, the wider the pulse (Figure \ref{pid}).

As the available time resolution further increases, other applications may become feasible.
For example, the depth measurement in several neighboring towers contributing to the shower signal may provide an indication of the {\sl direction} at which the particle(s) entered the calorimeter, thus allowing measurement of the entire four-vector.

\section{Operating calorimeters}

It is typically not easy to obtain meaningful information about difficulties/complications/problems encountered during the operation of calorimeter systems that are often the heart and soul of modern experiments in particle physics. Reluctance to admit that things are not always as easy/great as one would like or (worse) that crucial mistakes were made in the design, construction and/or commissioning of the detector system is of course fully understandable. However, it would be very helpful for the community to know which mistakes should be avoided in future experiments. For that reason, I include in this paper a section dedicated to some of the issues encountered in practice. For additional information in this context, I would like to refer to \cite{Liv17}.

\subsection{\it Calibration and non-linearity}

Many problems encountered in operating calorimeters derive from the calibration of the instruments, \ie the conversion of the measured signals into deposited energy.  A common, important and consequential misconception about calorimetry is that a shower is a collection of minimum ionizing particles (mips). Already in the early days, it was realized that the signal from a high-energy electron absorbed in a sampling calorimeter
is substantially different from that of a muon that traversed this calorimeter and deposited the same energy in it as the showering electron ($e/mip \ne 1$).
This is due to the fact that the composition of the em shower changes as a function of depth, or age. 
In the late stages, most of the energy is deposited by soft $\gamma$s which undergo Compton scattering or photoelectric absorption, and the sampling fraction for this shower component (\ie the fraction of the energy that contributes to the calorimeter signals) may be very different from that of the mips that dominate the early stages of the shower development. 
%
\begin{figure}[htbp]
\epsfysize=9cm
\centerline{\epsffile{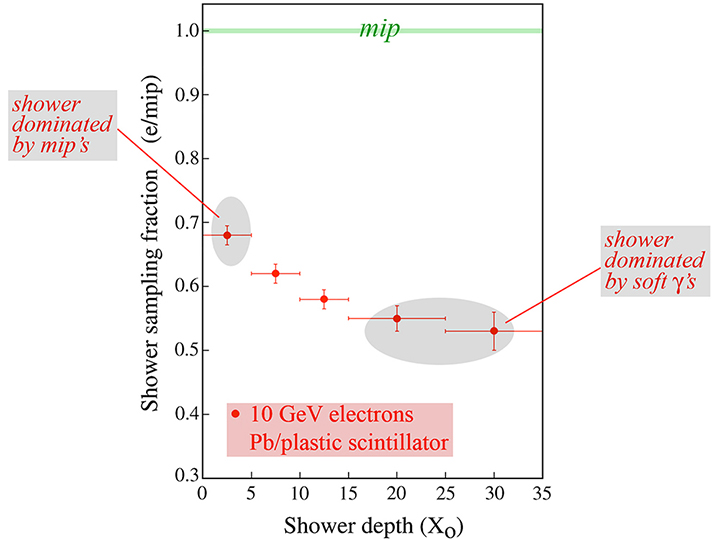}}
\caption{\footnotesize The sampling fraction changes in a developing shower. The local sampling fraction, normalized to that for a minimum ionizing particle, is shown as a function of depth for 10 GeV electrons in a Pb/scintillating-plastic calorimeter. Results of EGS4 calculations \cite{Argonne}.} 
\label{emipz}
\end{figure}

Figure \ref{emipz} illustrates how the sampling fraction of a given calorimeter structure depends on the stage of the developing showers. In calorimeters consisting of high-$Z$ absorber material (\eg lead) and low-$Z$ active material (plastic, liquid argon), the sampling fraction may vary by as much as 25 - 30\% over the volume in which the absorption takes place \cite{Argonne}.

A particular problem resulting from $e/mip \ne 1$ concerns the intercalibration of the different sections of a longitudinally segmented calorimeter system.
Since the longitudinal size of an em shower and the depth at which the shower maximum is reached depend on the energy (Figure \ref{emcont}a), and are also different for electrons and $\gamma$s of the same energy \cite{Zeyrek}, this phenomenon complicates the conversion of the measured signals from a longitudinally segmented calorimeter into deposited energy. These conversion factors (calibration constants) need to depend on the energy and on the nature of the showering particle. If this is done incorrectly, non-linearity is the inevitable consequence. Many experiments have experienced these problems in practice \cite{Aha06,Argonne,Olga,Cer02}. See Reference \cite{Wig17} for an extensive discussion of this issue.

An example of the pitfalls that this causes for calibrating a longitudinally segmented device is provided by the calorimeter for the AMS-02 experiment at the International Space Station \cite{Cer02}. 
This calorimeter has 18 independent longitudinal depth segments. Each segment consists of a lead absorber structure in which large numbers of plastic scintillating fibers are embedded, and is about $1 X_0$ thick. A minimum ionizing particle deposits 11.7 MeV upon traversing such a layer. The AMS-02 collaboration initially calibrated this calorimeter by sending muons through it and equalizing the signals from all 18 longitudinal segments. This seems like a very good method to calibrate this detector, since all layers have exactly the same structure.

\begin{figure}[htbp]
\epsfysize=8cm
\centerline{\epsffile{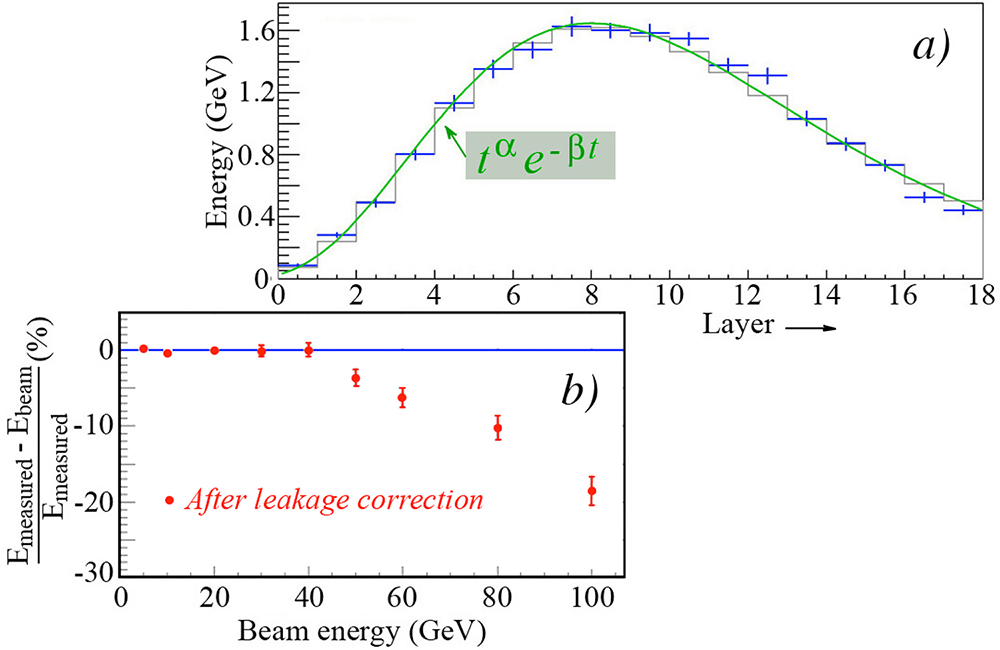}}
\caption{\footnotesize Average signals measured for 20 GeV electrons in the 18 depth segments of the AMS-02 lead/scintillating-fiber calorimeter ($a$). Average relative difference between the measured energy and the beam energy, after leakage corrections based on extrapolation of the fitted shower profile ($b$). Data from \cite{Cer02}.}
\label{ams}
\end{figure}

However, when this calorimeter module was exposed to beams of high energy electrons, it turned out to be highly non-trivial how to reconstruct the energy of these electrons.
Figure \ref{ams}a shows the average signals from 20 GeV electron showers
developing in this calorimeter. These signals were translated into energy deposits based on the described calibration.
The measured data were then fitted to a $\Gamma$-function and since the showers were
not fully contained, the average leakage was estimated by extrapolating this fit to infinity.
As shown in Figure \ref{ams}b, this procedure systematically underestimated this
leakage fraction, more so as the energy (and thus the leakage) increased. The reason for this is that a procedure in which the relationship between measured signals and the corresponding deposited energy is assumed to be the same for each depth segment will cause the energy leakage to be systematically underestimated, more so if that leakage increases. 

Based on these observations, AMS had to change its calibration procedure. Rather than just integrating the $\Gamma$-function that describes the measured 
shower profile to infinity, the energy of the particle that caused the shower is now determined on the basis of the fraction of the total signal measured in the last two detector segments. The deposited energy in the calorimeter derived from the $\Gamma$-function is subsequently corrected with a factor that is determined by this fraction. The value of this correction factor (needed to reproduce the actual electron energy) was established empirically in testbeam measurements \cite{Adl13c}, for the available energy range (up to 250 GeV).
\vskip 2mm
\begin{figure}[htbp]
\epsfysize=6cm
\centerline{\epsffile{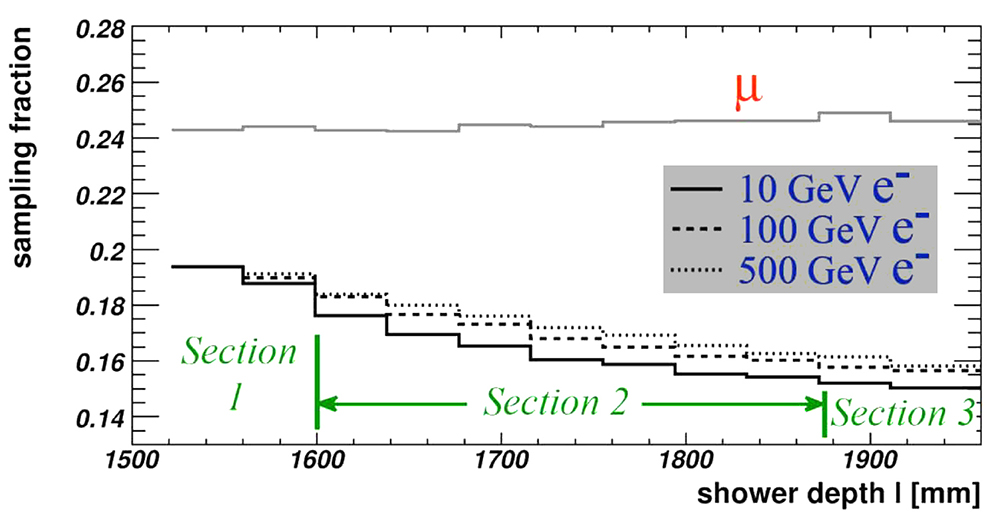}}
\caption{\footnotesize The evolution of the sampling fraction for electron showers of different energies in the three longitudinal segments of the ATLAS LAr calorimeter, at $\eta = 0$ \cite{Aha06}.} 
\label{ATLASemip}
\end{figure}

Similar complications were encountered by ATLAS, in the calibration of their Pb/LAr em calorimeter, which consists of three longitudinal compartments.
At $\eta = 0$, the depths of these segments are $4.3 X_0$, $16 X_0$ and $2 X_0$, respectively.
When the particles enter the barrel calorimeter at a non-perpendicular angle, the total depth of this calorimeter increases (from $22 X_0$ at $\eta = 0$ to $30 X_0$ at $|\eta| = 0.8$),
and so do the depths of these three segments. The sampling fraction for mips is the same in all three segments.

Figure \ref{ATLASemip} shows how the sampling fraction for em showers evolves as a function of depth, in an energy dependent way.
The sampling fraction for muons that traverse this detector does not change in this process and, therefore, the $e/mip$ value decreases by 20--25\%, depending on the 
electron energy. Initially, one tried to calibrate this detector with electron showers by minimizing the value of $Q$ in
\begin{equation} 
Q~=~\sum_{j=1}^N \biggl[E~-~A \sum_{i=1}^n S^{ij}_{\rm 1}~-~B \sum_{i=1}^n S^{ij}_{\rm 2}~-~C \sum_{i=1}^n S^{ij}_{\rm 3}
\biggr]^2
\label{calib3}
\end{equation}
It turned out that the resulting calibration constants  $A$, $B$ and $C$ for the three segments not only depended on the electron energy, but also on the location of the 
calorimeter module (the $\eta$ value). The latter dependence can be understood from the change in the effective depth of the longitudinal segments with
the angle of incidence of the particles. Moreover, any choice of the calibration constants resulting from such a minimization procedure
introduced a non-linearity.
\begin{figure}[htbp]
\epsfysize=10cm
\centerline{\epsffile{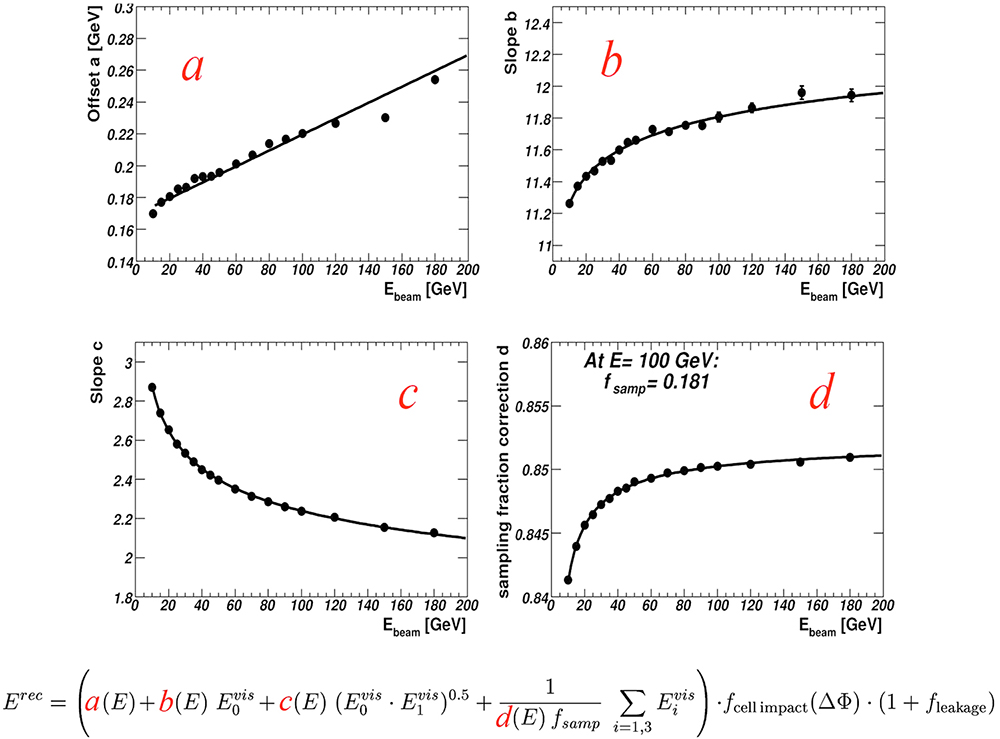}}
\caption{\footnotesize The formula used by ATLAS to determine the energy of a shower developing in the longitudinally segmented ECAL.
The energy dependence of the various parameters is shown in graphs $a - d$ \cite{Aha06}.} 
\label{ATLAS_ecal}
\end{figure}

Rather than intercalibrating the different longitudinal calorimeter segments with muons (as was done by AMS), ATLAS decided to approach this problem with Monte Carlo simulations, in an attempt to achieve (simultaneously) the best possible combination of energy resolution and linearity for showers induced by high-energy electrons \cite{Aha06}.
These elaborate simulations led to a very complicated procedure for determining the energy detected in the various segments of the calorimeter. This procedure was based on a variety of parameters that depended both on the energy and the $\eta$ value (Figure \ref{ATLAS_ecal}).
\vskip 2mm
The energy dependence of the various parameters derives from the change in the longitudinal shower profiles, and thus in the energy sharing between the three segments, with the energy of the incoming electron. As an aside, I mention that the optimum parameter values are likely to be somewhat different for showers induced by $\gamma$s. This is because of the differences between the longitudinal profiles of electron and $\gamma$ showers of the same energy \cite{Zeyrek}.

This very complicated problem will most definitely also affect calorimeters based on Particle Flow Analysis (PFA) \cite{Sef15}, which are all based on structures that are highly segmented, both longitudinally and laterally. 
\begin{figure}[b!]
\epsfysize=7.5cm
\centerline{\epsffile{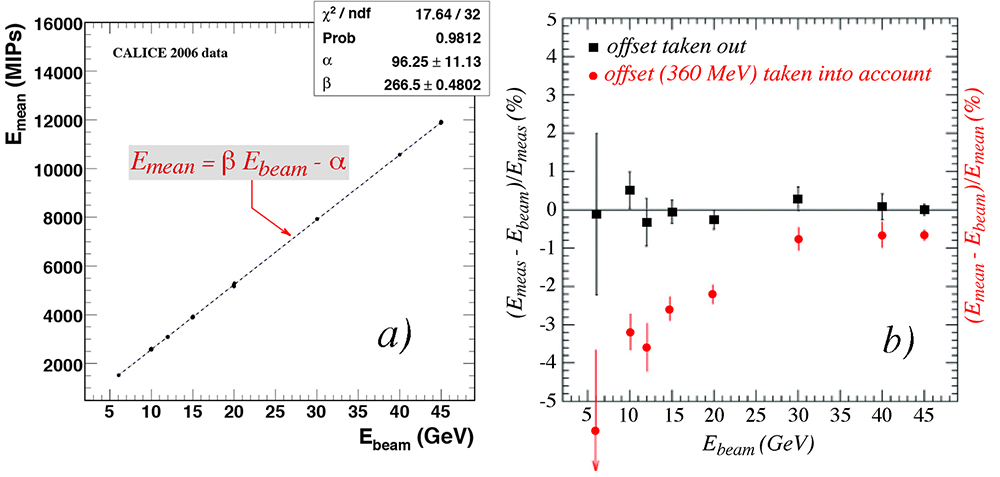}}
\caption{\footnotesize Average signal as a function of electron energy for the W/Si ECAL built by CALICE ($a$) \cite{Adl09}. Residual signals from this detector, before and after taking out a 360 MeV offset ($b$).} 
\label{nonlin}
\end{figure}

Figure \ref{nonlin} shows an example of non-linearity observed in a silicon based em calorimeter of this type \cite{Adl09}). The authors showed that the experimental data could be fit with a straight line (Figure \ref{nonlin}a) and concluded, incorrectly, that this is evidence for linearity.
Signal linearity means that the average calorimeter signal is {\sl proportional} to the deposited energy, \ie the {\sl response is constant}.
The straight line in Figure \ref{nonlin}a does not extrapolate to zero for zero deposited energy. The average signal measured per unit deposited energy gradually increased, by $\sim 5\%$ over the measured energy range from 6 to 45 GeV (Figure \ref{nonlin}b). Therefore, the correct conclusion is that this calorimeter was non-linear, and that the response changed by 5\% over an order of magnitude in deposited energy. 

One of the first full-scale calorimeters at a particle collider that will have to deal with this issue is the HGCAL which is envisaged to replace the radiation damaged endcap section of the CMS calorimeter system \cite{CMS15b}. 
The underlying problem in all these cases is that the relationship between deposited energy and resulting signal is not constant throughout a developing shower. As the composition of the shower changes, so does the sampling fraction. The figures shown in this subsection provide examples of the problems that this may cause. 

\subsection{\it The CMS ``spikes''}

\begin{figure}[htbp]
\epsfysize=4.5cm
\centerline{\epsffile{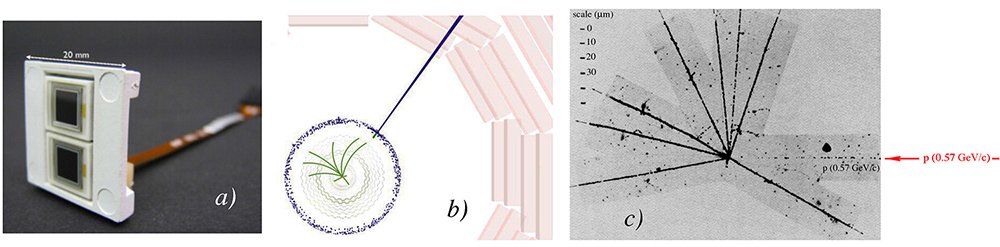}}
\caption{\footnotesize Photograph of two CMS APDs (active area $5\times 5$ mm$^2$) mounted in a capsule ($a$). CMS event display of a $pp$ collision event, showing an isolated ECAL spike (top-right) simulating a 690 GeV transverse energy deposit ($b$) \cite{Pet12}. A nuclear reaction induced by a proton with a kinetic energy of 160 MeV in a photographic emulsion ($c$). }
\label{spike}
\end{figure}

The CMS experiment at CERN's LHC encountered several specific problems upon starting the operations of its calorimeter systems. In this and the following subsection, two of these problems are discussed. When the CMS experiment was designed, much emphasis was placed on excellent performance for the detection of high-energy photons, in view of the envisaged discovery of the Higgs boson through its $H^0 \rightarrow \gamma \gamma$ decay mode. To that end, the Collaboration decided to use PbWO$_4$ crystals for the em section of the calorimeter, since these would provide $\cal{O}$(1\%) energy resolution for the $\gamma$s produced in this process. Since the crystals would have to operate in a strong magnetic field, Avalanche Photo Diodes were chosen to convert the light produced by these crystals into electric signals. Given the available sizes of APDs at that time, and in order to take full advantage of the available (small) light yield, each crystal was equipped with two APDs (Figure \ref{spike}a). However, in order to save some money, these APDs were ganged together and treated as one device in the data acquisition system.

The hadronic section of the CMS calorimeter consists of a sampling structure, based on brass absorber and plastic scintillator plates as active material. Both sections were developed completely independently, and tested separately in different beam lines at CERN. For the tests of the em section, high-energy electron beams were used, the hadronic section was exposed to beams of all available types of particles (electrons, pions, kaons, protons, muons). The performance of both sections was documented in detail, and found to be in agreement with expectations \cite{Adz07,Abd08}.

Yet, when the entire calorimeter system was assembled and exposed to high-energy hadrons, an unexpected surprise occurred \cite{Pet12}. 
In some fraction of the events, anomalously large signals were observed. An example of this phenomenon is shown in Figure \ref{spike}b.
What was going on?

The APDs that convert the light produced in the crystals are also extremely sensitive to ionizing particles. In fact, a mip traversing such an APD may create a signal equivalent to several thousand light quanta \cite{Hau94}. Measurements performed with muons traversing the CMS PbWO$_4$ crystals revealed that a muon that passed through the active layer of an APD generated a signal that was, on average, equivalent to the signal from scintillation photons created by an energy deposit of 160 MeV inside the crystal \cite{Ale97}.
The signals produced by densely ionizing charged particles are correspondingly larger.

In Figure \ref{spike}c, an example is given of interactions that are typical when a high-energy hadron strikes an atomic nucleus. Several densely ionizing nuclear fragments are visible in this picture, with $dE/dx$ values that are up to 100 times larger than that of a mip (see Figure \ref{pinSi}a).
If such a nuclear interaction would take place in the vicinity of an APD, the nuclear fragments traversing the active detector surface area could produce a very large signal. Given the relationship between the signals from mips and from scintillation photons,  such an event could well mimick an energy deposit of 100 GeV or more.

This phenomenon was only discovered when the em and hadronic calorimeter sections were assembled together and exposed to high-energy hadrons. Since the em section corresponds to about one nuclear interaction length, a substantial fraction of hadrons entering the calorimeter start the shower development process in the em section, and therefore the process that generates the ``spikes'' described above becomes a realistic possibility. 

As stated above, the nuclear reactions have to take place ``close to'' the sensor surface of the APD in order to produce this effect.
The scale on Figure \ref{spike}c clarifies what ``close'' means in this context, \ie within 100 $\mu$m or so. This also provides the answer to the question how this problem could have been avoided. Since the two APDs that read out each crystal are separated by several mm, a nuclear interaction of the type discussed here would never affect both APDs, but only one of them. Therefore, if the two APDs had been read out separately, instead of being treated as one detector unit in the data acquisition system, ``spike'' events would be easily recognized since the signal in only one of the APDs would be anomalously large, while the signal from the other one would still provide the useful information one would like to obtain from the event in question.

However, since the two groups that were responsible for the two calorimeter sections worked in their own individual, separate universes, this
problem was only discovered when it was too late to make the corrections needed to avoid it. I include this example here because it illustrates that it is important to realize that a calorimeter built for a given experiment is not a stand-alone device, but is part of an integrated system of detectors. The different components of this detector system may affect the performance of each other and it is important to realize and test this in the earliest possible stage of the experiment.

\vskip 5mm
Another example of an instrumental effect that is caused by one individual shower particle is the so-called ``Texas tower effect.''
This effect is a consequence of the very small sampling fraction of some calorimeters, and in particular those with a gaseous active medium that contains hydrogen. The sampling fraction of such calorimeters is typically 
$\cal{O}$(10$^{-5})$,
which means that the energy deposited by a showering 100 GeV hadron in the active material amounts to $\sim 1$ MeV. However, when one of the numerous MeV-type neutrons produced in this process scatters off a hydrogen nucleus, a comparable amount of energy may be deposited 
in the gas layer {\sl by that one recoil proton!}
\begin{figure}[htbp]
\epsfysize=6cm
\centerline{\epsffile{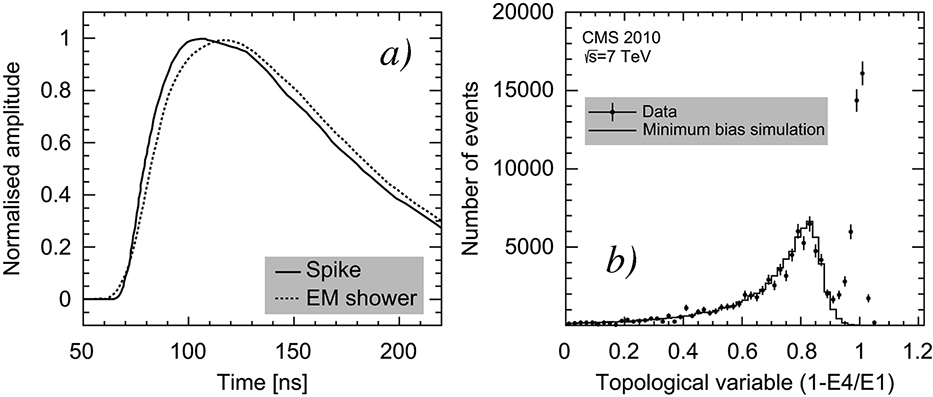}}
\caption{\footnotesize Distinguishing characteristics of CMS ``spike'' events. Shown are the average pulse shape ($a$) and the distribution of the so-called {\sl Swiss Cross} topological variable for the highest energy deposit in each event  ($b$). See text for details \cite{Pet12}.}
\label{spikerecognize}
\end{figure}

In beam tests, where a beam of mono-energetic particles is sent into the calorimeter, such anomalous events are easily recognized, and can be removed from the event samples. However, during the operation of such a calorimeter in an accelerator environment, this is not (always) possible, especially if no additional information about the event in question, \eg from a tracker system, is available. This difficulty led the CDF Collaboration to the decision to scrap their gas-based forward calorimeter, and replace it with a device that had a sampling fraction that was three orders of magnitude larger \cite{Alb02a}.

This phenomenon is very similar, at least in its consequences, to the ``spikes'' observed for hadron detection in the CMS calorimeter system, discussed above. Also there, one low-energy shower particle may cause an event in which an anomalously large amount of energy seems to be
deposited in the detector (see Figure \ref{spike}). Contrary to CDF, CMS has made an effort to deal with the real-life consequences of this phenomenon, and has succeeded in recognizing, and eliminating, affected events to a reasonable extent. This is illustrated in Figure \ref{spikerecognize}, which shows two characteristics that distinguish these ``spike'' events from the ``normal'' ones. One effective method is a cut on the so-called {\sl Swiss Cross} variable, in which the signal in each tower of the em calorimeter is compared to the sum of the signals in the four neighboring towers.
Monte Carlo simulations showed that a cut of events in which $E4/E1 < 0.05$ is an effective tool for eliminating the ``spike'' events \cite{Pet12}.
\begin{figure}[b!]
\epsfysize=8.5cm
\centerline{\epsffile{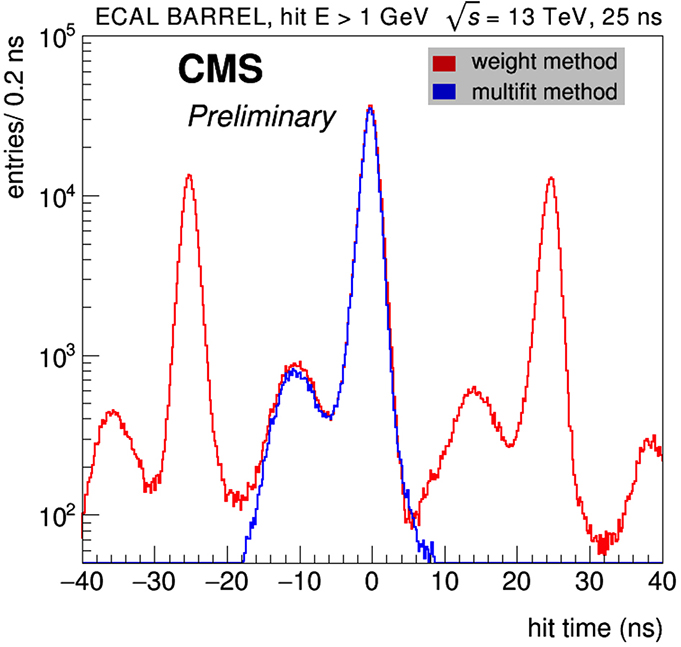}}
\caption{\footnotesize CMS ``spike'' events may be distinguished from regular (scintillation) events by means of the time structure \cite{CMS10}. See text for details.}
\label{spiketime}
\end{figure}

Also the time structure of the spike events is a differentiating characteristic. Since the signals in this case are caused by shower particles traversing the APDs, they are faster than the signals based on detection
of the scintillation light generated by the shower particles in the crystals. The latter are delayed because the molecules excited in the scintillation process take some time to decay ($\sim$10 ns). Figure \ref{spiketime} shows
the time distribution of the hits in the ECAL barrel with a reconstructed energy above 1 GeV, for an 80 ns time slice of the LHC operations. The 25 ns bunch structure is clearly visible. The events caused by the scintillation light are characterized by the peaks at -25 ns, 0 ns and + 25 ns in this plot. These peaks are preceded by smaller peaks that occur about 10 ns earlier. These are the
spike events, and the figure shows that these represent about  3\% of the total. It should be emphasized that this plot concerns the time characteristics of the events that {\sl survived} the Swiss Cross cuts.

Based on the understanding of the underlying cause of this phenomenon, it is expected that the frequency of these events will increase both with the luminosity and with the center-of-mass energy of the $pp$ collisions in the LHC. 
The reason why this example is included in this context is that it is an inherent feature of the CMS calorimeter system. Even though attempts to deal with the problem may seem successful, it is good to keep in mind that any data selection based on the calorimeter information alone 
may lead to a biased event sample. Also, it is inevitable that some fraction of the events in which the process that causes
a ``spike'' occurs will {\sl not} be eliminated by the cuts devised to deal with the problem.

Given the origin of the problems described in this subsection, other examples of catastrophic effects caused by a single shower particle may be expected in the High-Granularity calorimeter that is scheduled to replace the radiation damaged endcaps of the CMS calorimeter \cite{CMS15b}. For example, in the hadronic section of this instrument, 5 cm thick brass plates will be interleaved with 200 $\mu$m thick silicon sensors that act as active material. The sampling fraction for mips is thus $6\cdot 10^{-4}$ in this (FH) section, and even less for showers. A nuclear interaction such as the one shown in Figure \ref{spike}c may easily deposit 30 MeV in the silicon sensor if the event takes place in the vicinity of the boundary between the active and passive material. Since the silicon signals do not saturate for densely ionizing particles, this event will be interpreted as a 50 GeV highly localized energy deposit. The nuclear reactions responsible for this phenomenon are typically initiated by 
spallation neutrons with kinetic energies of a few hundred MeV. Such neutrons may travel tens of centimeters away from the shower axis before initiating the reaction that gives rise to the large signal. They will also be prolifically produced in the absorption of soft hadrons that constitute the ``pileup'' component of the events CMS is looking for, and one should thus expect very large signals at (multiple) random locations in the calorimeter, for essentially every bunch crossing \footnote{Note to CMS: Don't claim you were not warned.}.
Just like the Texas Tower effect, the described phenomena are a typical consequence of the development of hadron showers in a calorimeter with a small sampling fraction and a non-saturating active medium, and therefore play no role in calorimeters based on plastic-scintillator or liquid-argon readout.

\subsection{\it Signals from external sources}

\begin{figure}[htbp]
\epsfysize=7cm
\centerline{\epsffile{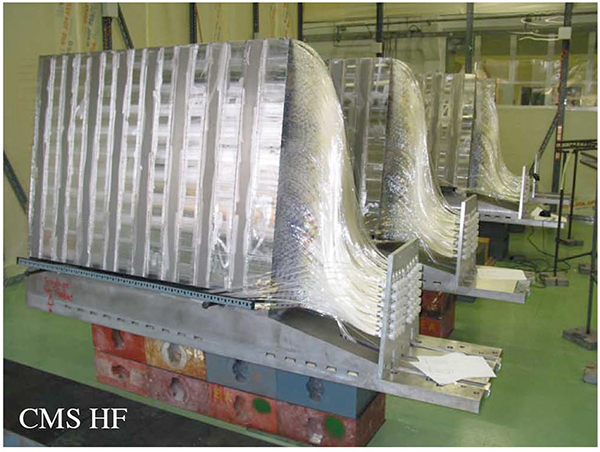}}
\caption{\footnotesize Three modules of the CMS forward calorimeter, which consists of $2\times 12$ such wedges. The quartz fibers that serve as the active material in this sampling calorimeter are bunched towards a readout box, where PMTs convert the \v{C}erenkov light signals into electric pulses.}
\label{CMSHF}
\end{figure}

A third class of problems encountered in operating calorimeters is caused by particles that are not supposed to contribute to the signals, since they have nothing to do with the event that is being studied. Also in this case, the CMS calorimeter system has provided a useful example. The very-forward section of the calorimeter system ($3 < \eta <  5$) is based on quartz fibers embedded in an iron absorber structure (Figure \ref{CMSHF}). High-purity quartz, chosen because of its radiation hardness, provides signals consisting of \v{C}erenkov light generated in it by charged relativistic shower particles, and the fiber structure makes it possible to transport this light to the outside world, where it is converted into electric signals by means of PMTs. 

In order to save money, the packing fraction of the quartz fibers was reduced to the point where one detected photoelectron corresponds to 5 GeV deposited energy. In other words, a 100 GeV particle absorbed in this calorimeter provides a signal of about 20 photoelectrons. However, relativistic charged particles that traverse the glass windows of the PMTs generate \v{C}erenkov light in this glass that leads to signals of the same (or larger) amplitude. There is no shortage of such particles during the LHC operations. The proton beams are surrounded by an intense halo of high-energy muons, which are barely affected by the material constituting the CMS detectors. When these muons traverse a PMT that is looking for light signals created by showers inside the calorimeter, they generate signals that in practice may make the information provided  by the very-forward calorimeters rather meaningless for the physics analyses, and most definitely renders these calorimeters useless for triggering purposes.

The solution envisaged for this problem is to replace the readout by sensors that are less subject to this effect \cite{Gul17}. In a first stage, the replacement sensors will be multi-anode PMTs with thinner glass windows. The idea is that light produced in the calorimeter will cause similar signals in all anodes of the PMT, while the signal from one single anode will dominate when (localized) light produced by a particle traversing the glass window of a PMT is detected.
In addition, algorithms intended to distinguish the stray muon signals from those generated by particles absorbed in the calorimeter are being developed. To that end, the time structure of the events is being used, just as was done to handle the spike events discussed in Section 6.2 (Figure \ref{spiketime}). In a later stage, CMS plans to replace the entire readout by a system based on silicon photomultipliers.

\section{Particle Flow Analysis}

In the past 15 years, a lot of effort has been spent on a completely different approach to achieve good performance for jet detection. This approach is
usually referred to as {\sl Particle Flow Analysis}, or PFA. 
This method is based on the combined use of a precision tracker and a highly-granular calorimeter. The idea is that the charged jet particles can be precisely measured with the tracker, while the energy of the neutral particles is measured with the calorimeter. 
Such methods have indeed been used with some success to improve the mass resolution of hadronically decaying  $Z^0$s at LEP \cite{Bus95},
and the jet energy resolution using $\gamma$-jet $p_T$ balancing events at CDF \cite{Boc01} and at CMS \cite{CMS09}. Two detector concepts studied in the context of the experimental program for the proposed International Linear Collider (ILC) are based on this method as well \cite{Beh13}.

The problem that limits the success of this method is that the calorimeter does not know or care whether the particles
it absorbs are electrically charged. Therefore, the detected calorimeter signals will have to be corrected for the contributions of the showering charged jet fragments. 
Proponents of this method have advocated a fine granularity as the key to the solution of this ``double-counting'' problem \cite{Sef15}. 
However, it has been argued by others that a fine granularity is in practice largely irrelevant in a compact 4$\pi$ experiment, because of the overlap between the showers from individual jet components, especially for jets with leading charged particles \cite{Lob02}.

Systems based on many millions of electronic channels have been proposed and are, in one specific case, being built \cite{CMS15b}. Interestingly,  
these millions of channels are primarily needed to discard the information they provide (on the showering charged particles, whose momenta are being measured more accurately by the tracker system).
 
In order to increase the spatial separation between showers induced by the various jet particles,
and thus alleviate the double-counting problem, the concept detectors for the ILC that are based on the PFA principle count on strong solenoidal magnetic fields (4 - 5T). Such fields may indeed improve the validity of
PFA algorithms, especially at large distances from the vertex, since they open up a collimated beam of particles. It is important to be quantitative in these matters. After having traveled a typical distance of 1 m in a 4~T magnetic field,
the trajectory of a 10 GeV pion deviates by 6 cm from that of a straight line, \ie less than one third of a nuclear interaction length (the characteristic length scale for lateral hadronic shower development)
in a typical calorimeter. The field is also not necessarily beneficial, since it may have the effect of bending jet particles with a relatively large transverse momentum with respect to the jet axis {\em into} the jet core. 

Of course, in the absence of reliable Monte Carlo simulations\footnote{Concern about the absence of 
reliable simulations for hadronic shower development was the main reason for a special workshop held at Fermilab in 2006 \cite{HSS06}. To my knowledge, the fundamental problems addressed at this workshop, \eg with regard to the hadronic shower widths that are crucial for PFA,  still exist.}
the only way to (dis)prove the advocated merits of the proposed PFA methods is by means of dedicated experiments in realistic prototype studies.

\subsection{\it The importance of calorimetry for PFA}

The first statement in any talk about PFA mentions that 2/3 of the final-state particles constituting a jet are electrically charged, and that the momenta of these particles can be measured extremely precisely. This is true, but the implication that the calorimeters of PFA based detector systems don't have to be very good, since they only have to measure one third of the jet energy, is incorrect.
In the absence of calorimeter information, based on tracker information alone, the jet energy resolution would be determined
by the {\sl fluctuations} in the fraction of the total jet energy that is carried by the charged fragments. This issue was studied by Lobban \etal, who found that these event-to-event fluctuations are very large. Depending on the chosen jet fragmentation function, the distribution of the energy fraction carried by the charged jet fragments was found to have a $\sigma_{\rm rms}$ equal to 25 -- 30\% of the average value, {\sl independent of the jet energy} \cite{Lob02}.

 One may wonder why these
fluctuations do not become smaller at higher energies, given that the {\em number} of jet fragments
increases. The reason for this is that the observed increase in multiplicity is uniquely caused by the
addition of more {\em soft} particles. The bulk of the jet energy is invariably carried by a small number of
the most energetic particles. This means that the fraction of the jet energy
carried by charged particles is strongly dependent on the extent to which these particular particles participate in the
``leading'' component of the jet. Therefore, the event-to-event fluctuations in this fraction are
large and do {\em not} become significantly smaller as the jet energy increases.

As an aside, I mention that the same argument thus necessarily also applies to the event-to-event fluctuations in the energy 
fraction carried by the neutral particles (mainly $\pi^0$s). These fluctuations are responsible for
the poor jet energy resolution of non-compensating calorimeters, especially at high energy, 
since the response of such calorimeters is usually considerably larger for the em jet components than for non-em ones.

In the absence of a calorimeter, measurements of
jet energy resolutions better than 25 -- 30\% can therefore not be expected on the basis of tracker information alone, {\em at any energy}. And since the 
contributions of showering charged particles to the calorimeter signals have to be discounted properly for the
PFA method to work, the quality of the calorimeter information is in practice very important for achieving
the energy resolution of 4 - 5\% needed to distinguish between hadronically decaying $W$ and $Z$ bosons, which is a major design goal of every experiment planned for a future high-energy $e^+e^-$ collider.

\subsection{\it PFA at LEP, the Tevatron and the LHC}

One of the conclusions of the analysis described in \cite{Lob02} was that the PFA approach may result in improving the jet energy resolution of ``poor'' calorimeter systems to become ``mediocre,'' but that it will do little for the performance of calorimeters with ``mediocre'' or ``good'' resolution. 
This can be understood by
considering the extreme cases: for a perfect calorimeter, there is nothing left for a tracker to improve upon, while for no
calorimeter at all, the tracker would still give 30\% resolution for the jets.
\begin{figure}[thbp]
\epsfysize=7.5cm
\centerline{\epsffile{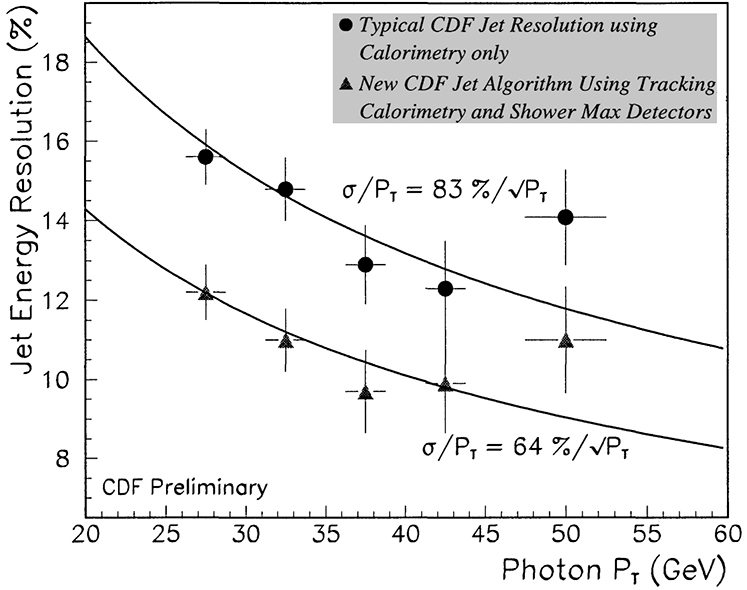}}
\caption{\footnotesize  The effect of including information from the tracking system and the shower max detectors on the jet energy resolution measured with the CDF detector, for jets in the central rapidity (barrel) region \cite{Boc01}.} 
\label{cdf_pfa}
\end{figure}

Practical experience so far seems to confirm this assessment. 
The first experiment in which PFA was elaborately applied was ALEPH \cite{Bus95}, one of the LEP experiments. The hadron calorimeter was not considered a very important component of the LEP detectors, which had, on the other hand, excellent tracking systems. 
Using a specific (biased) subsample of hadronically decaying $Z^0$s at rest\footnote{Any event in which energy was deposited within 12$^\circ$ of the beam line, as well as any event in which more than 10\% of the total energy was deposited within 30$^\circ$ of the beam line, was removed from the event sample used for this analysis.}, the authors exploited the properties of this tracking system to the fullest extent and achieved an energy resolution of 6.2 GeV, an improvement of about  25\% with respect to the reported hadronic energy resolution of the stand-alone calorimeter system.

The CDF experiment at the Tevatron also used PFA techniques to improve their jet energy resolution.
Figure \ref{cdf_pfa} shows the effects of including information from the tracking system and the shower max detectors on the measured jet energy, for jets produced at central rapidities, \ie fragments entering the calorimeter in the barrel region \cite{Boc01}.
\begin{figure}[b!]
\epsfysize=7cm
\centerline{\epsffile{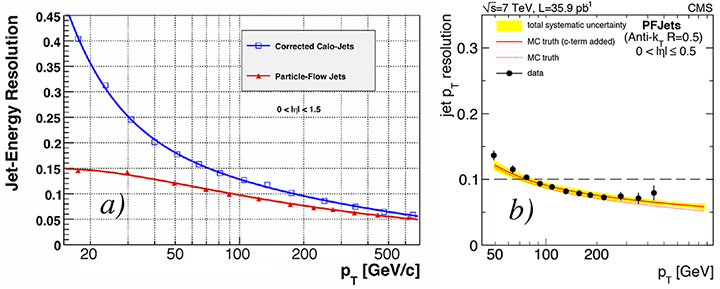}}
\caption{\footnotesize Simulated jet energy resolution in CMS as a function of transverse momentum, effect of PFA techniques ($a$). Measured  jet energy resolution using PFA techniques as a function of energy ($b$) \cite{CMS09}. } 
\label{cms_pfa}
\end{figure}

The CMS experiment took advantage of their all-silicon tracking system, plus a fine-grained ECAL, to improve their jet energy resolution. Figure \ref{cms_pfa}a shows the expected improvement of the jet energy resolution if PFA techniques would be used, which decreases from $\sim 30\%$ at 50 GeV to $\sim 20\%$ at 100 GeV and $\sim 10\%$ at 500 GeV in the energy range for which these predictions could be tested. The experimental data (Figure \ref{cms_pfa}b) show a somewhat smaller improvement at the lowest and highest energies measured for this purpose. However,
the improvement that resulted from the use of PFA was also here certainly significant. 

The ATLAS calorimeter system measures jets with much greater precision than CMS and, therefore, the replacement of the calorimeter information using the tracker data did not lead to significant benefits. However, the tracker data did help mitigating pileup effects. By subtracting the energy (momenta) of tracks
that did not point to the jet vertex from the measured calorimeter energy, a better measurement of the jet energy was obtained \cite{Aab17}.

Encouraged by the observed improvements in the jet performance, CMS has decided to replace its entire endcap calorimeter system with a dedicated PFA detector \cite{CMS15b}. This system, which is designed to comprise about six million electronic channels, is scheduled to replace the current endcap calorimeters in the forward region around 2025. It is intended to mitigate the problems of radiation damage and event pile-up, which are expected to have rendered the current system (consisting of PbWO$_4$ crystals, backed up by a brass/plastic-scintillator hadronic section) ineffective by then. The new system will consist of $5 \lambda_{\rm int}$ deep fine-grained calorimetry, 40 sampling layers with 1 to 1.5 cm$^2$ silicon 
pads as active material, backed up by another $5 \lambda_{\rm int}$ of ``conventional'' calorimetry. Also the tracking system upstream of this 
calorimeter will be replaced, with upgrades foreseen both in granularity and in $\eta$-coverage.

\subsection{\it PFA calorimeter R\&D}

A large collaboration, called CALICE, has set out to test the viability of the PFA ideas. In the past 10-15 years, they have constructed a variety of calorimeters, both for the detection of em showers as well as hadronic ones. The replacement of the CMS endcap calorimeters, mentioned in the previous subsection, is based on and inspired by the work of this collaboration \cite{Sef15}.

The calorimeters constructed by CALICE have one thing in common: a very high granularity. The calorimeter modules have a very large number of independent electronic readout channels, $\cal{O}$(10$^4$) in most modules, up to half a million in one specific case. The active elements are either:
\begin{enumerate}
\item Silicon pads, typically with dimensions of $1\times 1$ cm$^2$,
\item Small scintillator strips, read out by SiPMs,
\item Small Resistive Plate Chambers, operating in the saturated avalanche mode.
\item As an alternative, micromegas and GEMs are being tested.
\end{enumerate}
These readout elements are interspersed between layers of absorber material. Typically, tungsten is used for the detection of em showers.
Its Moli\`ere radius (9.3 mm) is the smallest of all practical absorber materials (only platinum is better!), so that the lateral development of the em showers is limited as much as possible. This is important for separating showers from several particles that enter the calorimeter in close proximity. For the deeper sections of the calorimeter, typically stainless steel is being used.
\begin{figure}[htbp]
\epsfysize=7.5cm
\centerline{\epsffile{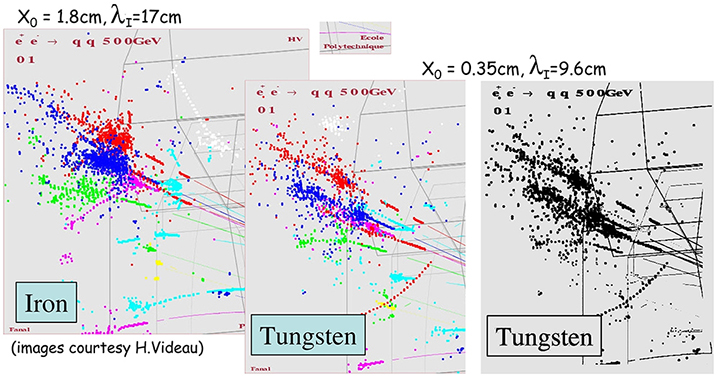}}
\caption{\footnotesize Simulated shower development of jet fragments in a calorimeter based on iron (left) or tungsten (center, right) as absorber material.}
\label{greypinkpions}
\end{figure}
Figure \ref{greypinkpions} is often shown to illustrate the advantage of using tungsten. The nuclear interaction length and especially the Moli\`ere radius, which determine the extent of the lateral shower development for hadronic and electromagnetic showers, respectively, are considerably smaller than for steel.
Of course, what really matters here is the {\sl effective} value of these parameters in the calorimeter, which also includes low-$Z$ materials such as plastic, silicon and air. A second thing to keep in mind is that the particles in reality are not colored. The difference between the colored and bitmap versions of the tungsten image illustrates that the task to assign calorimeter hits to individual jet fragments may in practice be quite daunting indeed, even in the densest possible absorber structures.

The largest calorimeter that was specifically designed for em shower detection is a tungsten/silicon device \cite{Rep08}. It has an active surface area of $18\times 18$ cm$^2$ and is 20 cm deep, subdivided longitudinally into 30 layers. The first 10 layers are $0.4 X_0$ thick, followed by 10 layers of $0.8 X_0$ and finally another 10 layers of $1.2 X_0$, for a total absorption thickness of $24 X_0$. The active layers consist of
a matrix of PIN diode sensors on a silicon wafer substrate. The individual diodes have an active surface area of $1\times 1$ cm$^2$, and there are thus $18\times 18 = 324$ calorimeter cells per layer, 9,720 in total. These are read out by means of a specially developed ASIC. Some results of measurements of em showers with this detector are shown in Figures \ref{hgcalres2} and \ref{nonlin}.
\begin{figure}[b!]
\epsfysize=6.5cm
\centerline{\epsffile{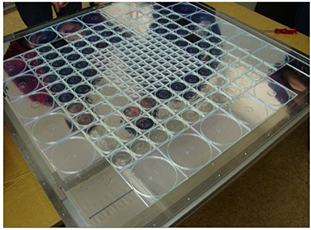}}
\caption{\footnotesize An active plastic-scintillator plane, used to detect the signals in the scintillator based CALICE hadron calorimeter \cite{Adl10}.}
\label{CAL_sci}
\end{figure}

CALICE also built and tested a large hadron calorimeter, a sandwich structure based on 38 layers of 5 mm thick plastic scintillator, interleaved with absorber plates \cite{Adl10}. For this instrument, they either used 17 mm thick steel or 10 mm thick tungsten plates. This absorber material thus represents a total thickness of about $4 \lambda_{\rm int}$ in both cases. The active layers are housed in steel cassettes with 2 mm cover plates on both sides. This increased the total depth of the instrumented volume to $\sim 5.3 \lambda_{\rm int}$. The transverse dimensions of the active layers are $90\times 90$ cm$^2$. Figure \ref{CAL_sci} shows a picture of one of the active layers. The layer is subdivided into tiles, small ones in the central region and larger ones in the outer regions (and also in the rear of the calorimeter module). The smallest tiles measure $3\times 3$ cm$^2$. Each tile has a circular groove in which a wavelength shifting fiber is embedded. This fiber collects the scintillation light produced in the tile, re-emits the absorbed light  at a longer wavelength and transports it to a SiPM, which converts it into an electric pulse.
In total, this calorimeter contains 7,608 tiles (\ie electronic channels). This was the first large-scale application of SiPMs in a particle detector.

\begin{figure}[htbp]
\epsfysize=8cm
\centerline{\epsffile{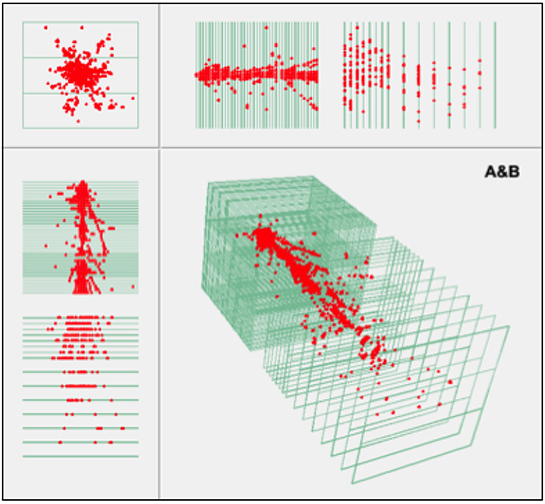}}
\caption{\footnotesize Event display for a 120 GeV $\pi^-$ showering in the CALICE digital hadron calorimeter \cite{Sef15,Rep13}.}
\label{CAL_evt}
\end{figure}

Another CALICE module has a lateral cross section of $\approx 1$ m$^2$ and a similar depth as the previous one. The effective depth can be varied through the choice of the absorber material and the thickness of the absorber plates. In between each two plates an array of RPC cells with dimensions of $1\times 1$ cm$^2$ is inserted, \ie about 10,000 per plane \cite{Dra07}.
In total, there are 54 independent longitudinal segments, so that the total number of active elements is about half a million. These RPCs operate in the saturated avalanche mode, and thus provide a ``yes'' or ``no'' signal when a particle develops a shower in this device. This is thus a ``digital'' calorimeter. An event display in this detector consists of a pattern of RPC cells that fired when the particle that created it was absorbed.
These patterns may be very detailed, as illustrated by the example shown in Figure \ref{CAL_evt} \cite{Rep13}.

The performance of the mentioned devices as stand-alone calorimeters is not particularly impressive. For example, the em energy resolution of the W/Si em calorimeter was reported as $16.5\%/\sqrt{E} \oplus 1.1\%$ \cite{Adl09}, which is almost twice as large as that of the ATLAS lead/liquid-argon ECAL \cite{Aha06}.
The proposed new endcap calorimeter for CMS, which is based on this CALICE design, has an envisaged em energy resolution of 20 -- 24\%/$\sqrt{E}$ \cite{CMS15b}, an order of magnitude worse than the resolution provided by the crystals it will replace. The reasons for these poor energy resolutions are discussed in Section 3.1 (Figure \ref{hgcalres2}).

\begin{figure}[htbp]
\epsfysize=7cm
\centerline{\epsffile{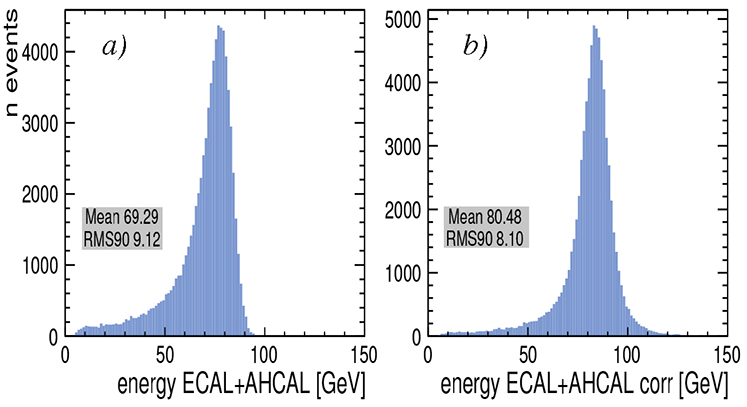}}
\caption{\footnotesize The line shape of the CALICE W/Si + Fe/plastic combination for 80 GeV pion showers, before ($a$) and after ($b$) correction procedures were applied based on the starting point of the showers and the estimated leakage \cite{Sef15}.} 
\label{CAL_lineshape}
\end{figure}

In stand-alone mode, the hadronic energy resolution of the iron/plastic-scintillator calorimeter was reported as $57.6\%/\sqrt{E} \oplus 1.6\%$, for a heavily biased event sample and after corrections that will not be applicable in a collider experiment \cite{Adl12}.
Since this calorimeter was not deep enough to fully contain high-energy hadron showers, the event sample used to obtain this result was limited to showers that started developing in the first few layers of the calorimeter module. When this device was combined with the high-granularity W/Si ECAL, the resolution deteriorated, as illustrated in Figure \ref{CAL_lineshape} \cite{Sef15}.
This figure shows the signal distribution for 80 GeV pion showers before ($a$) and after ($b$) a variety of corrections were applied. The resulting energy resolution is, even after these corrections, more than three times as large as the value reported by ZEUS \cite{Beh90}.

\begin{figure}[b!]
\epsfysize7.2cm
\centerline{\epsffile{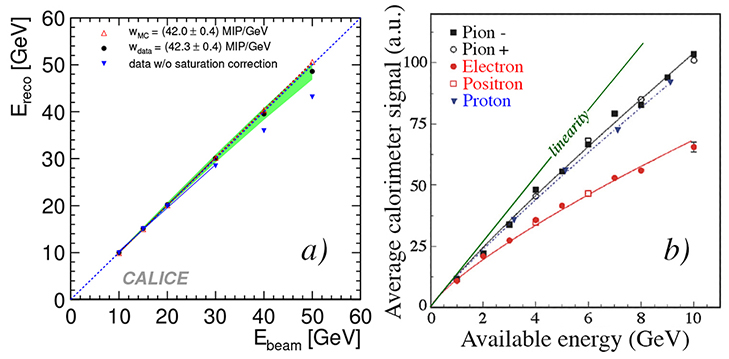}}
\caption{\footnotesize Non-linearity in the CALICE hadron calorimeters. Diagram $a$ shows  
the average signal for positrons in the CALICE analog hadron calorimeter, as a function of the beam energy. Shown are the measured data points, before and after corrections for saturation in the SiPM readout were applied, as well as the Monte Carlo prediction. The shaded area represents the systematic uncertainty in the corrections \cite{Adl11a}. Diagram $b$ shows the average signals of the ``digital'' calorimeter for electrons and hadrons as a function of energy, in the energy range of 1 -- 10 GeV. For comparison, the dependence for a linear calorimeter is given as well. Experimental data from \cite{Sef15,Rep12}.} 
\label{CALICElin}
\end{figure}

More worrisome than the unremarkable energy resolutions reported for these calorimeters is their non-linearity. Figure \ref{CALICElin}a shows the average signal of the Fe/plastic detector for positrons, as a function of energy \cite{Adl11a}. The measured data points exhibit a significant non-linearity, namely a $\sim 10\%$ decrease of the response in the energy range from 10 to 50 GeV. 

According to the authors, this is due to saturation of the SiPM signals,
and they expect that this may be remedied when SiPMs with a larger dynamic range become available. 
Unfortunately, not enough information is supplied to verify this explanation which, if true, would also invalidate the energy resolution reported by the authors. This is because, as a matter of principle, signal saturation implies that the fluctuations that determine the energy resolution are partially suppressed \cite{Liv17}.

The signal saturation phenomenon reaches very substantial proportions in the ``digital'' calorimeter built by CALICE. The resulting non-linearity (Figure \ref{CALICElin}b) is even so large (already for particle energies that are much smaller than expected in the experiments for which this device is intended), that it leads to apparent {\sl overcompensation} \cite{Sef15,Rep12}. Because of the large suppression of fluctuations in the shower development process, the quoted energy resolutions are not very meaningful \cite{Liv17}. The large signal-nonlinearity observed for small signals is important since high-energy jets, such as the ones from the hadronic decay of intermediate vector bosons and the Higgs boson, consist of a considerable number of low-energy final-state particles, which together represent a significant fraction of the total jet energy. Quantitative information on this point is given in Figure \ref{webber}.

The CALICE Collaboration has apparently also realized these problems and has embarked on equipping the RPCs with a 2-bit readout system.
This provides the possibility to subdivide the signals into three categories, on the basis of different threshold levels. This is called the ``semi-digital'' option \cite{Ste14}. 
However, the RPCs still operate in avalanche mode, and the relationship between the different thresholds (corresponding to bit settings 1/0, 0/1 and 1/1, respectively) and the deposited energy is not a priori clear. 

\begin{figure}[htbp]
\epsfysize=6.5cm
\centerline{\epsffile{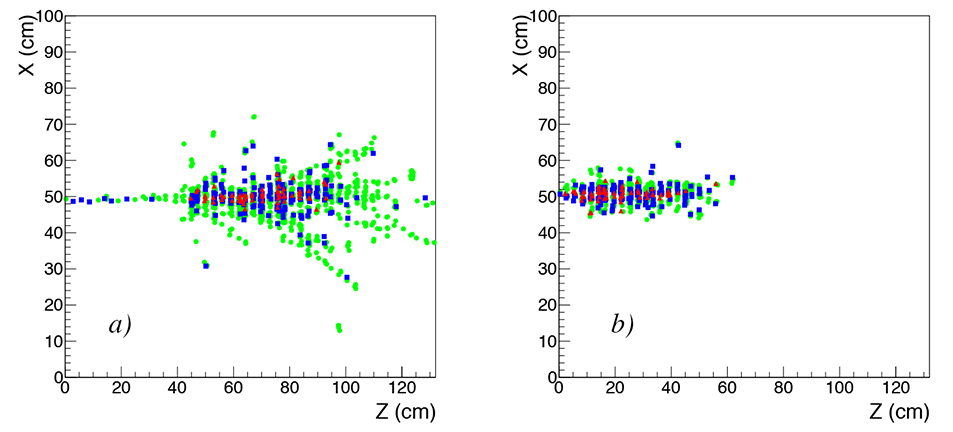}}
\caption{\footnotesize Event displays for a 70 GeV pion (left) and a 70 GeV electron (right) shower in the semi-digital hadron calorimeter built by CALICE \cite{Den16}. The different colors indicate the amplitude of the RPC signals (see text for details).}
\label{SDHCALevt}
\end{figure}

Some results on beam tests performed with this so-called SDHCAL are reported in \cite{Den16}. The three thresholds were set at 0.11, 5 and 15 picoCoulombs
in these tests. Since the average signal produced by a mip in the RPCs was 1.2 pC, the second and third thresholds corresponded to the simultaneous passage of at least 4 and 12 mips, respectively. 
Figure \ref{SDHCALevt} shows event displays for a 70 GeV pion and a 70 GeV electron in which the cells exceeding the different threshold levels are indicated with different colors, red for level 3, blue for level 2, green for level 1. Not surprisingly, the average multiplicity of tracks is clearly larger for the 
electron shower, and the highest concentration of red cells in the pion event is found near the shower axis, where most of the $\pi^0$ production takes place.

\begin{figure}[htbp]
\epsfysize=6cm
\centerline{\epsffile{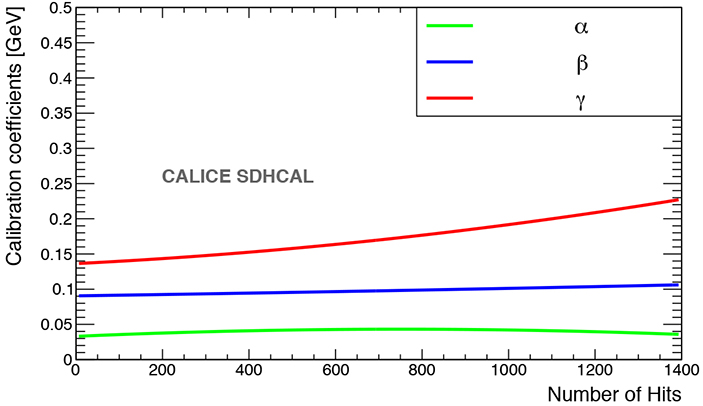}}
\caption{\footnotesize Evolution of the coefficients $\alpha$, $\beta$ and $\gamma$ as a function of the total number of RPC hits recorded in the CALICE digital calorimeter \cite{Den16}.}
\label{SDHCAL2}
\end{figure}

\begin{figure}[htbp]
\epsfysize=6.5cm
\centerline{\epsffile{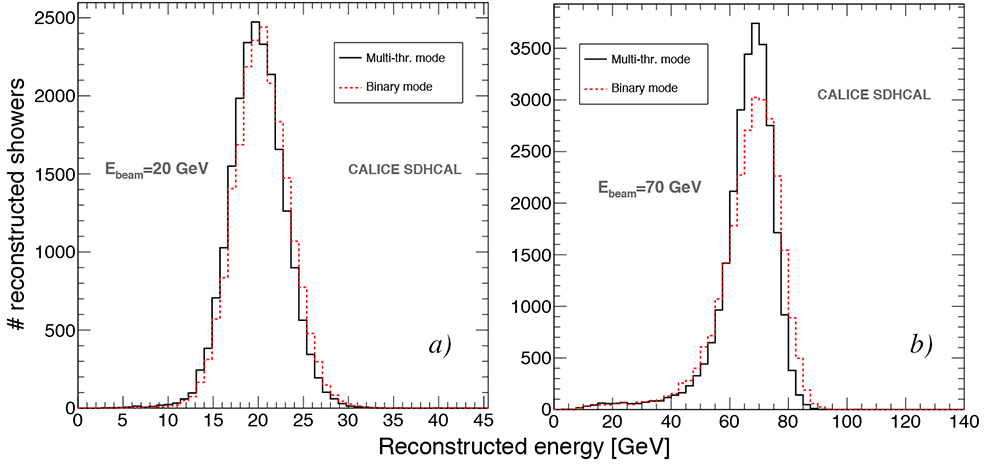}}
\caption{\footnotesize Distribution of the reconstructed energy with the binary mode (red dashed line) and with the three-threshold mode (solid black line), for pions of 20 GeV ($a$) and 70 GeV ($b$) in the semi-digital hadron calorimeter built by CALICE \cite{Den16}.}
\label{SDHCAL1}
\end{figure}

The additional information provided in this way was used to reconstruct the shower energy. To that end the signals observed in the RPCs were given weight factors
intended to compensate for the saturation effects:
\begin{equation}
E_{\rm reco}~=~\alpha N_1 + \beta N_2 + \gamma N_3
\label{eq:sdcal}
\end{equation}
in which the values $N_i$ represent the number of hit cells with signals above the thresholds $i$, and the weight factors $\alpha < \beta < \gamma$.
It turned out that the values of the weight factors had to vary with energy in order to reconstruct the pion energy correctly. This is illustrated in Figure \ref{SDHCAL2}. Moreover, the weight factors had to be given different values to reconstruct the energy of other types of particles (especially electrons).
Figure \ref{SDHCAL1} shows the effects of the additional information from the RPC signals on the distribution of the reconstructed energy for event samples of pions at 20 GeV (Figure \ref{SDHCAL1}a) and 70 GeV (Figure \ref{SDHCAL1}b). The solid lines represent the distributions obtained with this information, while the dashed lines were obtained in the ``binary mode'', \ie without information on the size of the RPC signals. The difference is only significant at the highest energy. 

Proponents of the PFA approach argue that all these shortcomings of their calorimeters are not very important, because of the limited role played by the calorimeter in the measurement of jet properties. The most crucial feature of the calorimeters developed in this context is the very high granularity, intended to unravel the structure of the jets. Events such as the ones shown in Figures \ref{CAL_evt} and \ref{SDHCALevt}, which were obtained with a cell size of $1\times 1$ cm$^2$, are used to illustrate this point. 

\begin{figure}[htbp]
\epsfysize=5.5cm
\centerline{\epsffile{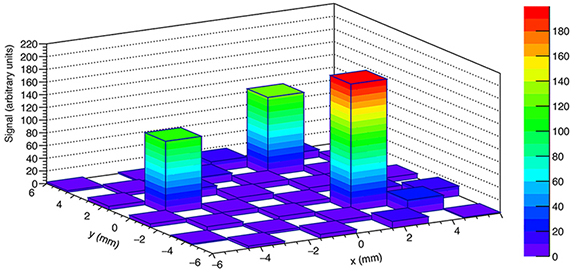}}
\caption{\footnotesize Event display for an event in which three particles enter a dual-readout calorimeter equipped with SiPM readout simultaneously \cite{Pez18}. Shown are the signals from the scintillating fibers, each of which was connected to a SiPM sensor. The \v{C}erenkov fibers were fed through holes in the white fields and read out by a second SiPM array located directly behind the one used for the scintillation signals. The calorimeter area covered by this event display has transverse dimensions of only $1.2 \times 1.2$ cm$^2$ \cite{rd52sipm}.}
\label{sipm_evt}
\end{figure}

\begin{figure}[b!]
\epsfysize=5.5cm
\centerline{\epsffile{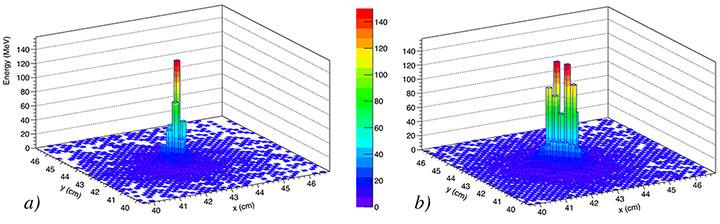}}
\caption{\footnotesize Simulated event displays of a 50 GeV electron shower ($a$) and the showers produced by the $\gamma$s from the decay of a 100 GeV $\pi^0$ ($b$) produced 2 m upstream of a calorimeter of the type that was used to measure the event from Figure \ref{sipm_evt}  \cite{Pez18}.}
\label{sipmevents}
\end{figure}

However, this is by no means a feature that is unique for calorimeters of this type.
Calorimeters that were developed to achieve excellent performance in stand-alone mode are also capable of measuring the detailed structure of events in which several particles enter the calorimeter in close proximity. 
Figure \ref{sipm_evt} shows an event display in which three particles enter a dual-readout calorimeter equipped with SiPM readout simultaneously. In an area with transverse dimensions of $1.2 \times 1.2$ cm$^2$ (\ie almost the same as the size of one RPC cell in the CALICE digital calorimeter), the three peaks are clearly separated from each other. This feature, which is a consequence of the extremely narrow shower profile (Figure \ref{emprofiles}a) illustrates the benefits of a lateral detector granularity that is much finer than one would choose based on the value of the Moli\`ere radius (31 mm in this particular calorimeter). 
These benefits include the possibility to distinguish between em showers caused by (the two $\gamma$s from) a $\pi^0$ and by a single electron or $\gamma$.
Figure \ref{sipmevents} shows simulated event displays of a 50 GeV electron shower and of a 100 GeV $\pi^0$ produced in a vertex 2 m upstream of the entrance window of such a calorimeter. The two $\gamma$s, which develop showers that are separated by only a few mm, are clearly recognizable as such.

Because of this feature and other unique advantages of the dual-readout approach (Section 5.2), several experiments that are being planned for future $e^+e^-$ colliders are considering a calorimeter system of this type \cite{Fer18}. 

\section{Energy resolution}

I want to finish the review of calorimeters that are intended for experiments at particle accelerators with some general comments about energy resolution, since this is usually one of the most important criteria for selecting a certain type of calorimeter for a particular application.
The energy resolution is defined as the precision with which the energy of an unknown object can be determined from the signals it produces in the calorimeter. Typically, this resolution is determined as the relative width of the signal distribution measured for a beam of mono-energetic particles from an accelerator.
However, it was recently shown \cite{slee17} that the properties derived in that respect from a collection of identical particles are not necessarily valid for the detection of one individual particle out of that collection. 

Yet, in the absence of a better method, let us assume that the width of the signal distribution for a beam of mono-energetic particles is a measure of the energy resolution of the calorimeter. This leads me to the following comments.
\begin{itemize}
\item {\bf The line shape}.
Often, the measured signal distributions exhibit non-Gaussian tails. In that case, one should quote the $\sigma_{\rm rms}$ value as the energy resolution. However, some authors 
use another variable, in order to make the results less dependent on the tails of the signal distributions they measure, and thus look better. This variable, called $\rm rms_{90}$, is defined as the root-mean-square of the values located in the smallest range of reconstructed energies that contains 90\% of the total event sample (see Figure \ref{CAL_lineshape} as an example).
For the record, it should be pointed out that for a Gaussian distribution, this variable gives a 21\% smaller value than the true $\sigma_{\rm rms}$ (\ie $\sigma_{\rm fit}$).
Of course, one is free to define variables as one likes. However, one should then not use the term ``energy resolution'' for the results obtained in this way, and compare results obtained in terms of $\rm rms_{90}$ with genuine energy resolutions from calorimeters with Gaussian response functions. This misleading practice is followed by the proponents of PFA \cite{Tho09}.
\item {\bf Saturation effects}
in the sensors will reduce the width of the signal distribution. Since a source of fluctuations is suppressed/eliminated because of this,
the energy resolution of such calorimeters is in reality worse than suggested by the width of the signal distribution. This phenomenon affects, for example, the measurements of calorimeters in which the light signals are sensed by SiPMs. An extreme case of saturation occurs in digital calorimeters such as the one discussed in Section 7.3. Saturation effects go hand in hand with non-linearity. One may use weighting schemes to restore a semblance of linearity (Figure \ref{SDHCAL2}), but such schemes are of no consequence for the mis-measured energy resolution, since they do not address the fluctuations about the mean signal values that are modified in such procedures.
\item  {\bf Non-linearity}.
In general, one would want to know the energy resolution of a calorimeter over a certain range of energies.
Even if the width of the signal distribution is determined in a statistically correct way, and if saturation effects do not play a role, the assumption that this width represents the energy resolution is only correct if the average value of that measured signal distribution corresponds indeed to the correct energy of the particles.
Signal non-linearities tend to invalidate that assumption. Let me give two examples, taken from calorimeters discussed elsewhere in this review, one electromagnetic and the other hadronic. The first example concerns the em W/Si calorimeter built by CALICE. As shown in Figure \ref{nonlin}b, the response of this calorimeter to em showers of 30 GeV differs by about 2\% from that to showers of 15 GeV. This means that the average signal of this calorimeter for 30 GeV $\gamma$s will differ by 2\% from that for 30 GeV $\pi^0$s, since the latter produce two showers of 15 GeV each. If these two showers are not spatially resolved, this phenomenon leads to a systematic energy mis-measurement. \hb
The second example concerns the SDHCAL discussed in Section 7.3. The energy measured by this calorimeter is reconstructed by means of weight factors that depend on the total number of recorded hits (Figure \ref{SDHCAL2}). For example, when a hadron shower consists of 1400 hits, then the RPCs with a signal above the highest of the three thresholds are attributed an energy of 0.23 GeV each. However, that hadron shower may also have been produced by three jet fragments, each responsible for 1/3 of the total number of hits. According to Figure \ref{SDHCAL2}, RPCs that produce a signal above the highest threshold should in that case be attributed
an energy of 0.16 GeV each in order to reconstruct the correct energy. If the three jet fragments are not recognized as such by the calorimeter, the event will be  
interpreted as being caused by a single hadron, and by applying the weight factor of 0.23 GeV, the energy of the event will be overestimated.

\item {\bf Biased event samples}.
One of the most common mistakes that are made when analyzing the performance of a calorimeter derives from the selections
that are made to define the experimental data sample. This selection process may easily lead to biases, which distort the performance characteristics
one would like to measure. In some extreme cases, this may lead to very wrong conclusions, such as the claim that uranium/liquid-argon calorimeters are 
compensating \cite{Wig17}. As a general rule, the calorimeter data themselves should {\sl not} be used to apply cuts and thus select the event sample to be analyzed. Any such cuts should be based on external detectors, such as upstream \v{C}erenkov counters and/or preshower detectors, downstream leakage detectors and muon counters located behind substantial amounts of absorber material.\hb
Almost every analysis of test beam data I know of suffers from bias problems, the question is only {\sl to what extent} the obtained results are affected by this. 
As an extreme case, I mention the analysis of the performance of the SDHCAL discussed in Section 7.3 \cite{Den16}.
In order to select pure samples of pion induced showers, the authors relied completely and exclusively on the energy deposit pattern in the tested calorimeter itself. This pattern was used to eliminate perceived contamination of electrons and muons to the data samples. Amazingly, the authors even removed events that were
believed to be caused by {\sl neutral} beam particles, which thus apparently had somehow managed to sneak into the beams that were momentum selected with magnets.

\item {\bf Misinterpretation of experimental data}. 
Another mistake that is not uncommon concerns the extrapolation of measurement results far beyond their region of validity.
A classic example concerned the ``determination'' of the energy resolution for
high-energy em shower detection in liquid xenon, based on convolving the signals from
large numbers of low-energy electrons (100 keV) recorded in a small cell \cite{Seg92}.
These measurements only revealed something about the energy resolution for the
detection of these low-energy electrons. In a high-energy em shower, a variety of new
effects, absent or negligible in the case of these electrons, affect the signals and their
fluctuations. For example, in a large detector volume, the 174 nm scintillation light is subject to light attenuation, \eg through self-absorption.
\hb
\begin{figure}[htbp]
\epsfysize=6cm
\centerline{\epsffile{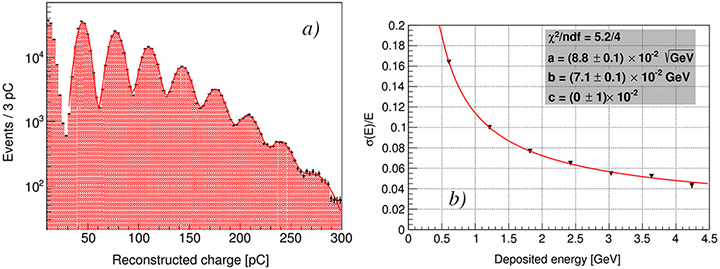}}
\caption{\small
Distribution of the total collected charge in beam tests of a prototype calorimeter for the NA62 experiment at CERN ($a$). These tests were carried out with a beam of 606 MeV electrons  \cite{Ant18}. The peaks are the result of several beam particles entering the detector simultaneously. The energy resolution derived from these measurements ($b$). The data points were fit with a curve of the type
$\sigma/E = a E^{-1/2} \oplus b E^{-1} \oplus c$. The resulting values of the coefficients $a$, $b$ and $c$ are shown in the legend.}
\label{frascati}
\end{figure}

A more recent example concerns beam tests of a calorimeter intended to detect $\gamma$s with energies up to 5 GeV at small scattering angles in the NA62 experiment \cite{Ant18}. The authors tested this detector at the DA$\Phi$NE linac in Frascati in a 606 MeV electron beam. They obtained energy resolutions at higher energies by using signals in which $n$ such electrons entered the calorimeter simultaneously (Figure \ref{frascati}). As before, such tests only reveal something about 606 MeV electrons.\hb
It should be stressed that, as a matter of principle, measurements made for low-energy particles cannot be
used to determine/predict the high-energy calorimeter performance.  
\end{itemize}

Finally, it should be pointed out that often times a good energy resolution is only part of the requirements for obtaining the desired physics sensitivity.
As an example, I mention the Higgs boson, discovered in 2012 by two experiments at the Large Hadron Collider through its decay mode 
$H^0 \rightarrow  \gamma \gamma$ \cite{Aad12,Chat12}.
The invariant mass of a particle decaying into two $\gamma$s is given by
\vskip 1mm
\begin{equation}
M~=~\sqrt{2 E_1 E_2 (1 - \cos{\theta_{12}})}
\label{invmas2g}
\end{equation}
The precision with which the mass can be measured is thus not only determined by the energy resolution,
\ie the measurement uncertainty on the $\gamma$ energies $E_1$ and $E_2$, but also by the 
relative uncertainty on the angle ($\theta_{12}$) between the directions of these $\gamma$s.
A good localization of the $\gamma$s is thus very important to identify the parent particle.
While CMS emphasized excellent energy resolution for em showers in its design of the experiment, at the expense of degraded
hadronic performance, ATLAS concentrated its efforts also on the localization issue. As a result, the mass resolution for the
Higgs bosons turned out to be very similar in both experiments.

In preparing this review, I have read dozens of different papers and hundreds more for my recent book \cite{Wig17}. Unfortunately, I have to conclude that the reviewers of manuscripts in which tests of calorimeters are described do not take the issues listed above very seriously. As a result, it is very difficult, if at all possible, to compare the performance of different calorimeters in an objective way. I sincerely hope, but doubt that this will be different when decisions involving large sums of taxpayer money have to be made in the context of experiments planned for future colliders.

\section{Calorimeters for non-accelerator experiments}

In the past three decades, we have witnessed the development of a number of non-accelerator experiments that employ calorimetric methods to detect particles, typically of extraterrestrial origin. A trailblazer in this context was the Japanese Kamiokande experiment, which was originally designed to detect proton decay (which turned out not to be detectable), but whose moment of glory came in 1987 when neutrinos from a Supernova explosion in the Large Magelhanic Cloud were observed. Redesigned and considerably upgraded, Superkamiokande was also the first experiment to detect solar neutrinos in real time \cite{Abe11}. These successes have inspired the development of several other large instruments intended to look for astrophysical phenomena using calorimetric methods. These instruments were either built from scratch, \eg the SNO detector in Sudbury (Canada) \cite{Ahar06}, or used the natural environment as the absorbing medium. Examples of the latter approach can be found at the South Pole (IceCube) \cite{Hal10}, in the Mediterranean (KM3) \cite{Adr16}, in Lake Baikal \cite{Ayt08}, or in the Argentinian pampas (Auger) \cite{Aug16}. In the latter case, thousands of cubic kilometers of the atmosphere serve as the detector, in the other cases water is the medium of choice. In the water detectors, as well as in some atmospheric detectors, \v{C}erenkov light constitutes the signals. In other cases, atmospheric scintillation light and/or charged shower particles that reach the Earth's surface provide the experimental information, while the water-based detectors are experimenting with acoustic signals as well \cite{Agu11}. Apart from extraterrestrial (or even extragalactic) neutrinos, these experiments are also active in detecting extremely-high-energy cosmic rays, studying phenomena such as the knee(s) in the cosmic-ray spectrum and the GZK cutoff.

Some of the experiments have extended the scope of their research program by looking for interactions by neutrinos produced in far-away
accelerators or reactors. Superkamiokande (in the process of being transformed into Hyperkamiokande) is studying neutrino oscillations in a beam produced at the J-PARC accelerator, 700 km away, while KAMLAND looked at anti-neutrinos produced in a large number of nuclear reactors located 
at various distances in Japan and Korea \cite{Gan11}. The main Fermilab program nowadays also focusses on detecting neutrinos produced by the proton accelerator in calorimetric detectors located at different distances from the source, varying from a few km ($\mu$BOONE) to 1,300 km (DUNE) \cite{Kem17}. 
\begin{figure}[htbp]
\epsfysize=6cm
\centerline{\epsffile{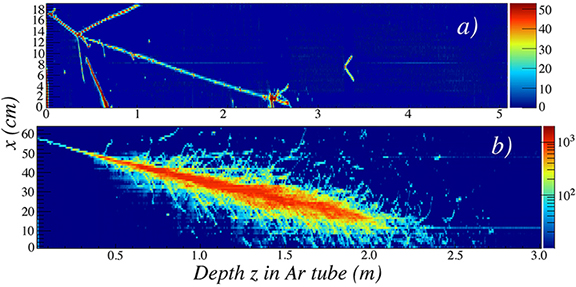}}
\caption{\footnotesize Examples of cosmic ray events in a large liquid argon TPC \cite{Ere13}. Diagram $a$ shows an interaction by a charged particle, in which three charged particles are produced, plus possibly some neutral objects. Several $\delta$-rays are also visible along the track of the particle that re-interacts after about 2 m. Diagram $b$ shows a textbook example of a developing em shower in argon.}
\label{Antonio}
\end{figure}
Several of the these detectors are based on liquid-argon technology, instrumented as Time Projection Chambers, which produce event images of developing showers that rival those from bubble chambers (Figure \ref{Antonio}b). Whereas measuring the energy deposited by particles absorbed in the detector is an important goal of all these experiments, they typically emphasize complete imaging of the detected events (Figure \ref{Antonio}a).

\subsection{\it Neutrino detectors}

The most important scientific successes of this class of detectors have been obtained by the calorimeters installed in the Kamioka mine in Japan, and the Nobel prize for physics was awarded both in 2002 and in 2015 for discoveries made there.These discoveries concerned supernova neutrinos (SN1987a),
solar neutrinos and neutrinos produced in the Earth's atmosphere in showers initiated by high-energy cosmic rays. Some of these discoveries are of fundamental importance for the understanding of the neutrino properties, \eg the rest mass issue. The upgrade to Hyperkamiokande is, among other things, intended to make the detectors sensitive to neutrinos from supernova explosions in Andromeda and beyond. 
Dispersion in the arrival times of such neutrinos might increase the sensitivity of the detectors to neutrino mass values smaller than 1 eV/$c^2$.

\begin{figure}[htb]
\epsfysize=12cm
\centerline{\epsffile{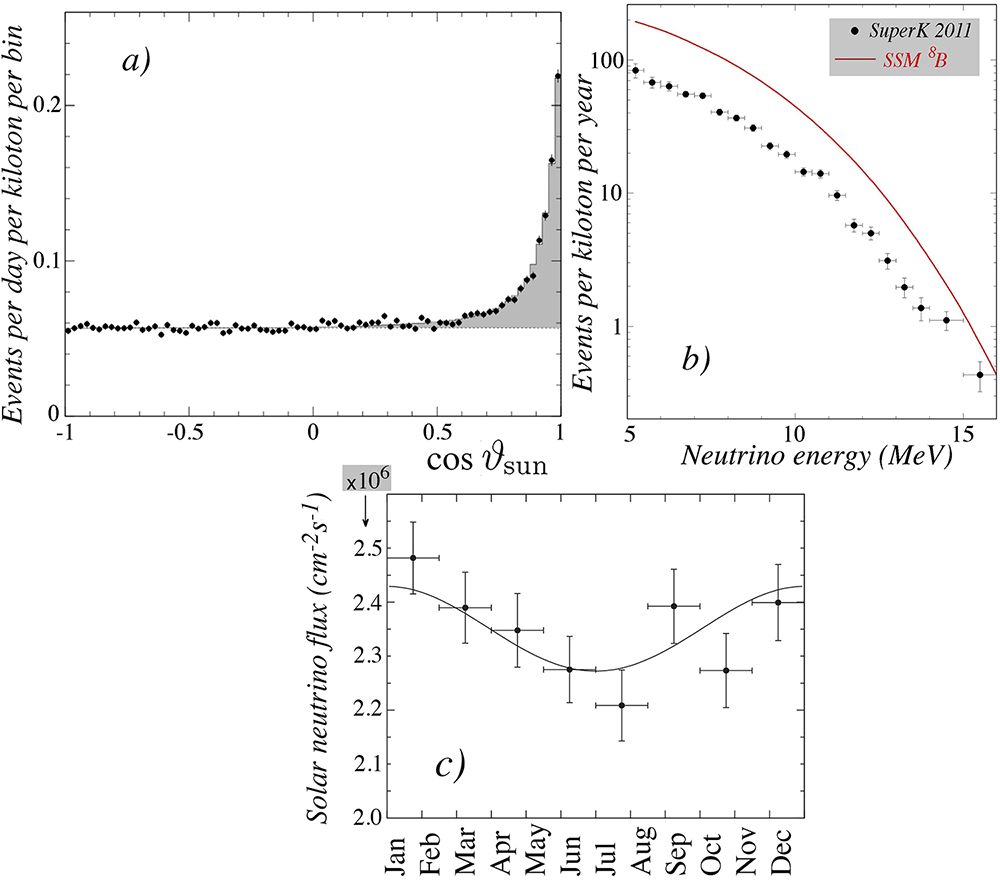}}
\caption{\small
Angular distribution of the electron direction of the $\nu_e + e^- \rightarrow \nu_e + e^-$
candidate events, measured by the Superkamiokande detector with respect to the direction of the Sun ($a$), and the measured spectrum of the solar neutrino
component of this sample ($b$). For comparison, the predicted spectrum of the $^8$B component of the solar
neutrinos is shown as well \protect\cite{Abe11}. Seasonal variation in the measured solar neutrino flux 90$c$). The solid line is the prediction based on the eccentricity of the Earth's orbit \protect\cite{Hos06}.}
\label{SKsolnu}
\end{figure}

Figure \ref{SKsolnu} illustrates to what extent neutrino physics has become a precision science in the Kamioka mine.
This figure concerns solar neutrinos, which are detected by means of \v{C}erenkov light from electrons produced in neutrino-electron scattering.
The energies of the detected solar neutrinos range from 5 - 16 MeV.

The energy resolution in the solar-$\nu$ energy range is limited by
photoelectron statistics, $\sim$7 photoelectrons per MeV (which is 
some five to ten times larger than in a typical lead-glass calorimeter). 
Because of this high light yield, the recoil electrons from the $\nu_e + e^- \rightarrow
\nu_e + e^-$ reactions induced by solar neutrinos produce significant signals. The limiting
factor in the analysis is not so much the size of the signals, but rather the extremely small
event rates (15 events per day) and thus the sensitivity to background processes that produce similarly
significant signals. Even though its levels have been reduced by some seven orders of magnitude, natural radioactivity, and in particular $^{222}$Rn, has remained the main source of this background.

The crucial factor for distinguishing this background from the genuine solar neutrino
interactions turned out to be SuperK's capability to reconstruct the direction of the
electrons that caused the signals. Even though the angular resolution was only modest in the
energy range of interest (20$^\circ$-- 40$^\circ$), this made an enormous
difference, as illustrated by Figure \ref{SKsolnu}a.     

For each candidate solar-$\nu$ event, the angle ($\vartheta_{\rm sun}$) between the reconstructed
track and the line pointing from the Sun to the SuperK detector at the time of the event was
calculated. Figure \ref{SKsolnu}a shows a distribution of candidate events recorded after the detector was refurbished 
(2006 -- 2008) as a function of
this angle. This distribution is mainly flat in $\cos {\vartheta_{\rm sun}}$, as expected
for background events, for which the position of the Sun is irrelevant.
However, the distribution also shows a very clear excess of events near $\cos {\vartheta_{\rm sun}}
= 1$, \ie events in which the electron seemed to come from the direction of the Sun. 
Out of a total of some 30,000 candidate events, recorded over a period of 548 days, 8,148 could be attributed to the Sun on this basis,
with an uncertainty of 2.7\%, dominated by systematic effects. 

Compared with detailed model predictions of
the solar neutrino flux, the measured rate is a factor of two to three too small (Figure \ref{SKsolnu}b). This experimental result, earlier 
observed in radiochemical measurements and known as
the ``solar neutrino puzzle,'' was historically the first indication of the occurrence of neutrino oscillations. 
Figure \ref{SKsolnu}b shows the calculated spectrum of neutrinos produced in $\beta$ decay of $^{8}$B, the dominant process through which the
Sun produces neutrinos in this energy range \cite{Bah04}. The figure shows that the discrepancy tends to become a bit smaller near the high-energy endpoint.
This is believed to be evidence for the contribution of an even more rare process to the solar neutrino production \cite{Bah98}:

$$^3{\rm He} + p \rightarrow ~ ^4{\rm He} + e^+ + \nu_e$$
commonly referred to as the {\sl hep} process.
The expected {\sl hep} neutrino flux is three orders of magnitude smaller than the $^8$B solar neutrino flux. However, since the end point of the {\sl hep} neutrino spectrum is about 18.8 MeV, compared to about 16 MeV for the $^8$B neutrinos, the high-energy end of the measured SuperK spectrum should be relatively enriched with {\sl hep} neutrinos. 

During the initial period of solar neutrino measurements (1,496 days), before the 2001 accident in which a large fraction of the PMTs were destroyed, in total 22,404 solar neutrinos were detected,
with an uncertainty of 3.6\%. This high-statistics run made it even possible to observe a small, but significant, seasonal variation in the solar neutrino flux,
caused by the elliptic nature of the Earth's orbit around the Sun (Figure \ref{SKsolnu}c). 
\vskip 2mm
The data shown in this subsection illustrate the enormous 
progress that has been made since the days that Ray Davis (Nobel laureate 2002) was counting $^{37}$Ar atoms in his cleaning liquid,
and how far we have come in understanding the details of the mechanism through which our star generates the energy that makes life on our
planet possible.
\vskip 2mm
A different class of neutrino detectors uses the natural environment as a calorimeter. Also these experiments are now reaching a level of sophistication and precision that makes it possible to probe an energy regime that is (and maybe forever will be) inaccessible to manmade particle accelerators.
Neutrinos with energies of more than 1 PeV have been observed in IceCube, currently the world's largest neutrino detector, encompassing a cubic kilometer of ice near the South Pole. 
Since these {\sl astrophysical} neutrinos travel from their source to Earth with essentially no attenuation and no deflection by magnetic fields,
they might reveal details about the mechanism through which they are produced. This sets them apart from extragalactic charged particles,
even those at the very highest energies.

\subsection{\it The cosmic-ray spectrum}

The instruments discussed in this section have also made major contributions to understanding some features of the spectrum of the ultrahigh-energy particles that reach us from outer space: the ``knees'' at 5 PeV and 200 PeV, and the GZK cutoff in the EeV region. The latter phenomenon, which implies that the mean free path of extremely-high-energy protons is limited (to a few hundred Mpc) because of photo-pion production off the (2.7 K) cosmic microwave background, was unambiguously demonstrated by the Auger experiment in Argentina \cite{Abr11, Ber14}.

\begin{figure}[htbp]
\epsfysize=6.5cm
\centerline{\epsffile{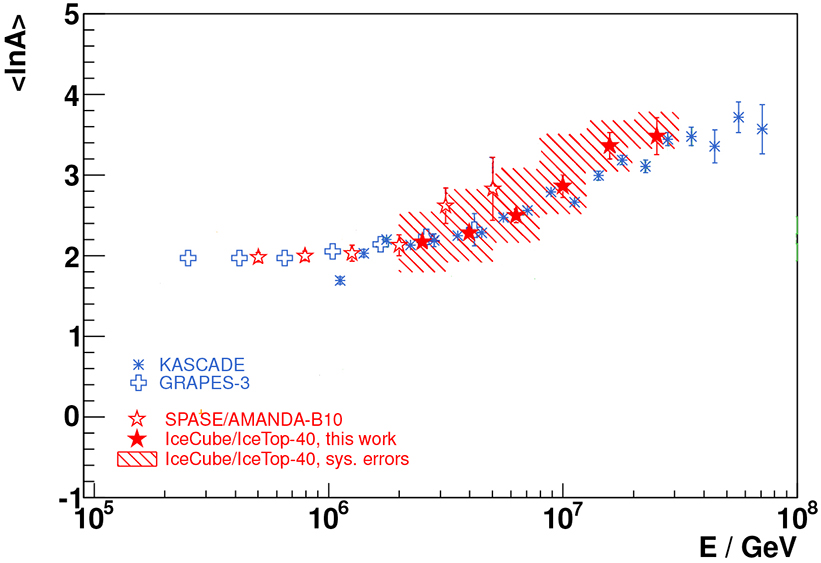}}
\caption{\small
The average atomic number ($A$) of the  particles registered as cosmic rays by the IceCube and KASCADE-Grande Collaborations \cite{Abba13}.}   
\label{lnA}
\end{figure}

Other important contributions to understanding the high-energy cosmic-ray spectrum have been made by the KASCADE-Grande experiment near Karlsruhe, Germany \cite{KAS16}. This experiment does not detect any light produced in the atmosphere, but relies completely on the shower particles that reach the Earth's surface, which are detected with a relatively finely spaced grid of muon detectors, as well as a large sampling calorimeter based on tetramethyl pentene as active material. This experiment collected evidence that the ``knee'', a sudden steepening near 4-5 PeV that has also been observed by many other detectors, is an exclusive feature of the proton component of the cosmic-ray spectrum, and thus does not occur for the heavier nuclei (such as helium, carbon, oxygen, iron). This is illustrated in Figure \ref{lnA}, which shows the average $A$ dependence of the nuclei that initiated the atmospheric showers detected by KASCADE-Grande \cite{Abba13}.

The IceCube Observatory has recently been equipped with a surface detector (called IceTop), which has made it possible 
to perform competitive studies of extremely-high-energy cosmic rays. Thanks to its high altitude (2,835 m above sea level) IceTop
is sensitive to the em component of cosmic air showers, while IceCube measures the muonic component for energies above $\sim 1$ TeV.
Combination of the signals measured for muons in both IceTop and IceCube has made it possible to determine the track direction of such muons with a resolution of about 0.5$^\circ$.
The IceCube Observatory is sensitive to cosmic rays of a very large energy range, from $\sim 1$ PeV, just below the {\em knee}, to the EeV regime. In one study, they have measured the change in the chemical composition of the cosmic rays in the energy range from 1--30 PeV (Figure \ref{lnA}), and confirmed the conclusion of KASCADE-Grande that the knee is a feature that only affects the proton component of the spectrum.

Other experiments concentrate on measuring showers created by particles absorbed in the Earth's atmosphere. An ambitious new project is the \v{C}erenkov Telescope Array,  that will comprise a Northern Hemisphere and a Southern Hemisphere component with in total more than 100 telescopes \cite{CTA11}. Of these, 99 will be installed at the Cerro Armazones site in Chile, and 19 at La Palma (Canary Islands).
It is expected that CTA will allow the detection of $\gamma$-ray induced air showers over a large area on the ground, and increase the number of detected $\gamma$-rays dramatically, while at the same time providing a much larger number of views of each cascade. This would result in both improved angular resolution and better suppression of cosmic-ray background events. 
Precursors of CTA, such as VERITAS (Arizona \cite{Ver16}), MAGIC (Canary Islands \cite{Ale16}) and HESS (Namibia \cite{HES16})  have already produced impressive scientific results. For example, HESS recently demonstrated that the center of our galaxy acts as a {\sl PeVatron}, \ie a source of extremely high-energy protons and nuclei \cite{Abr16}. Until now, it was generally believed that supernova explosions are the accelerators in the PeV regime. 
\vskip 2mm

The examples shown in this section illustrate that calorimetric particle detection, combined with the precise imaging capabilities of the instruments that have been developed for this purpose, has started to pay big dividends in the study of phenomena in an energy range inaccessible to particle accelerators. In addition, the instruments discussed here have made it possible to gain crucial insight into some astrophysical phenomena. Having seen how fast this field has developed in the past 20 years, I am convinced that this has been only the beginning of a very important new development in our understanding of the physical world.

\section{Cryogenic calorimeters}

At the other extreme end of the energy spectrum, 
there is a class of highly specialized detectors that employ calorimetric methods to study a series
of very specific phenomena in the boundary area between particle physics and astrophysics:
dark matter, solar neutrinos, magnetic monopoles, nuclear double $\beta$-decay, \etc 
~All these issues require precise measurements of very small energy deposits. In order to achieve that
goal, these detectors exploit phenomena that play a role at temperatures close to 
zero, in the few milli-Kelvin to 1 Kelvin range. These phenomena include the following:
\begin{description}
\item[$a)$] Some elementary excitations require very little energy. For example, Cooper
pairs in superconductors have binding energies in the $\mu$eV--meV range and may be broken by phonon 
absorption.
\item[$b)$] The specific heat for dielectric crystals and for superconductors decreases to
very small values  at these low temperatures.
\item[$c)$] Thermal noise in the detectors and the associated electronics becomes very
small.
\item[$d)$] Some materials exhibit specific behavior (\eg a change in magnetization, 
latent heat release) that may provide detector signals.
\end{description} 
Many of the devices that have been proposed in this context are still in the early phases of the 
R\&D process. In many cases, this R\&D involves fundamental research in solid-state physics and
materials science. However, some devices have reached the stage where practical applications 
have been successfully demonstrated. Among these, I mention
\begin{itemize}
\item {\sl Bolometers}, which are based on principle ($b$). These are calorimeters in the true
sense of the word, since the energy deposited by particles (in an insulating crystal at very low
temperature) is measured with a resistive thermometer. 
\item {\sl Superconducting Tunnel Junctions}, in which the quasi-particles and -holes (Cooper pairs)
excited by incident radiation tunnel through a thin layer separating two superconducting materials.
\item {\sl Superheated Superconducting Granules}, which are based on the fact that certain type I
superconductors can exhibit metastable states, in which the material remains superconducting in
external magnetic fields exceeding the critical field. These detectors are usually prepared as a
colloid of small (diameter 1--100 $\mu$m) metallic granules suspended in a dielectric matrix 
(\eg paraffin). Heat deposited by an interacting particle may drive one or several granules from
the superconducting to the normal state. The resulting change in magnetic flux (disappearance of the
Meissner effect) may be recorded by a pickup coil.
\end{itemize}
\begin{figure}[hbtp]
\epsfysize=7cm
\centerline{\epsffile{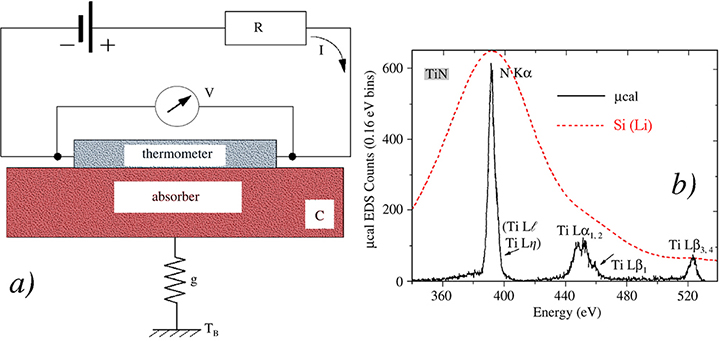}}
\caption{\small
The principle on which a cryogenic calorimeter is based ($a$). Spectrum of X-rays of titanium nitrate, measured with a cryogenic calorimeter ($\mu$cal), and with a standard Si(Li) semiconductor detector ($b$) \cite{Pre00}.}
\label{pretzl}
\end{figure}

Figure \ref{pretzl}a shows the operating principle of a typical cryogenic calorimeter. It consists of an absorber with heat capacity $C$, a thermometer and a thermal link with a heat conductance $g$ to a reservoir with temperature $T_B$. The thermometer is typically a thin superconducting strip that operates very close to the transition temperature between the superconducting and normal phases. A small local increase in the temperature may dramatically increase the electric resistance, and thus lower the current flowing through this circuit. The temperature increase needed for this to happen may be caused by phonons created by particles interacting in the absorber, which travel to the surface and break Cooper pairs. The extremely low temperature at which these detectors are operating is needed to reduce the thermal noise, which limits the size of the measurable signals.
The use of superconductors as cryogenic particle detectors is motivated by the very small binding energy of the Cooper pairs, $\sim 1$ meV, compared to 3.6 eV needed to create an electron--hole pair in silicon. Thus, compared to a semiconductor, several orders of magnitude more free charges are produced, which leads to a much higher intrinsic energy resolution (Figure \ref{pretzl}b).

Because of their sensitivity to very small energy deposits, cryogenic calorimeters are widely used in the search for dark matter, and in particular for low-mass 
WIMPs\footnote{Weakly Interacting Massive Particles.}. 
As a matter of fact, the most stringent limits on WIMPs with masses less than 5 GeV/$c^2$ come from cryogenic experiments such as CDMS \cite{Agn18}, EDELWEISS \cite{Arm17} and CRESST \cite{Pet17}, which all operate bolometric detectors with masses in the 10 kg range at temperatures well below 1 K, in tunnels or deep mines. Plans for upgrades to 100 kg or more exist in all cases. An example of a cryogenic calorimeter looking for neutrinoless $\beta \beta$ decay is CUORE \cite{Ald17}, which is building a detector containing 740 kg worth of TeO$_2$ crystals (\ie 240 kg of the isotope of interest, $^{130}$Te), which will operate at  a temperature of 15 mK in the Gran Sasso Laboratory. Superconducting Tunnel Junctions have found useful applications in astronomy, where they are used to detect radiation in the sub-mm wavelength domain.

There is a considerable amount of effort going into the development of these and many related, 
similarly ingenious devices. However, this highly specialized work falls somewhat outside the
scope of this paper. The interested reader is referred to reviews of this field that can be found in 
\cite{Pre00,Ens05}. A recent review on dark matter searches is given in \cite{Rot17}.

\section{Outlook}

The history of physics in general, and of
nuclear and particle physics in particular, is filled with examples that prove that
measurement precision pays off. A better, more accurate instrument allows more precise
measurements. More precise measurement results make it possible to discover new phenomena,
and/or to better understand old ones. Better understanding of the physical world has always been 
a crucial element in the evolution of mankind. 

The history of calorimetry itself illustrates this process in a nutshell. In particle physics, calorimeters were
originally designed as crude, cheap instruments for some specialized applications
(for example, detection of neutrino interactions).
The original literature is testimony to the fact that their performance was often perceived as
somewhat mysterious by their users.
Only after the physics on which calorimeters are based was better understood did it become
possible to develop these detectors into the precision instruments that they are nowadays
and which form the centerpiece of many modern experiments in particle physics and related fields such as astroparticle physics.

In the past half-century, the study of the fundamental structure of matter has greatly benefited from the enrichment of the arsenal of experimental techniques brought about by calorimetry. There are many examples of discoveries in which different features
of the calorimetric performance have made a crucial difference:
\begin{itemize}
\item The discovery of the intermediate vector bosons \cite{Arn83a,Ban83} was possible thanks to the capability of the
calorimeters to recognize and select events in which large-$p_\perp$
particles were produced
on-line, amidst a background that outnumbered these events by some seven orders of magnitude.
\item The observation of $\nu$-oscillation effects in SuperKamiokande \cite{Kaj99} was possible thanks to some
special features that derived from the use of \v{C}erenkov light as the
source of experimental information in this calorimeter, namely ($a$) the possibility to distinguish between low-energy
electrons and muons, and ($b$) the capability to reconstruct the direction of the particles that
generated the light cones.
\item The experimental observation of the Higgs boson \cite{Aad12,Chat12} became possible
thanks to the capability of the calorimeters to detect neutral particles ($\gamma$s) and to measure
their four-vectors with very high precision.
\item The discovery of the top quark by D0 was made possible by the capability of the calorimeter
and its auxiliary systems to recognize charged leptons that were part of jets \cite{Abb99}.
\item The unraveling of the details of direct 
$CP$-violation in the $K^0$ system was made possible by the excellent mass resolution for $\pi^0$s in the 
calorimeters of the NA31 \cite{Barr93}, KTeV \cite{Ala99} and KLOE \cite{Des09} experiments.  
\end{itemize}

As the energy at which the study of the structure of matter is further increased,
exclusive hadronic decay modes become very complicated (they involve many final-state particles) and each
represents only a tiny fraction of the decays. For example, the decay of some charmed and bottom mesons proceeds 
through more than one hundred different channels that have a measurable branching fraction. 
Increasingly, the primary process at the quark level is of great interest, rather than the
precise composition of the final state. For example, the hadronic decay of a $W$ boson proceeds through the
process $W \rightarrow q\bar{q}$. In order to demonstrate that hadronically decaying $W$s
were produced in certain reactions, it is much more important to be able to reconstruct these two
jets and determine their invariant mass than to know the characteristics of all individual
particles that make up the final state.
   
Thirty years ago, intermediate vector bosons were the objects of intense experimental searches.
Now, they may become the key to discovering new phenomena at the 0.1--1.0 TeV mass scale, and
multi-quark spectroscopy may turn out to be an invaluable tool for this.
The experimental successes listed above relied primarily on the precision with which calorimeters could identify and measure the properties
of particles that developed em showers. The contribution of hadron calorimeters was mainly limited to the measurement of energy flow variables, such as (missing) transverse energy. However, for multi-quark spectroscopy the hadron calorimeters will need to provide information on the four-vectors of fragmenting 
quarks and gluons with a similar precision as we have become used to for electrons and $\gamma$s. The calorimeter systems used in the present generation of (LHC) experiments are falling far short in that respect. New concepts to improve the precision show encouraging results (\eg Figures \ref{rd52}, \ref{sehwook4}), but there is a long way to go before these concepts can be incorporated in a $4\pi$ detector at a colliding-beam experiment. The new endcap detectors of CMS will provide the first opportunity to test claims that the PFA approach is capable of delivering the required precision in practice, and not only in simulations.

The quality of the hadronic calorimetry may thus well turn out to be very important for the study of
phenomena beyond the Standard Model, assuming that there is physics beyond this Model that
can be studied with the available tools.

Calorimeters are also very important detectors for experiments at existing and newly emerging facilities for hadron physics research in the intermediate energy range (\ie up to $\sim 100$ GeV). 
This research area includes facilities like CEBAF (Jefferson Lab, Newport News) and RHIC (Brookhaven) in the US, the facility for antiproton and ion research ( FAIR) at GSI in Germany, the proton accelerator complex J-PARC in Japan, the ion collider NICA at Dubna in Russia, and the new accelerator laboratories planned in China. 
Experiments at these facilities may also greatly benefit from the improvements in calorimeter quality described in this review. 

In the other mentioned fields in which calorimetric particle detection is used (Sections 9,10), the future looks bright.
The techniques used in cryogenic calorimeters are now maturing to the point where ton-scale experiments are feasible, and important contributions in the search for dark matter and neutrinoless $\beta \beta$ decay may be envisaged.

The upgrade of SuperKamiokande to Hyperkamiokande will make this detector the prime instrument for studying neutrinos in the world, both from astrophysical sources (solar, supernova and atmospheric neutrinos), as well as for manmade neutrinos (from particle beams or reactors). In this process, we may expect to learn a lot, not only about astrophysical phenomena, but also about the neutrinos themselves (masses, mass differences, mass hierarchy) and address some of the most important unresolved mysteries of nature. Hyperkamiokande will also become by far the most sensitive detector to study proton decay and, if unobserved, increase the limits on this process by two orders of magnitude.

Detectors such as IceCube and HESS are already detecting neutrinos and $\gamma$s with energies in the PeV domain. This is interesting since the trajectory of these particles is not affected by magnetic fields,
and therefore one may hope to localize the origin of these objects and, possibly, the mechanism through which they acquire these extremely high energies. Further progress in this field may be expected when large new projects, such as CTA and KM3, take off.

In summary, calorimetric detection techniques continue to provide the keys to further progress in our understanding of the most fundamental aspects of the physical world. However, a word of caution is in place with regard to the extremely complicated and costly calorimeter systems that form the heart and soul of 4$\pi$ experiments at modern particle colliders. As illustrated by some of the examples in Section 6, mistakes made in the design, testing or commissioning of the instruments can be very costly and have serious implications. Many of these mistakes are avoidable, or can be corrected when discovered sufficiently early. 
The importance of extremely thorough and unbiased prototype testing can thus not be overstated.

\section*{Acknowledgements}

This study was carried out with financial support of the United States
Department of Energy, under contract DE-FG02-12ER41783.
The author would like to thank John Hauptman and Michele Livan for their thorough reading of this manuscript 
and for their invaluable suggestions which have found their way into it.

\end{document}